\documentclass[12pt,a4paper,twoside]{book}
\usepackage[utf8]{inputenc}
\usepackage{amsmath}
\usepackage{amsfonts}
\usepackage[english]{babel}

\allowdisplaybreaks 
\usepackage{lmodern}
\usepackage[nottoc]{tocbibind} 	
\usepackage{pdfpages}			
\usepackage{todonotes}
\usepackage{epigraph}			
\usepackage{subfigure}
\usepackage{fancyhdr}			
\usepackage[babel, titelside, en, nat]{ku-forside} 
\usepackage{wallpaper}			

\usepackage{geometry}			

\usepackage[format=plain, indention=0cm ,labelsep=period,singlelinecheck=false,margin=10pt, font={small}, labelfont={bf}]{caption} 
\usepackage{booktabs}
\usepackage{tabularx}

\definecolor{kugreen}{RGB}{50,93,61}
\renewcommand{\headrulewidth}{0.4pt}

\renewcommand{\footrulewidth}{0pt}
\newcommand{\fancypages}{
\renewcommand{\headrulewidth}{0.4pt}
\fancyhead[RO,LE]{\slshape }
\fancyhead[RE]{\slshape \leftmark}
\fancyhead[LO]{\slshape \rightmark}
\fancyfoot[C]{}
\fancyfoot[LE]{\thepage}
\fancyfoot[RO]{\thepage}
}
\fancypagestyle{plain}{%
\fancyhf{} 
\fancyfoot[C]{}
\fancyfoot[LE]{\thepage}
\fancyfoot[RO]{\thepage}
\renewcommand{\headrulewidth}{0pt}
\renewcommand{\footrulewidth}{0pt}}

\newcommand{\paperpages}[1]{
\fancyhead[RE]{\slshape \leftmark}
\fancyhead[RO,LE]{{}  {}  	\thepage}

\renewcommand{\footrulewidth}{0pt}
\fancyfoot[]{}  
}

\newcommand{\blankpage}{\newpage
\thispagestyle{empty}
\mbox{}}

\newcommand{\blurbpage}[1]{\blankpage \chapter*{#1}   \thispagestyle{empty}}

\newcommand{\abstractP}[2]{  \pagestyle{empty} \cleardoublepage
\begin{center} \begin{Large} \bf
#1
\end{Large} \end{center}
\vspace{50pt}
\noindent 
#2
\clearpage
}

\newcommand{\secondtitle}{
{
\ThisURCornerWallPaper{1}{KUNATWmark_edit.pdf}
\begin{center}
\Huge \bf
\tyk \TITEL  \\ 
\vspace{30pt}
\Large 
\UNDERTITEL
\normalfont \normalsize \\ \vspace{50pt}
\tynd by \\ \vspace{50pt}
\Large 
\FORFATTERET \\ 
\end{center}
\begin{flushright}
\vfill 
 \large \it
This thesis has been submitted to the \\
Niels Bohr Institute\\
Faculty of Science, University of Copenhagen\\ 
Blegdamsvej 17, DK-2100 Copenhagen \O, Denmark
\thispagestyle{empty}
\normalfont \normalsize 
\end{flushright}
}\myfont
 }

\newcommand*{\myfont}{\fontfamily{lmr}\selectfont}

\includepdfset{pagecommand=\thispagestyle{empty},frame=false,noautoscale=true,offset= 10 0 }

\usepackage{etoolbox}
\usepackage{tikz}

\usepackage{framed}

  \newcommand*\quotefont{\fontfamily{lmr}} 

\newcommand*\quotesize{60} 
\newcommand*{\openquote}
   {\tikz[remember picture,overlay,xshift=-4ex,yshift=-2.5ex]
   \node (OQ) {\quotefont\fontsize{\quotesize}{\quotesize}\selectfont``};\kern0pt}

\newcommand*{\closequote}[1]
  {\tikz[remember picture,overlay,xshift=4ex,yshift={#1}]
   \node (CQ) {\quotefont\fontsize{\quotesize}{\quotesize}\selectfont''};}

\colorlet{shadecolor}{white}

\newcommand*\shadedauthorformat{\emph} 

\newcommand*\authoralign[1]{%
  \if#1l
    \def\authorfill{}\def\quotefill{\hfill}
  \else
    \if#1r
      \def\authorfill{\hfill}\def\quotefill{}
    \else
      \if#1c
        \gdef\authorfill{\hfill}\def\quotefill{\hfill}
      \else\typeout{Invalid option}
      \fi
    \fi
  \fi}
\newenvironment{shadequote}[2][l]%
{\authoralign{#1}
\ifblank{#2}
   {\def\shadequoteauthor{}\def\yshift{-2ex}\def\quotefill{\hfill}}
   {\def\shadequoteauthor{\par\authorfill\shadedauthorformat{#2}}\def\yshift{2ex}}
\begin{snugshade}\begin{quote}\openquote}
{\shadequoteauthor\quotefill\closequote{\yshift}\end{quote}\end{snugshade}}

\usepackage[draft=false]{hyperref}   

\newcommand{\md}{\mathrm{d}}
\newcommand{\tZ}{\tilde Z}
\newcommand{\pg}{\pi^{\gamma}}
\newcommand{\g}{\gamma}
\newcommand{\DT}{\text{DT}}
\newcommand{\CDT}{\text{CDT}}
\newcommand{\CS}{\text{CS}}

\newcommand{\mc}[1]{\mathcal{#1}}

\newcommand{\av}[1]{\left\langle #1 \right\rangle}
\newcommand{\bra}[1]{\langle #1|}
\newcommand{\ket}[1]{ |#1 \rangle}

\newcommand{\pd}[2]{\frac{\partial #1}{\partial #2}}

\newcommand{\triI}{\begingroup
\setbox0=\hbox{\includegraphics[height=12pt]{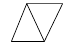}}%
\parbox{\wd0}{\box0}\endgroup}

\newcommand{\triII}{\begingroup
\setbox0=\hbox{\includegraphics[height=12pt]{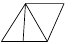}}%
\parbox{\wd0}{\box0}\endgroup}

\newcommand{\triIII}{\begingroup
\setbox0=\hbox{\includegraphics[height=12pt]{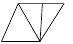}}%
\parbox{\wd0}{\box0}\endgroup}

\newcommand{\p}{\prec}
\newcommand{\peq}{\preceq}
\newcommand{\pl}{<}

\newcommand{\Nmd}{\langle N_m^d \rangle}

\newcommand{\Nmdv}[2]{\langle N_{#1}^{#2} \rangle}

\newcommand{\Nzerod}{\langle N_0^d \rangle}

\newcommand{\suml}[2]{\sum\limits_{#1}^{#2}}
\newcommand{\intl}[2]{\int\limits_{#1}^{#2}}

\newcommand{\dtwo  }{\frac{d}{2} }
\newcommand{\twod  }{\frac{2}{d}  }
\newcommand{\dodd}{{d_{\text{odd}}}}
\newcommand{\deve}{{d_{\text{even}}}}

\newcommand{\G}[1]{\Gamma\left(#1 \right)}
\newcommand{\mFm}[4]{\, _{#1}F_{#1}\!\left( \genfrac{}{}{0pt}{}{#2}{ #3  }  \bigg| #4 \right) }

\newcommand{\Poch}[2]{\left( #1 \right)_{#2}}

\newcommand{\Od}[1]{\mathcal{O}_{#1}}
\newcommand{\bigspace}{\ \ \ \ \ \ \ \ \ \ \ \ \ \ \ }
\newcommand{\bigspaced}{\bigspace \bigspace }

\newcommand{\NmCC}{N_m(C)}

\newtheorem{claim}{Claim}
\newtheorem{conjecture}{Conjecture}
 
\newtheorem{definition}{Definition}

\graphicspath{ {papers/}{figures/}{figures/CDT/}{figures/EDT/}{figures/Dimer/}{figures/HL/}{figures/Local/}{figures/Factor/}{figures/CST/} }

\title{Endeavours in Discrete Lorentzian Geometry} 
\undertitel{A thesis in five papers}
\opgave{Ph.D. Thesis} 
\author{Lisa Glaser}
\forfatter{}
\forfatterET{Lisa Glaser}
\forfatterTO{}
\dato{June 20,2014}%
\vejleder{Supervisor: Jan Ambj\o rn } 

\begin{document}
\pagenumbering{Alph}
\maketitle

\blankpage
\blankpage

\secondtitle

\abstractP{Abstract}{
To solve the path integral for quantum gravity, one needs to regularise the space-times that are summed over.
This regularisation usually is a discretisation, which makes it necessary to give up some paradigms or symmetries of continuum physics.

Causal dynamical triangulations regularises the path integral through a simplicial discretisation that introduces a preferred time foliation.

The first part of this thesis presents three articles on causal dynamical triangulations.
The first article shows how to obtain a multicritical $2$d model by coupling the theory to hard dimers.
The second explores the connection to Ho\v{r}ava-Lifshitz gravity that is suggested by the time foliation and establishes that in $2$d the theories are equivalent.
The last article does not directly concern causal dynamical triangulations but Euclidian dynamical triangulations with an additional measure term, 
which are examined to understand whether they contain an extended phase without the need for a preferred time foliation.

Causal set theory uses an explicitly Lorentz invariant discretisation, which introduces non-local effects.

The second part of this thesis presents two articles in causal set theory.
The first explicitly calculates closed form expressions for the d'Alembertian operator in any dimension, which can be implemented in computer simulations.
The second develops a ruler to examine the manifoldlikeness of small regions in a causal set, and can be used to recover locality.

}

\blurbpage{Declaration}
This thesis presents my original research work.
Wherever contributions of others are involved, every effort is
made to indicate this clearly, with due reference to the literature,
and acknowledgements of collaborative research and discussions.
The work was done under the guidance of Professor Jan Ambj\o rn at the Niels Bohr Institute.

\vspace{150pt}
\hfill\parbox{7cm}{
      \flushright
      \rule{5cm}{1pt}\\
       Lisa Glaser
    } 

\blurbpage{Acknowledgements}

I would like to thank my supervisor Jan Ambj\o rn for giving me the freedom to work on anything I chose and travel to any conference I liked.
I would not be where I am today if not for Fay Dowker, who convinced me that space-time could be discrete and then continued to nurture my interest in this.
Sumati Surya deserves thanks for inviting me to India for two fantastic and productive stays.

Thanks to Rafael Sorkin and Joe Henson for agreeing to assess this dissertation.

I would also like to thank Renate Loll, David Rideout, David Meyer, Bianca Dittrich, Danielle Oriti, Jerzy Lewandowski, Astrid Eichhorn and all those I forgot now for inviting me for visits and great discussions.
Special thanks go to Andrzej G{\"o}rlich and David Rideout for explaining their computer codes to me and in course making me a better programmer.

I also want to thank the whole HET group at NBI for lunch time discussions, cookie breaks and all the other interactions that fill out a working day.

I owe a bubble tea to Laura and Joyce for our weekly bubble tea meetings, which were occasion to vent about big and small frustrations, and sometimes even to enjoy the weather.

Thomas I owe thanks for moral support, and last but not least I thank my parents,  for their encouragement and patience.

This version of the thesis contains some corrections and comments made to me by my comitee Charlotte Kristjansen, Joe Henson and Rafael Sorkin.
To reduce the length of the document I replaced the original papers at the end with links to their arXiv versions.

\vspace{50pt}
\begin{flushright}
\begin{minipage}[l]{0.55\textwidth}
\begin{shadequote}[c]{Faust I, Vers 382 }
Dass ich erkenne, was die Welt \\ Im Innersten zusammenh\"{a}lt. 
\end{shadequote}
\blankpage
\end{minipage}
\end{flushright}

\setcounter{tocdepth}{1} 
\tableofcontents
\pagenumbering{arabic}
\pagestyle{fancy}
\fancypages

\chapter{Quantising space and time}
One of the big open challenges in contemporary physics is to discover a theory of gravity that is valid at the smallest lengths and highest energies.
Einstein's theory of general relativity describes our experiments very well on scales from below a millimetre up to the size of the universe, assuming the existence of dark matter  \cite{PDG}. 
Gravitational measurements at smaller scales are complicated by gravity's relative weakness compared to the other forces.
However, theoretical consistency demands that the theory changes at smaller lengths / higher energies. 
At the latest at the Planck scale $E_{pl}=1.22\cdot 10^{19} \mathrm{GeV}$ we expect to detect effects of quantum gravity.
Two particles of Planck mass $m_{pl}$ and elementary electric charge $e$ are attracted to each other by gravity as strongly as they are repelled from each other by electromagnetism.
Thus at this energy scale gravity and the other forces should be treated on equal footing.

In general relativity, space and time are the coordinates of a $3+1$ dimensional manifold.
The shape of this space-time manifold is determined by the distribution of matter and energy.
In this world view, the passage of time is just perceived because we travel through this fixed manifold along a timelike geodesic.

In quantum field theory, the space-time manifold is usually treated as a background.
Most often it is assumed to be flat.
Even when quantum field theory considers curved space-time manifolds, these are treated as an unchanging background on which the fields propagate, not a dynamic object they interact with.

This backreaction of quantum fields on the space-time is very hard to describe.
At lower energies one can describe an effective theory where gravity couples to the expectation value of the stress-energy tensor of quantum matter\cite{Hack:2012qf}, but this description will break down at higher energies and one will need to quantise space-time.
Yet we have no idea what a quantised space-time should look like.
Many approaches to solve this problem are being explored, and most of them postulate that quantised space-time is a complicated object.

Some problems in quantum gravity can be traced back to the static nature of space-time in general relativity.
A canonical quantisation of general relativity will lead to a fixed universe.
After all, that is what the classical theory describes.

Other problems arise through our understanding of quantum theory.
How can we apply quantum theory, most often formulated in terms of observers and repeated independent measurements, to the universe as a whole?
We only have one universe, which makes it hard to repeat the experiment.
And we are part of the universe we cannot step out of it to take a measurement as an outside observer.

The combination of gravity and quantum theory thus shines a spotlight on issues that arise in both theories but can be circumvented within each theory itself.
The full magnitude of these issues only becomes apparent in our attempts to combine quantum theory and gravity.

Despite, or maybe because of, these and other problems, the search for a quantum theory of gravity is a very active area of research.
There are many approaches, starting from various points and using different tools, all vying for a theory of reality.

Some examples are:
\begin{itemize}
\item {\bf String theory} 

Quantum field theory is based on the concept of point particles.  
One idea to alleviate the need for renormalisation is to replace these points by strings.
Quantising the vibrational modes of these strings gives rise to a whole spectrum of particles, among them the spin-2 graviton  \cite{Polchinski:1998rq,Polchinski:1998rr}. 
A quantised spin 2 field is a quantum theory of gravity, which kindled the hope that this theory might describe our world. 
To make the theory consistent, one needs to introduce supersymmetry and postulate $9+1$ space-time dimensions.
It was projected that the LHC would detect either supersymmetry or extra dimensions.
Unfortunately, during the first run of the LHC, there were no signs of either\cite{Lust:2013koa,PDG}. 

\item {\bf Loop quantum gravity} 

In this approach, gravity is quantised in the Hamiltonian formalism.
The Einstein Hilbert action is rewritten in so-called Ashtekar connection variables  \cite{Ashtekar:2004eh} and then canonically quantised  \cite{lrr-2008-5}. 
The diffeomorphism symmetry of gravity gives rise to a system of constraints that does not allow for evolution in time.
Ideas for solving this problem are to quantise a reduced phase space  \cite{Thiemann:2004wk} or to use matter fields as clocks  \cite{Brown:1994py}, there is no consensus on these questions.
A spin-off of loop quantum gravity is loop quantum cosmology, which attempts to use simplified models of loop quantum gravity to make cosmological predictions \cite{Ashtekar:2006uz}. 
\pagebreak
\item {\bf Spin foams}

Spacelike hypersurfaces in loop quantum gravity can be described as spin network states.
A spin network is a triangulation of a three-dimensional hypersurface that is decorated with additional spin labels.
Barrett and Crane quantised BF-theory, a topological field theory, by introducing spin labels on four-dimensional triangulations \cite{Barrett:1997gw}. 
It was quickly pointed out that these four-dimensional triangulations with spin labels can be interpreted as time developments of the three-dimensional spin-networks of loop quantum gravity.
Even though this connection has not been made precise yet, spin foams are a very active area of research  \cite{lrr-2013-3}.

\item {\bf Dynamical triangulations} 

In  $d$-dimensional dynamical triangulations, the space-time geometry is approximated through equilateral $d$-simplices. 
The interior of these consists of flat $d$-dimensional space.
Curvature can arise through the number of simplices meeting at a face of codimension two.
In $2$ dimensions, such a face is a point. 
If six triangles meet at one point, it is flat, while more or less triangles encode negative or positive curvature respectively. 
These piecewise linear geometries are then used to regularise the path integral for gravity. 
In $2$d, dynamical triangulations can be solved through matrix models, and in higher dimensions the phase space has been examined through Monte Carlo simulations.
These simulations have shown that dynamical triangulations in four dimensions do not contain a $2$nd order phase transition and thus no continuum phase  \cite{Ambjorn:1999nc}.

\item {\bf Tensor track}

Tensor models are a generalisation of matrix models to higher dimensions \cite{PROP:PROP201300032}. 
Expanding the path integral for a $d$-tensor potential gives rise to Feynman graphs that are duals to a $d$-dimensional space-time. 
In their pure form, tensor models are an attempt to solve the partition function for random geometries, like dynamical triangulations, by analytic means.
These analytic expressions allow for scaling limits which have lead to the conjecture that a continuum phase might be obtained by repeated scaling  \cite{Bonzom:2011zz}.
Tensor models can be generalised by adding group variables to the faces of the geometries these group field theories are then closely related to spin foams.

\item {\bf Causal dynamical triangulations} 

Causal dynamical triangulations are a modification of dynamical triangulations.
In this approach, the simplices have different edge lengths for spacelike or timelike edges, and only configurations that allow for a time-foliation are included in the path integral.
This makes it possible to recover an average geometry that shows continuum characteristics \cite{Ambjorn:2012jv}.

\item {\bf Ho\v{r}ava-Lifshitz gravity} 
\nopagebreak

The Einstein-Hilbert action is not perturbatively renormalisable.
Newton's coupling has positive mass dimension and this introduces new divergences at every loop order.
In Ho\v{r}ava-Lifshitz gravity a Lifshitz-type scaling, which is an anisotropic scaling in time and space, is introduced.
One can tune this scaling to make the coupling dimensionless.
The theory is then power counting renormalisable. 
The scaling also breaks the diffeomorphism symmetry of general relativity to the group of foliation preserving diffeomorphisms. 
This new, smaller symmetry group allows a new potential for gravity which introduces Lorentz violating effects.

\item {\bf Asymptotic safety}

The asymptotic safety programme approaches the problem of the renormalisability of gravity from a Wilsonian point of view.
The idea is that although gravity is not perturbatively renormalisable its renormalisation group flow might lead to a non-Gaussian fixed point.
Such a non-Gaussian fixed point cannot be investigated using the perturbative methods that are mostly used in quantum field theory, but the development of improved renormalisation group techniques makes is possible to examine it further \cite{Reuter:2012id}.

\item {\bf Causal set theory} 

The idea of causal sets as a fundamental structure is based on the proof by Hawking and Malament  that space-time can be almost completely characterised by its causal structure  \cite{Hawking:1976fe,malament:1399}. 
Based on this one can use a partially ordered set, with the partial order induced through the causal relations, to describe space-time  \cite{Bombelli:1987aa}.
\end{itemize}

This list is not exhaustive or ordered by merit there are many other interesting approaches. 
Yet, there are so far no indications that any approach truly solves the problem of quantum gravity in a way considered satisfying by a majority of physicists.
The main reason is that none of these has yet given rise to quantitative predictions that have been verified by experiments
\footnote{There is the prediction of the cosmological constant made in Causal Set Theory, however it predicts  just an order of magnitude and does need several assumptions.
It is thus not considered as a smoking gun signature by a majority of physicists.}.
There are a number of ideas how to test quantum gravity, but until now they all came up negative.

Each of these theories is worthy of study and to decide which to pursue is not a trivial task.
One may call upon symmetries and principles of physics to guide this decision.
Guided by the search for the largest possible symmetry and a unified description, one might chose to work on string theory.
While someone attempting to stay close to known canonical quantisation  schemes  would be drawn to loop quantum gravity.

The overarching idea I pursued is that space-time should be Lorentzian.
Only in Lorentzian manifolds can one define a causal structure.
The causal structure of space-time has important implications, some of which will be examined in the course of this thesis.
I thus decided to investigate causal dynamical triangulations and causal set theory, with some projects building bridges to dynamical triangulations and Ho\v{r}ava-Lifshitz gravity.

A Wick rotated, Euclidian theory is easier to handle, so it has merit as a tool for calculations.
Any result from such a calculation would, however, need to be Wick rotated back.
If no such `inverse' Wick rotation exists, it seems unlikely that the result should be connected to phenomenology.

In causal dynamical triangulations, the explicit time foliation does allow one to define a Wick rotation.
It is then possible to compare the results obtained in this Euclidianised theory with the Lorentzian reality.
Causal dynamical triangulations is a prime example of how to use Wick rotations as a calculational tool.

Causal set theory does not allow for a Wick rotation. 
The theory is fundamentally Lorentzian, as the partial order is constructed from timelike distances. 
In an Euclidian space, there are no timelike and spacelike distances and the causal order cannot be defined.
In causal set theory, one thus works completely in the Lorentzian description. 

This thesis consist of four chapters.
This first one gives an overview of different approaches to quantum gravity and motivates the choice of approaches in this thesis.

In the second chapter, I lay out my work on causal dynamical triangulations, it is split into five sections.
In the first section, I introduce causal dynamical triangulations. 
I also sketch the basics of critical phenomena and Monte Carlo simulations, which will  be used to examine the theory.
I demonstrate how $2$d causal dynamical triangulations can be solved using the transfer matrix approach and illuminate the phase diagram of causal dynamical triangulations.

In the second section, I present my work on multicritical causal dynamical triangulations  \cite{Ambjorn:2012zx}. 
There, I explain how we coupled dimers to causal dynamical triangulations using a bijection between causal dynamical triangulations and rooted tree graphs. 
I further give account of how these results have been verified using matrix models.

In the third section, I establish the connection between causal dynamical triangulations and Ho\v{r}ava-Lifshitz gravity.
After a short introduction to Ho\v{r}ava-Lifshitz gravity, I explain the reasons why such a connection was conjectured and present my own work that shows that causal dynamical triangulations and projectable Ho\v{r}ava-Lifshitz gravity in $2$ dimensions are the same \cite{Ambjorn:2013joa}.

The fourth section is about simulations in Euclidian dynamical triangulations.
There we examine whether a new measure term leads to a new phase \cite{Ambjorn:2013eha}.
I conclude the chapter with a short summary and a view to the future of the field of causal dynamical triangulations.

In my third chapter, I present my work on causal set theory.
In the first section, I recount the development of the field and introduce key results.
I focus, in particular, on the Hauptvermutung of causal set theory, the path integral as a sum of partial orders, and particle propagation on causal sets. 

In the second section, I introduce the definition of a d'Alembertian operator in higher dimensions and how to construct a causal set theory action using it.
I then sketch my calculations on this problem  \cite{Glaser:2013xha}.

In the third section I introduce the problem of locality in causal set theory, after which I present my work on defining a local region in causal set theory  \cite{Glaser:2013pca}.
I again conclude with a summary and an outlook to the prospects of the field.

In the last chapter, I synthesise the lessons learned and try to anticipate developments in the field of quantum gravity.


\cleardoublepage

This thesis is based on the following five peer-reviewed publications, which are collected in Appendix \ref{app:papers}.
\begin{itemize}
\item[\ref{app:Dimer}] 
\newblock {\bf New multicritical matrix models and multicritical 2d CDT}.\\
Jan Ambj{\o}rn, Lisa Glaser, Andrzej G{\"o}rlich, and Yuki Sato. \\
\newblock {\em Phys.Lett.}, B712:109--114, 2012.
\item[\ref{app:HL}]
\newblock {\bf 2d CDT is 2d Ho\v{r}ava-Lifshitz quantum gravity}.\\
Jan Ambj{\o}rn, Lisa Glaser, Yuki Sato, and Yoshiyuki Watabiki.\\
\newblock {\em Phys.Lett.}, B722:172--175, 2013.
\item[\ref{app:EDT}]
\newblock {\bf Euclidian 4d quantum gravity with a non-trivial measure term}.\\
J.~Ambj{\o}rn, L.~Glaser, A.~G{\"o}rlich, and J.~Jurkiewicz.\\
\newblock {\em Journal of High Energy Physics}, 2013(10):1--24, 2013.
\item[\ref{app:Factor}]
\newblock {\bf A closed form expression for the causal set d’Alembertian}.\\
Lisa Glaser.\\
\newblock {\em Classical and Quantum Gravity}, 31(9):095007, 2014.
\item[\ref{app:Local}]
\newblock {\bf Towards a definition of locality in a manifoldlike Causal Set}.\\
Lisa Glaser and Sumati Surya.\\
\newblock {\em Phys.Rev.}, D88:124026, 2013.
\end{itemize}

\chapter{Causal dynamical triangulations}
\section{\label{sec:CDTintro}Introduction to causal dynamical triangulations}
One way to approach the problem of quantum gravity is to solve the gravity path integral, as introduced by Feynman  \cite{Feynman:1996kb}
\begin{align}
\mc{Z} &= \int \mc{D}[g_{\mu \nu}] \;  e^{\frac{i}{\hbar} \mc{S}_{EH}[g_{\mu \nu}]}
\intertext{with}
\mc{S}_{EH}&=\kappa \int \md^{d+1} x \sqrt{-\det{g_{\mu \nu}}} (R -2 \Lambda) \;.
\end{align}
$R$ is the Ricci scalar, containing derivatives of the metric $g_{\mu \nu}$, $\Lambda$ is the cosmological constant, $\kappa= \frac{1}{16 \pi G_N}$ is the gravitational coupling constant, and $\hbar$ is Planck's constant.

This integral is difficult to solve, so several manipulations have to be made.
Our first step is a Wick rotation. 
Under a Wick rotation, time is transformed $t \to i \tau$.
This transformation is used to different effects in general relativity and quantum mechanics.

In general relativity, a Wick rotation changes the $3+1$-dimensional Lorentzian space-time metric into a four-dimensional Riemannian metric.
This procedure is not in general well defined, and its results can be coordinate dependent. 
In the cases where it is well defined, the positive-definite Riemannian metrics are often easier to handle than Lorentzian metrics and lead to new insights into the theory.
One example is the Wick rotation of the Schwarzschild black hole, which allows for an easy derivation of the black hole temperature\cite{Wald:1984rg}.

For a path integral formulation of quantum mechanics, the Wick rotation has an other effect.
Under this transformation, the factor of $i$ in the exponential is cancelled leading to
\begin{align}
\mc{Z}_E &= \int \mc{D}[g_{\mu \nu}] \;  e^{-\frac{1}{\hbar} \mc{S}_{EH}[g_{\mu \nu}]}
\intertext{with}
\mc{S}_{EH}&=-\kappa \int \md^{d+1} x \sqrt{\det{g_{\mu \nu}}} (R -2 \Lambda) \;.
\end{align}
This can be interpreted as the partition function of a statistical ensemble of geometries.
The Wick rotation turns our quantum theory of Lorentzian manifolds into a statistical theory of Riemannian manifolds.
In the remainder of this chapter, we mostly examine Wick rotated theories and will specifically remark whenever this is not the case.
Even after the Wick rotation, the path integral over geometries remains hard to solve.

\subsection{\label{sub:critic}Critical phenomena}
\begin{figure}
\subfigure[\label{fig:mag1}]{\includegraphics[width=0.45\textwidth]{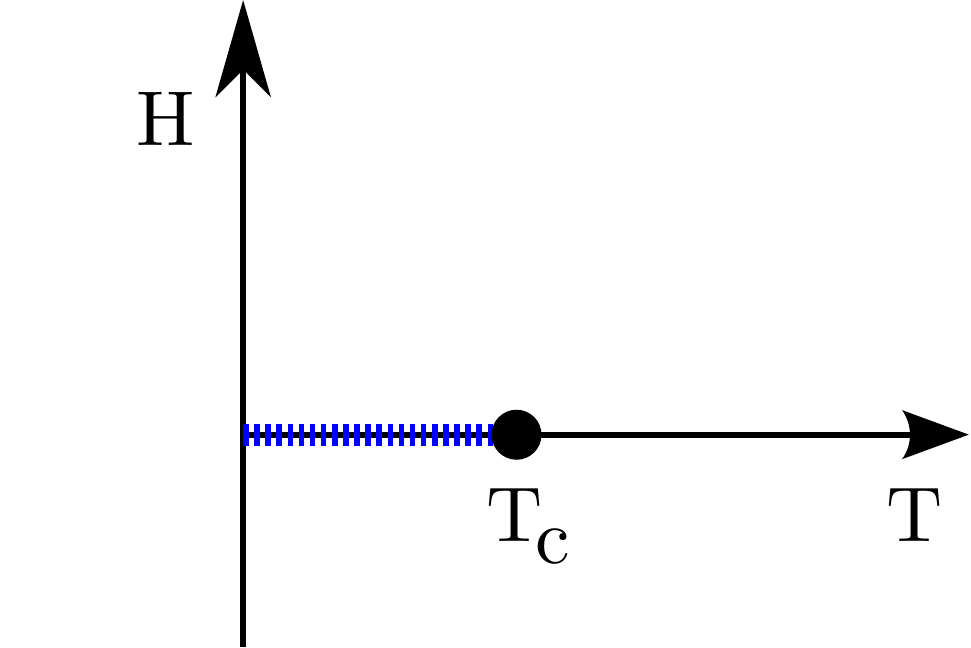}} 
\subfigure[\label{fig:mag2}]{\includegraphics[width=0.45\textwidth]{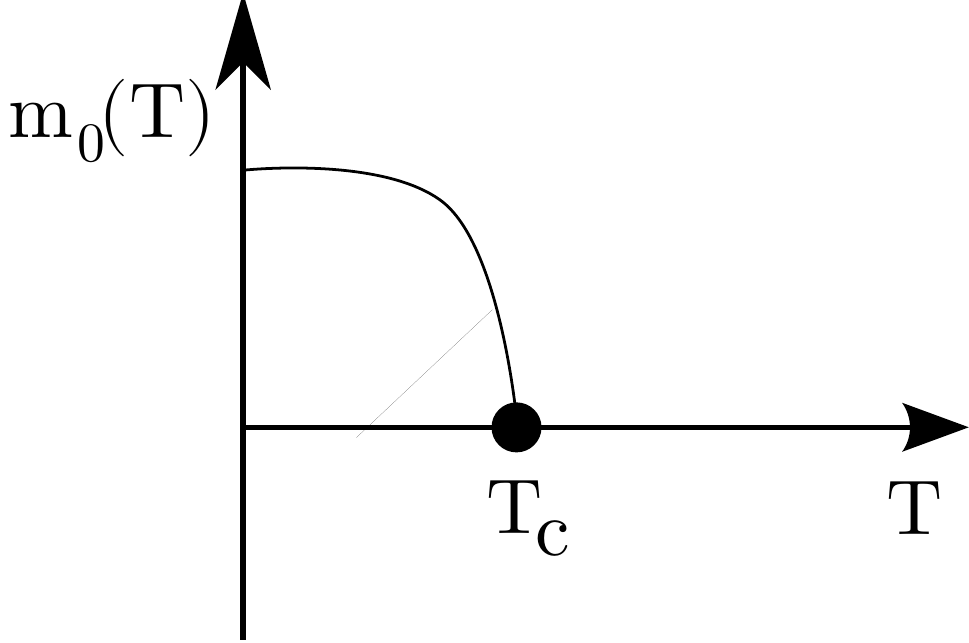}}
\caption{\label{fig:magneticphases} The left figure shows the phase diagram of a magnetic theory. The blue dashed line marks a discontinuity in the magnetisation $m$. This discontinuity ends at the critical point $T_c$. We could also define $H_c=0$. The right figure shows the absolute value of the spontaneous magentisation $m_0$ as a function of temperature. }
\end{figure}

To solve this, we discretise the geometries in our ensemble. 
To examine those discretised ensembles, we use methods from statistical physics, which will be introduced here.
A better overview over this topic is given in  \cite{CardyRG}.
We will use a magnetic system, with free energy
\begin{align}
F(V,H,T)=V f(H,T),
\end{align}
as an example.
Here $H$ is an external magnetic field, $T$ is the temperature, and $V$ is the volume of the system.
The phase diagram of the $(H,T)$ plane of such a system is shown in Figure \ref{fig:mag1}.
The blue line represents a cut from $T=0$ to $T=T_c$. 
This cut marks the phase transition, and its endpoint $(0,T_c)$ is a critical point.
In Figure \ref{fig:mag2}, we can see the spontaneous magnetisation $m_0(T)=m(H=0,T)$ as a function of $T$.

In the disordered high temperature phase, the correlation function, which measures the correlation between the orientations of the magnet in different regions of the system, is
\begin{align}
G (r) \approx \frac{1}{r^{d-2+\eta}} e^{-\frac{r}{\xi}},
\end{align}
where we introduce the correlation length $\xi$ and the anomalous scaling $\eta$. Both $\xi$ and $r$ are numbers, counting multiples of the lattice length $a$.

For the magnetic system close to the critical temperature $T_c$, we can define the critical exponents.
The critical exponents describe the simple power law behaviour physical quantities show in highest order close to the critical point.
We will expand in $t=\frac{T-T_c}{T_c}$ and $H$, where if we expand in $t$ we assume that we take $H=0$ after the derivative, and if we expand in $H$ we assume $t=0$ after the derivative
\begin{align}
c_v&=T \pd{^2 f}{T^2} \approx |t|^{-\alpha} && \text{from the heat capacity $c_v$, }\\
m&=\pd{ f}{H} \approx |t|^{-\beta}&& \text{and }\\
m &\sim H^{\sigma} && \text{from the magnetisation $m$,}\\
\chi &=\pd{^2 f}{H^2} \approx |t|^{-\gamma}&& \text{from the susceptibility  $\chi$, }\\
\xi & \sim |t|^{-\nu} && \text{from the correlation length $\xi$.}
\end{align}
These equations define the critical exponents $\alpha, \beta, \gamma, \sigma,\nu$.
Important for us is that for geometric systems, like random walks or random surfaces, $\nu$ is related to the Hausdorff dimension $1/\nu=d_h$.
The Hausdorff dimension of a geometry is defined as $V(r) \propto r^{d_h}$, where $V(r)$ is the volume of an open ball of radius $r$.

One interesting observation about these exponents is that the same numerical values arise in disparate systems.
So on the first glance very diverse models behave in the same way close to their critical point.
This universality can be explained using renormalisation group methods.
One can show that the universality class of a system only depends on its dimension and symmetries.

At the phase transition, the correlation length of the system diverges, and thus the theory becomes insensitive to the properties of the system at a smaller scale.
Therefore at the critical point, theories in the same universality class will approach the same continuum theory.
In the geometric systems of dynamical and causal dynamical triangulations, the scale of discretisation $a$ is taken to zero while the volume $V$ is kept fixed.
This continuum limit corresponds to taking the number of degrees of freedom $N \to \infty$.
Under this transformation, the physical correlation length $l$ of the system transforms as
\begin{align}
l \sim a \cdot \xi \;.
\end{align}
For any point other than the critical point, the correlation length $\xi$ is finite. 
Thus the physical correlation length goes to zero, and the physics washes out.
Only at the critical point, we have $\xi \to \infty$ and can then tune our continuum limit such that $ l \to const$.
This is why, we need a phase transition of at least second order to take a continuum limit.

If the phase transition is of $m$-th order, then along the phase transition line the first $m-1$ derivatives of the singular part of the free energy are zero
\begin{align}
\pd{^k f}{T^k}&=0 &\text{for} & & k &=1,\dots,m-1 \;,
\end{align}
and the $m$-th derivative of the singular part diverges.
This definition is true for $m>1$. A first order phase transition is characterised by a discontinuity of the free energy.
In general, a critical point can be of higher order than the phase transition on which it lies.
These multicritical points give rise to new critical exponents, and thus lie in a different universality class than the rest of the phase transition line.

\subsection{Dynamical triangulations}
In dynamical triangulations, space is discretised using equilateral simplices.
To distinguish dynamical triangulations from causal dynamical triangulations (CDT), they are often referred to as Euclidian dynamical triangulations (EDT).
Instead of integrating over all smooth geometries, one sums over all geometries built out of equilateral simplices.
These piecewise linear geometries are coordinate free, thus solving the problems associated with a choice of coordinate gauge.
The simplices are not considered as fundamental, but as regulators.

If one were to equip the space of continuous geometries with a suitable distance measure, the ensemble of piecewise linear manifolds, in the limit of vanishing edge length, should be dense in the space of smooth metrics.
In $2$d, there is a regular tiling of flat space with six triangles meeting at each vertex.
No such tiling exists in higher dimensions, since the dihedral angle $\theta_d$ of each d-simplex meeting at the $(d-2)$-simplex is given as $\theta_d=\arccos{\frac{1}{d}}$ which is not a rational fraction of $2 \pi$ for $d>2$.
Thus for $d>2$ we cannot build flat space out of equilateral simplices.

The interiors of the simplices are flat space, and all curvature is  concentrated at the $(d-2)$-simplices where they meet.
\begin{figure}
\centering
\includegraphics[width=0.8\textwidth]{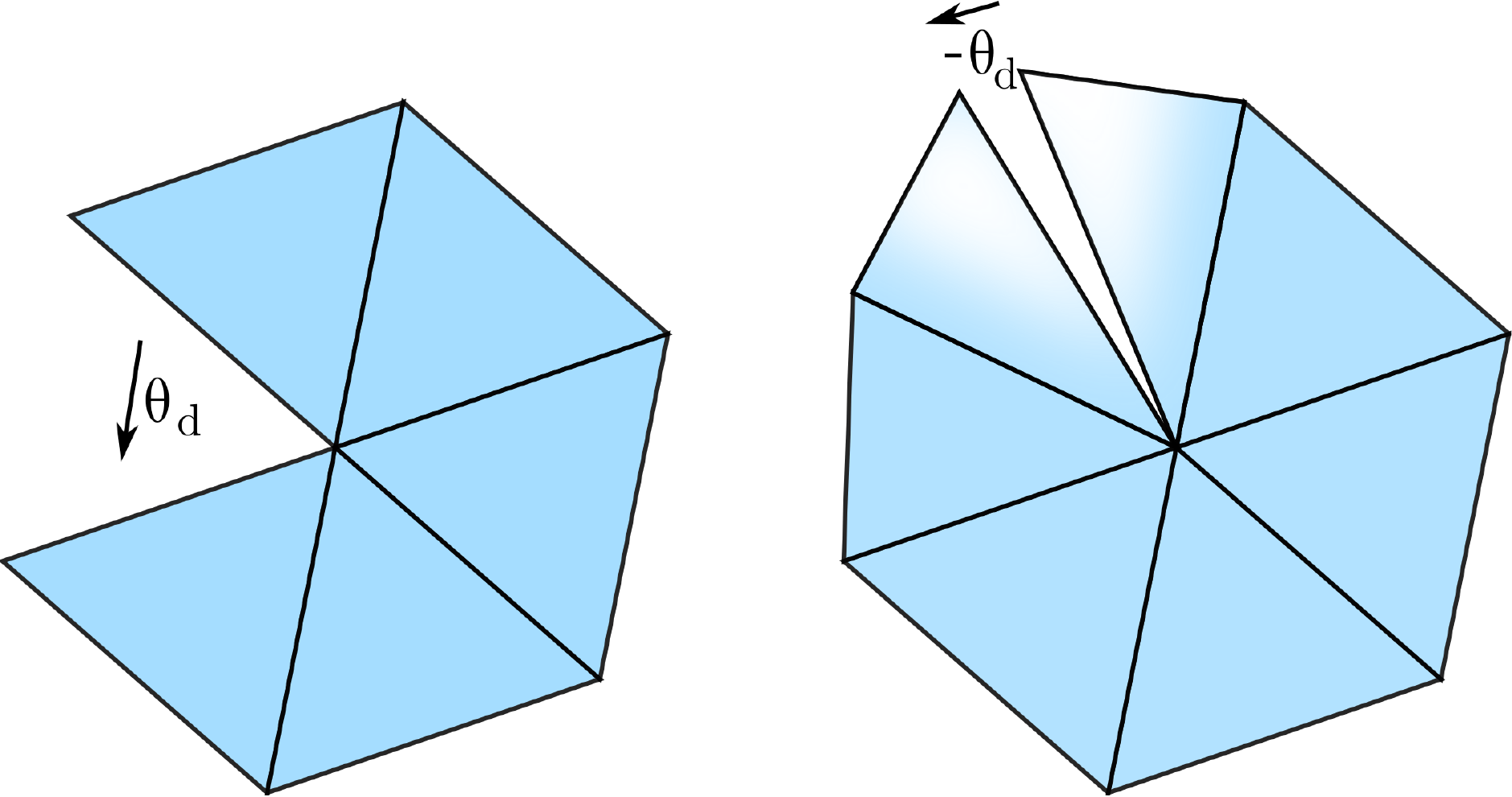}
\caption{\label{fig:deficit}In the left figure, only five triangles meet in one point. 
The result is a deficit angle $\theta_d$ and positive curvature at this point.
On the right hand side, we have seven triangles meeting in one point, two of them are `pointing up' to indicate that it is not possible to place seven equilateral triangles adjacent to one point in flat two-dimensional space. 
This leads to a deficit angle of $-\theta_d$ and negative curvature.}
\end{figure}
The deficit angle at simplex $v$ is defined as $\vartheta_d= 2 \pi -n \theta_d$ for a $(d-2)$ vertex where $n$ $d$-simplices meet. 
For $2$d, the deficit angle of triangles meeting at one vertex is illustrated in Figure \ref{fig:deficit}.

Using this simplicial discretisation, the integrated curvature can be determined by counting
\begin{align}
\int \md x^{d+1} \sqrt{\det{g_{\mu \nu}}} R  &\propto N_{d-2}  \\
\int \md x^{d+1} \sqrt{\det{g_{\mu \nu}}}    &\propto  N_{d} \;.
\end{align}
The integrated curvature of space is proportional to the number of $(d-2)$-simplices $N_{d-2}$, while the total volume of space-time is proportional to the number of $d$ simplices.
We can use this to write down the actions,
\begin{align}
\mc{S}_{\DT\, 2d}&=  N \Lambda \label{eq:2dCDTaction}\\
\mc{S}_{\DT\, 4d}&= - \kappa_2 N_2 + \kappa_4 N_4 \;. \label{eq:DTaction}
\end{align}
Clearly the two-dimensional action is a special case, since in $2$d the integral over the curvature is a topological invariant.
Thus, for fixed topology the only contribution is a volume term.

The space of $2$d surfaces is well enough understood to reduce the path integral to a counting problem.
This allows us to solve the path integral in $2$d in several ways (see  \cite{jansbook} for a pedagogical review).
In $4$d, the space of geometries is not well enough understood to obtain analytic solutions to the path integral; instead we run computer simulations to examine the phase space of dynamical triangulations.

In fact, one of the reasons why discretised actions are interesting is that it is possible to examine them through Monte Carlo (MC) simulations.
MC simulations can be used to implement an importance sampling on the phase space of a Wick rotated theory.
The weight of each configuration is determined by the Boltzmann weight the partition function 
\begin{align}
\mathcal{Z}_{\DT}= \sum_{\mc{T}}\frac{1}{C(T_d)}  e^{-\mc{S}_{\DT}(\mc{T})}\;
\end{align}
prescribes. 
The sum runs over the ensemble of all labelled triangulations $\mc{T}$, and the combinatorial factor $C(T_d)$ arises to correct for the over counting the labelling leads to.

The simulations then explore the ensemble of all labelled triangulations using the Pachner moves  \cite{Gross1992367}.
Pachner moves are ergodic moves that split or merge simplices to obtain new configurations.
For a longer introduction into the subject of Monte Carlo simulations for (causal) dynamical triangulations see  \cite{Gorlich:2011ga}.

Investigating EDT through MC codes produces  many interesting results.
In practice, one tunes the coupling $\kappa_4$ to obtain ensembles at fixed volume, to avoid the problem of simulating systems of infinite size. 
Then the phase space in $DT$ effectively consists of only one coupling $\kappa_2$, and is split into two phases.
For small $\kappa_2$ the geometry collapses, so that all $d$-simplices are very close to each other, and the effective Hausdorff dimension is $\infty$.
The other phase is at large $\kappa_2$. 
In this phase, the $d$-simplices arrange themselves in long chains. 
This phase is called the Branched Polymer phase (BP), and it has Hausdorff dimension $2$.
The phase transition between these two is of first order, and thus no continuum limit exists.
There have, however, been recent reinvestigations of some aspects of this, which will be reported on in more detail in section \ref{sec:EDT}.

\subsection{CDT and the transfer matrix}

One possible reason for the failure of EDT to capture real world physics is, that not all geometries which can be build out of Riemannian simplices can be Wick rotated back into Lorentzian geometries.
Restricting the geometries to those that could be Wick rotated is one motivation for causal dynamical triangulations \cite{Ambjorn:1998xu}.
This thesis introduces many facets of this theory, yet for a more comprehensive review see \cite{Ambjorn:2012jv}.

\begin{figure}
\centering
\includegraphics[width=0.8\textwidth]{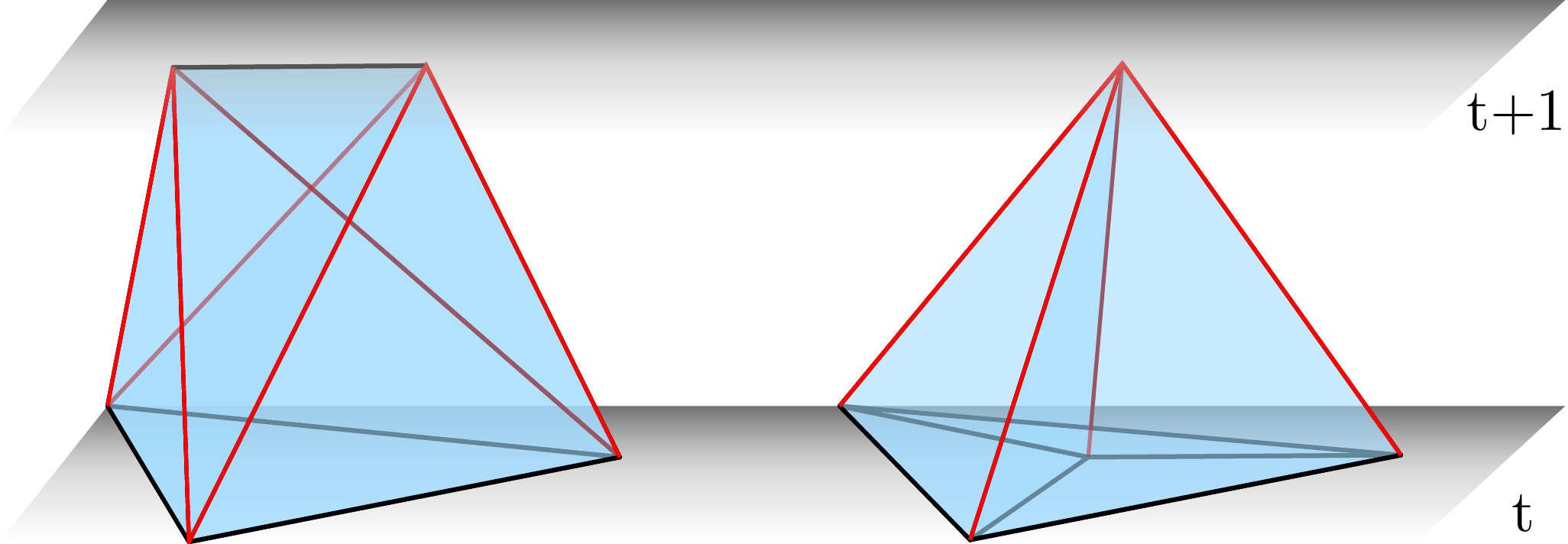}
\caption{\label{fig:CDT_simplices} An illustration of the simplices in 4d CDT, the timelike edges are marked in red.}
\end{figure}

In CDT, the simplices are not equilateral. 
Spacelike edges have a length $l_s^2=a^2$, and timelike edges have a length of $l_t^2=-\alpha a^2$ before the Wick rotation.
After the Wick rotation, $l_t^2 = \alpha a^2$.
The parameter $\alpha$ makes it possible to interpret this as an analytic continuation in the parameter $\alpha \to i \alpha$ \cite{Sorkin:2009ka}.
At the same time, the parameter allows us to keep track of timelike edges, even after Wick rotation to an Euclidian theory.
We then introduce a fixed time slicing and restrict the simplices to those that connect adjacent time slices.
In $2$d CDT, there are two possible building blocks, triangles with two timelike edges that either point up or down.
In $4$d CDT, we allow for four building blocks, the $(2,3),(3,2),(4,1)$ and $(1,4)$ simplices.
The numbers stand for the number of vertices that lie in one spacelike slice. 
The $(3,2)$ and the $(4,1)$ simplices are illustrated in Figure \ref{fig:CDT_simplices}, and the $(2,3)$ and $(1,4)$ simplices are time reflected versions of this.
The difference in length $\alpha$ between timelike and spacelike edges leads to an altered volume for $(3,2)$ and $(4,1)$ simplices in $4$d.
Since the $2$d triangles both have two timelike edges they both have the same volume.
The parameter $\alpha$ can then be absorbed into the cosmological constant $\Lambda$, which we will henceforth do.
The Wick rotation to compare Euclidian results with Lorentzian results  then becomes an analytic continuation in $\Lambda$.
The hope in this approach is that the time foliation is just a choice of gauge and that in the continuum limit the full gauge freedom can be reconstructed.

In $2$ dimensions, CDT can be solved analytically. 
Possible ways are the transfer matrix method, as used in  \cite{Ambjorn:1998xu}, matrix models, or an isomorphism between CDT and branched polymers. 
We will discuss the matrix models and the isomorphism in section \ref{sec:dimers}, in which they are used to solve a model of $2$d CDT coupled to matter.

The time foliation of CDT makes the transfer matrix approach an especially apt description.
Using the time foliation, we can interpret the transfer matrix as the operator that evolves the triangulation from one time slice to the next.
Taking the configuration at time $t$ to be represented as $\bra{T_t}$ and at time $t+1$ as $\ket{T_{t+1}}$ the transfer matrix $\hat T$ is defined as
\begin{align}
\bra{T_t} \hat T \ket{T_{t+1}}=\sum_{ T_d \in \mc{T}_d: T(t) \to T(T+1)} \frac{1}{C(T_d)} e^{-\Delta S(T_d)}\; ,
\end{align} 
where $\mc{T}_d$ denotes the intervening triangulations.
$\Delta S(T_d)$ denotes the contribution to the action through a given $T_d$.
Using the transfer matrix, we can then define the propagator from an initial configuration $\bra{T_0}$ to a final configuration $\ket{T_f}$ in time $t_f$ as
\begin{align}
G(T_0,T_f;t_f)&= \bra{T_0} \hat T^{t_f} \ket{T_f}\;.
\end{align}
This propagator satisfies the semi group property
\begin{align}
G(T_0,T_f;t_1+t_2)&= \sum_{T_s} C(T) G(T_0,T_s;t_1) G(T_s,T_1;t_2)\;.
\end{align}
Here, $C(T)$ is a combinatorial factor that corrects for the factors $C(T_d)$ before. 
The fact that these symmetry factors are different makes the composition rule non-obvious.
But, checking the symmetries of the boundary triangulations, we can convince ourselves that it leads to the correct counting.
In $2$d, we can calculate the transfer matrix and thus the entire propagator.
Fixing the topology of our boundary to be $S_1$, the spacelike surfaces at a given time step are just segmented loops.
The combinatorial factor is $C(T)=l$ with $l$ the number of segments or equivalently the number of vertices.
The weight added to the action by one time-step is $\Delta S= \lambda n(T)$.
To simplify the calculations, we can mark a point on the entrance loop, and thus eliminate the factors of $l$.
Then
\begin{align}\label{eq:Gcomp}
G_\lambda(l_1,l_2;t)&=\sum_{l=1}^\infty G(l_1,l;t-1)  G(l,l_2;1) \\
G_\lambda(l_1,l_2;1)&= \sum_{k_1, \dots , k_{l_1}} e^{-\lambda \sum_{i=1 \dots l_1} k_i}\;, 
\end{align}
with $k_i$ being the number of timelike links connected to the $i$-th vertex at the entrance loop.

It is then useful to introduce the generating function
\begin{align}
G_\lambda(x,y;t)&= \sum_{k,l} x^k y^l G_\lambda(l,k;t) \\
G_\lambda(x,y;t_1+t_2)&= \oint \frac{\md z}{2 \pi i z} G_\lambda(x,\frac{1}{z};t_1)G_\lambda(z,y;t_2) \;,\label{eq:gencomp}
\end{align}
where the contour should include the singularities of $ G_\lambda(x,\frac{1}{z};t_1)$ but not those of $G_\lambda(z,y;t_2) $.
We can interpret $x=e^{\lambda_1}, y=e^{\lambda_2}$  as lattice boundary cosmological constants.
It is then natural to also write $g=e^{-\lambda}$ and recast
\begin{align}
G(x,y,g;t)=G_\lambda(x,y;t) \;.
\end{align}
The generating function is a sum over all possible edge lengths and intervening geometries with a factor $g$ for each triangle, $x$ for each link in the lower, and $y$ for each link in the upper edge,
\begin{align} 
G(x,y,g;1)&=\triI + \triII +\triIII+ \dots\\
&=\sum_{k=0}^{\infty}\left( g x \sum_{l=0}^\infty (gy)^l \right)^k - \sum_{k=0}^\infty (gx)^k = \frac{g^2 x y}{(1-gx)(1-gx-gy)}\;.\label{eq:genfun}
\end{align}
This expression is not symmetric under $x \leftrightarrow y$, because we marked a point on the entrance loop but not on the exit loop. 
It is necessary to subtract the second sum to remove the singular cases, where either the entrance or the exit loop is of length zero.
We can now compose equation \eqref{eq:gencomp} for the case $t_2=1$ with equation \eqref{eq:genfun} and obtain
\begin{align}\label{eq:GG}
G(x,y,g;t)=\frac{gx}{(1-gx)} G(\frac{g}{1-gx},y,g;t-1)\;.
\end{align}
This solution can be iterated leading to
\begin{align}\label{eq:Grecurse}
G(x,y,g;t)&=\frac{g^2 x y F_1^2(x) F_2^2(x) \dots F_{t-1}^2(x)}{[1-g F_{t-1}(x)][1-g F_{t-1}(x) -gy]} \\
F_t(x) &= \frac{g}{1-g F_{t-1}(x)}, \qquad F_0(x)=x \;.
\end{align}
The fixed point of this equation is 
\begin{align}
F=\frac{1-\sqrt{1-4g^2}}{2g}\;,
\end{align}
and we can use this to write
\begin{align}\label{eq:recurse}
F_t(x)= F\frac{1-x F +F^{2t-1}(x-F)}{1-x F +F^{2t+1}(x-F)}\;.
\end{align}
Combining \eqref{eq:recurse} and \eqref{eq:Grecurse} we obtain
\begin{align}\label{eq:Gtotdisc}
G(x,y,g;t)=\frac{F^{2t} (1-F^2)^2 xy }{(A_t-B_t x)(A_t-B_t(x+y)+C_t xy)} \;, \\
\intertext{where}
A_t=1-F^{2t+2} \qquad B_t= F(1-F^{2t})\qquad C_t= F^2(1-F^{2t-2}) \;.
\end{align}
The power expansion in $x,y,g$ is convergent for all $t$ if
\begin{align}
|g|<\frac{1}{2} \qquad |x|<1 \qquad |y|<1 \;.
\end{align}
To compute $G_{\lambda}(l_1,l_2;t)$ from \eqref{eq:Gtotdisc}, we use a (discrete) inverse Laplace transformation
\begin{align}
G_{\lambda}(l_1,l_2;t) &= \oint \frac{\md x}{2 \pi i x} \oint \frac{\md y}{2 \pi i y}  \frac{1}{x^{l_1}}\frac{1}{y^{l_2}} G(x,y,g;t) \\
			&= l_1 \left( \frac{B_t}{A_t}\right)^{l_1+l_2}\quad \sum_{k=0}^{\mathrm{min}(l_1,l_2)} \frac{(l_1+l_2-k-1)!}{k!(l_1-k)!(l_2-k)!} \left(\frac{A_t C_t}{B_t^2} \right)^k \;.
\end{align}
As introduced for critical phenomena, we take the continuum limit of this theory such that the edge length $a \to 0$ while the physics remains constant.
To this end, we take the scaling around the critical point $\lambda_c,\lambda_{1,c},\lambda_{2,c}$.
It is known that dimensionfull couplings undergo additive renormalisation, while the propagator undergoes a multiplicative wavefunction renormalisation.
The expected behaviour for the cosmological constants is then
\begin{align}
\lambda &= \lambda_c +\Lambda a^2 & \lambda_1 &= \lambda_{1,c} + X a & \lambda_2 &= \lambda_{2,c} + Y a \;,
\intertext{which lets us express the coupling constants as}
g&=g_c e^{-\Lambda a^2}&x&=x_c e^{-X a}&y&=y_c e^{-Y a}\\
g_c&=e^{-\lambda_c} & x_c&= e^{-\lambda_{1,c}} & y_c&= e^{-\lambda_{2,c}} \;,
\end{align}
where we introduced the critical coupling constants.
The renormalised propagator is then
\begin{align}\label{eq:Glimit}
G_\Lambda(X,Y;T)= \lim_{a\to 0} a^\eta \, G(x,y,g;t) \;.
\end{align}
The wave function is renormalised with $a$ to the power of the anomalous dimension $\eta$, and discrete time rescales into continuum time $T=a t$.
We determine the anomalous dimension using equation \eqref{eq:Gcomp}. 
If we require that this relation survive under the limit \eqref{eq:Glimit}, we find that $\eta=1$.
From equation \eqref{eq:recurse}, we see that for a non-trivial limit $|F| \to 1$, thus the critical value is $g_c =\pm \frac{1}{2}$.
Without loss of generality, we can then restrict the discussion to $g_c=\frac{1}{2}$.
To recover macroscopic loops in the limit $a \to 0$, we need to take $x,y\to 1$ thus giving us the critical values $x=y=1$.
We then have
\begin{align}
g&=\frac{1}{2} e^{-\Lambda a^2}&x&=e^{-X a}&y&=e^{-Y a}\;,
\end{align}
and can take the continuum limit of \eqref{eq:Glimit}
\begin{align}
G_\Lambda(X,Y;T)=& \frac{4 \Lambda e^{- 2 \sqrt{\Lambda} T}}{(\sqrt{\Lambda}+X)+e^{-2 \sqrt{\Lambda}T}(\sqrt{\Lambda}-X)} \nonumber \\
&\times \frac{1}{(\sqrt{\Lambda}+X)(\sqrt{\Lambda}+Y)-e^{-2 \sqrt{\Lambda}T}(\sqrt{\Lambda}-X)(\sqrt{\Lambda}-Y)} \;.
\end{align}
From this we calculate the propagator $G_\Lambda(L_1,L_2;T)$ through an inverse Laplace transformation.
\begin{align}
G_\Lambda(L_1,L_2;T)&=\int_{- i\infty}^{i \infty} \md X \int_{- i\infty}^{i \infty} \md Y e^{X L_1} e^{Y L_2} G_\Lambda(X,Y;T) \\
					&=\frac{e^{- [\coth{\sqrt{\Lambda}T}] \sqrt{\Lambda}(L_1+L_2)}}{\sinh{\sqrt{\Lambda }T}} \frac{\sqrt{\Lambda L_1 L_2}}{L_2} I_1\left( \frac{2 \sqrt{\Lambda L_1 L_2}}{\sinh{\sqrt{\Lambda}T}}\right)\;.
\end{align}
This is equivalent to the limit $a \to 0$ of equation \eqref{eq:Gcomp}, and $I_1$ is a Bessel function of the first kind.
The asymmetry between $L_1$ and $L_2$ can be removed by dividing by $L_1$ to obtain the unmarked amplitude, or by multiplying with $L_2$ to obtain the amplitude with points marked on entrance and exit loop.

Using the transfer matrix approach, we can also derive a Hamiltonian for 2d CDT.
We can regard the propagator as the matrix element between two boundary states of a Hamiltonian evolution in Euclidian time $T$ 
\begin{align}\label{eq:therightpropagator}
G_\Lambda(L_1,L_2;T) =\bra{L_2} e^{-\hat H T} \ket{L_1} \;.
\end{align}
This evolution has to fulfil the Heat kernel equation 
\begin{align}
\pd{}{T}G_{\Lambda}(L_1,L_2;T)=-\hat H(L_1) G_{\Lambda}(L_1,L_2;T) \;,
\end{align}
or equivalently its  Laplace transform
\begin{align}
\pd{}{T} G_{\Lambda}(X,Y;T)=-\hat H(X) G_{\Lambda}(X,Y;T)\;.
\end{align}
In the limit $a\to 0$, we can read off the Hamiltonian from equation \eqref{eq:GG}
\begin{align}
\hat H(X)= \pd{}{X} (X^2-\Lambda)\;,
\end{align}
and an inverse Laplace transform leads us to
\begin{align}
\hat H(L,\pd{}{L})=- L\pd{^2}{L^2} +\Lambda L \;.
\end{align}
We should, however, be careful to keep track of our boundary conditions. 
Marking a point on the entrance boundary corresponds to using a different measure on the states of this boundary than on the states of the other boundary. 
Depending on the measure we use, we calculate distinct Hamiltonians.
Three Hamiltonians that will be of special interest in section \ref{sec:HL} are
\begin{align}
-L \pd{^2}{L^2} &+\Lambda L	& \text{with measure} &\quad \frac{\md L}{L} \;, \label{eq:CDTham1}\\
-\pd{^2 }{L^2}L &+\Lambda L	& \text{with measure} &\quad L \md L \;, \text{ and} \label{eq:CDTham2}\\
-\pd{ }{L}L\pd{ }{ L}&+\Lambda L 	& \text{with measure}& \quad \md L \;.\label{eq:CDTham3}
\end{align}
The first of these corresponds to the above case of a marked entrance loop, while the second one is the Hamiltonian for the propagator of unmarked loops.
The last one describes the propagation of an open boundary without marked points, corresponding to wave functions derived with the flat measure $\md L$.
These continuum Hamiltonians for CDT can be used to compare the theory to Ho\v{r}ava-Lifshitz gravity, which will be done in section \ref{sec:HL}.

In $4$d, the transfer matrix cannot be calculated analytically.
But it can be measured in MC simulations and allows us to make a fit for the effective action of the theory  \cite{Ambjorn:2014mra,Ambjorn:2012pp}.

\subsection{Phases of $4$d CDT}

For higher dimensional CDT, no exact analytic solutions are know.
However, both three and four-dimensional models are extensively studied in Monte Carlo simulations \cite{Ambjorn:2014mra,Ambjorn:2012pp,Ambjorn:2010fv,Ambjorn:2004qm,Ambjorn:2012ij,Ambjorn:2011cg}.

Treating CDT analytically we could always Wick rotate the results back to Lorentzian space-times, by analytic continuation in the parameter $\alpha$.
In numerical simulations, we can not do that. 
Then to ensure that our results are applicable for  Lorentzian space-times we have to carefully chose the observables we examine.
For example, in flat space the mass of a particle can be measured in Minkowski space as the pole of the propagator or in flat Euclidian space as the fall-off of the correlation length.

\begin{figure}
\centering
\includegraphics[width=0.85\textwidth]{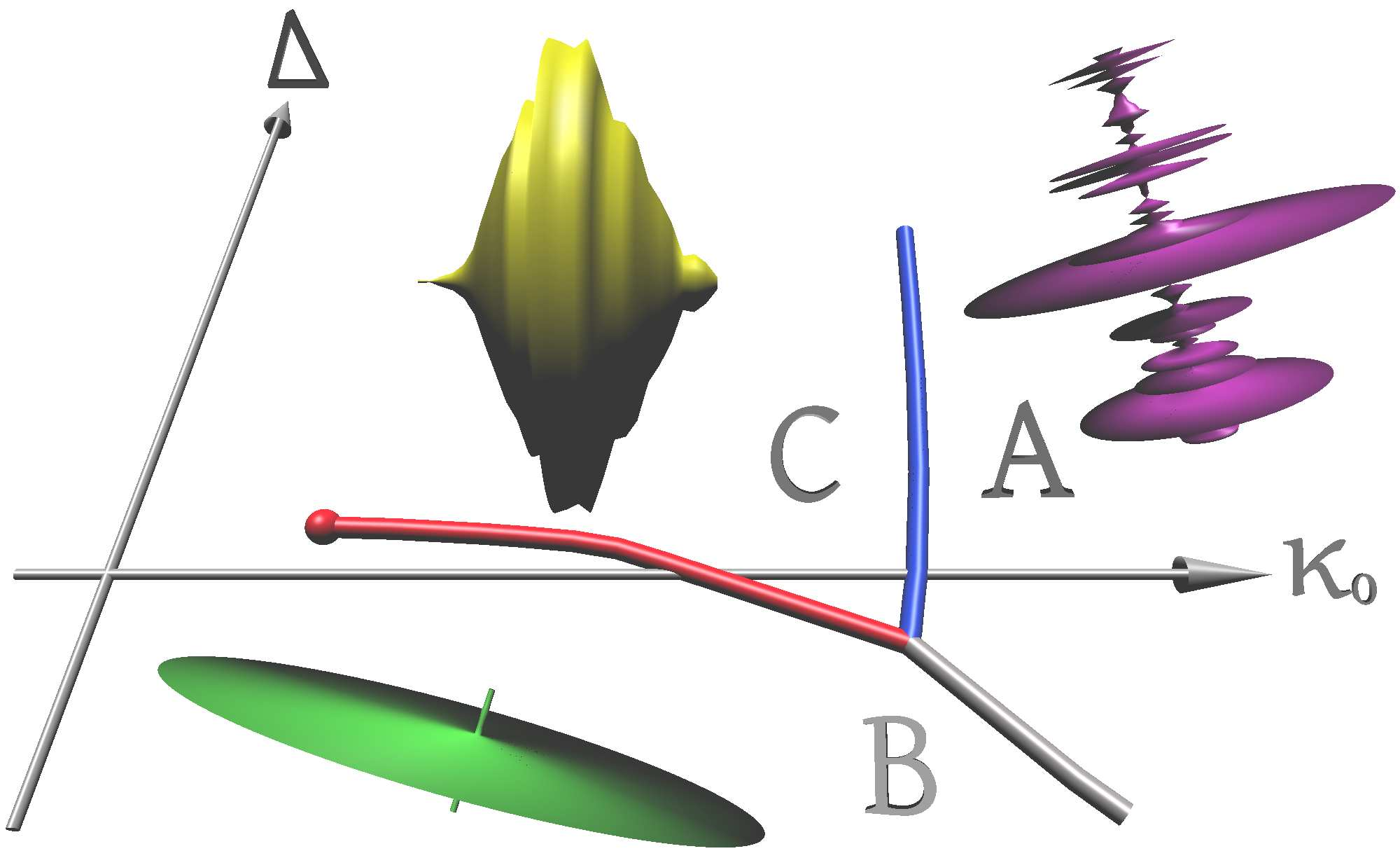}
\caption{\label{fig:CDT_phases}The phase diagram of CDT, depicting the three phases. This figure is taken from  \cite{Ambjorn:2012ij}.}
\end{figure}

We restrict our review to the $4$d case, since $3$d models are mostly considered a test bed for the $4$d theory.
In $4$d, the parameter $\alpha$ leads to altered volumes and deficit angles for the two classes of simplices, and thus modifies the action \eqref{eq:DTaction}. 
The action in $4$d depends on the number of $(4,1)$-type four-simplices $N_4^{(4,1)}$ and on the number of $(3,2)$ type four-simplices $N_4^{(3,2)}$
\begin{align}
\mc{S}_{\text\CDT\, 4d}&=- (\kappa_0+ 6 \Delta) N_0 + \kappa_4 (N_4^{(4,1)}+ N_4^{(3,2)})+ \Delta (2 N_4^{(4,1)}-N_4^{(3,2)}) \;,
\end{align}
where $\Delta$ depends on $\alpha$.
As in EDT, we fix the volume of our simulations by tuning the coupling $\kappa_4$ to its critical value.
This leaves us with two effective coupling constants $(\Delta, \kappa_0)$.
Examining the $(\Delta,\kappa_0)$ plane, we measure three distinct phases \cite{Ambjorn:2012ij} as shown in Figure \ref{fig:CDT_phases}.

Phase $A$ corresponds to the BP phase found in dynamical triangulations.
The geometry in this phase is best described as fluctuating, with the three-volume of adjacent spacelike slices completely uncorrelated.

Phase $B$  corresponds to the crumpled phase, in this phase almost the entire volume of the universe is concentrated in a single time slice.

\begin{figure}
\centering
\includegraphics[width=0.8\textwidth]{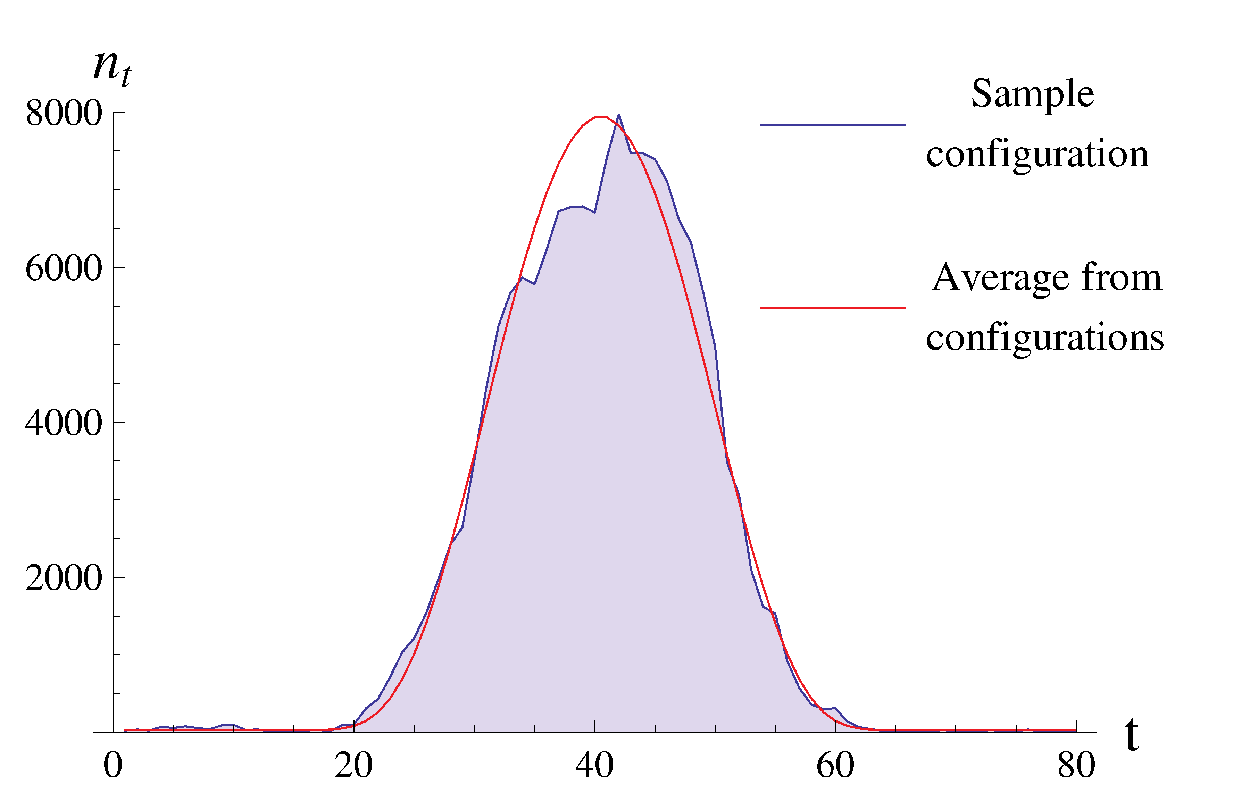}
\caption{\label{fig:volprof}The volume profile of a CDT in phase $C$. The blue curve is the volume profile of a single configuration, while the red line shows the average over all configurations.
This figure is taken from  \cite{Ambjorn:2013tki}.}
\end{figure}

Phase $C$, on the other hand, shows genuine four-dimensional behaviour. 
The average geometry in this phase corresponds to an Euclidian de Sitter universe.
In Figure \ref{fig:volprof}, we see the volume profile of a single geometry in this phase vs the average over the Monte Carlo ensemble.
The volume profile $N_3(t)$ is a measurement of the number of tetraheda that lie in time step $t$.
The total length of the simulations that underlie this plot was fixed to $80$ time steps. 
The time extent of the universe scales like $N_4^{1/4}$ while the volume scales like $N_4^{3/4}$, consistent with macroscopically four-dimensional geometries \cite{Ambjorn:2004qm}.
At first glance this result might seem trivial, after all, we are using four-dimensional building blocks. 
Despite this, it is important to realise that the dimensionality of the building blocks does not determine the dimensionality of the space-time. 
At no point in the non-perturbative path integral over geometries of CDT, is any background geometry assumed.

It is thus indeed remarkable that a continuum phase is found.
The phase diagram of CDT has striking similarities with the phase-diagram of a Lifshitz theory.
This and other similarities hint at the connection between Ho\v{r}ava-Lifshitz gravity and CDT.
We will examine this connection in more detail in section \ref{sec:HL}.

Another observable that shows that phase $C$ gives rise to a large scale four-dimensional geometry, while highlighting the quantum behaviour at smaller scales, is the spectra dimension.
The spectral dimension on an Euclidian manifold can be characterised through a diffusion process.
On the discrete geometries that are considered in CDT this diffusion process can be implemented as a random walk. 
We can measure the return probability $P_g(\sigma)$ of the random walk to its starting point, as a function of the diffusion time $\sigma$.
Diffusion processes on fractal geometries have been studied in statistical physics  \cite{ben2000diffusion}.
The return probability is then given as
\begin{align}
P_g(\sigma) & \sim \frac{1}{\sigma^{\frac{D_s}{2}}} F\left( \frac{\sigma}{N^{\frac{2}{D_s}}}\right) \;,
\end{align}
where $N$ is the volume of the geometry and $D_s$ the spectral dimension.
Here we introduce the function $F(x)$ which goes to $1$ for $x\to0$.
For all known cases $F(x)$ falls off at least exponentially for $x>1$.

We expect the same behaviour on CDTs and can thus measure the logarithmic derivative 
\begin{align}
-2 \frac{d \log{P_g(\sigma)}}{d \log{\sigma}}= D_s(\sigma)
\end{align}
to obtain the spectral dimension on CDTs.
In $4$d CDT, the spectral dimension at large scales is measured as $D_s(\sigma=\infty)=4.02 \pm 0.01$, but at short scales it reduces to $D_s(\sigma=0)=1.80\pm 0.25$ \cite{Ambjorn:2005db}.
After this was measured in CDT, the same spectral dimension was calculated for Ho\v{r}ava-Lifshitz gravity \cite{Horava:2009if} and asymptotic safety  \cite{Lauscher:2005qz}.

While it is encouraging that phase $C$ has continuum properties, the goal in CDT is to take the continuum limit.
Since the continuum limit depends on a second order phase transition, the phase transitions have been examined in detail \cite{Ambjorn:2012ij}. 

The transition between phases $A,B$ divides the two non-physical phases and is hard to measure, thus no special effort has been expended to examine it in detail. 
The transition between $A,C$ was shown to be of first order, and thus not a candidate for a continuum theory. 

The $B,C$ transition, on the other hand, is of at least second order  \cite{Ambjorn:2012ij}.
There are two interesting points along the $B,C$ transition line.
The first is the triple point where all three phases meet, and the second is the end-point of the phase transition.
Special points along a phase transition line can be multicritical points of higher order than the phase transition itself.
It is thus plausible to expect that if any point in the CDT phase diagram describes real world physics it could be one of these.
One conjecture is that the continuum theory at the triple point is a Ho\v{r}ava-Lifshitz type theory, while the endpoint of the phase transition describes general relativity at its asymptotic fixed point \cite{Ambjorn:priv}.

\section{\label{sec:dimers}Multicritical CDT models}
In subsection \ref{sub:critic} above, we saw how we can take the continuum limit of a discrete theory at a phase transition of at least second order.
In the last section we showed that $2$d CDT has a continuum limit at the critical point $g_c=1/2$, the radius of convergence of the generating function for the propagator.
The critical exponents at this point are
\begin{align}
d_h&=2 & \gamma &=\frac{1}{2} \;.
\end{align}
In a model that successfully couples this theory to matter, we expect these exponents to change, thus leading to a new continuum theory.
It is, for example, well established that multicritical matrix models are equivalent to quantum gravity coupled to matter  \cite{Kazakov:1986hu,Staudacher:1989fy}.

One type of matter that is easy to couple to CDT are hard dimers\footnote{The work described in this section is published \cite{Ambjorn:2012zx} and is reprinted in Appendix \ref{app:Dimer}}.
Hard dimers are links with an extra weight $\xi$, that are placed on the edges of the triangulation such that they cannot touch.
On a random graph, hard dimers are equivalent to the high temperature expansion of the Ising model on the same graph \cite{PhysRevLett.24.1412,Kurtze:1979zz}.

To couple the hard dimers to CDT, we use an isomorphism between $2$d CDT and rooted tree graphs \cite{Durhuus:2009sm} (often also referred to as branched polymers (BP) in the physics literature).

\begin{figure}
\centering
\includegraphics[width=0.5\textwidth]{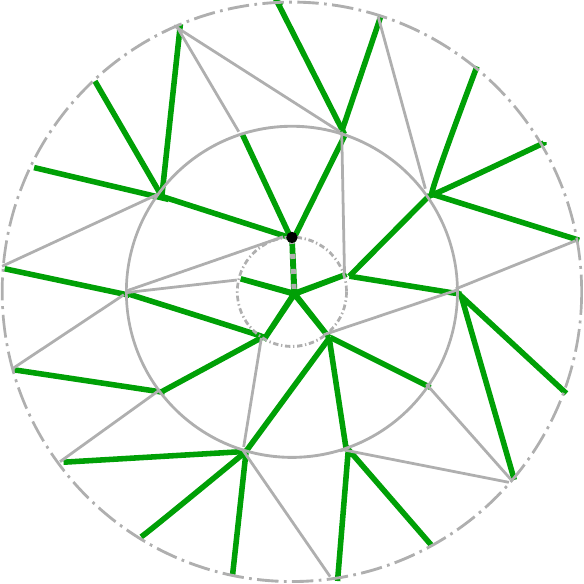}
\caption{\label{fig:CDTtoBP}Mapping a CDT to a branched polymer. The edges of the triangulation that are part of the branched polymer are marked in green. This figure is taken from  \cite{Ambjorn:2012zx}.}
\end{figure}

To transform a CDT into a BP, we first delete all spacelike links.
Then the leftmost forward pointing link at each vertex is deleted.
This algorithm assigns a BP uniquely to a CDT, except at the initial surface. 
At the initial surface the definitions have to be changed, see  \cite{Durhuus:2009sm}.
This algorithm is illustrated in Figure \ref{fig:CDTtoBP}.

Thus, if we can show multicriticality for hard dimers coupled to BPs we have a multicritical system that couples a restricted class of hard dimers to CDT.
The class is restricted, since not all edges that are part of the CDT are included in the BP.

\begin{figure}
\centering
\includegraphics[width=0.7\textwidth]{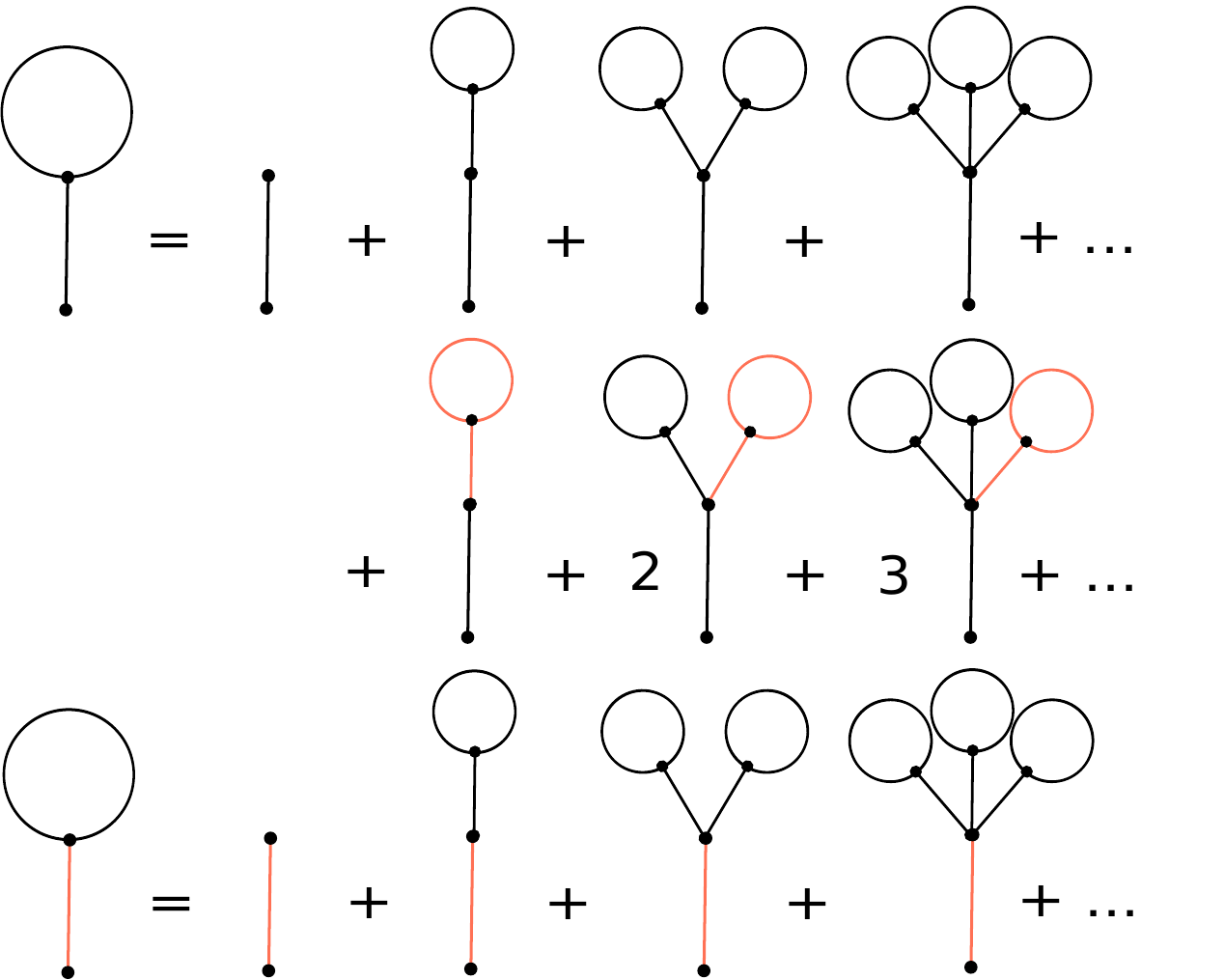}
\caption{\label{fig:DimersBP} Graphical representation of the recursive equation for the partition function including hard dimers. The dimers are marked in red. This figure is taken from \cite{Ambjorn:2012zx}. }
\end{figure}

The partition function for BPs can be defined as
\begin{align}
Z&=\sum_{BP} \prod_{i} v_i \prod_l e^{-\mu}\;.
\end{align}
We associate a weight $v_i$ to each vertex and a fugacity $e^{-\mu}$ to each link.
In standard BP models, the weight of the vertices depends only on the order of the vertex.
The partition function for CDT with the action \eqref{eq:2dCDTaction} is equivalent to the weights $v_i=1$ for all $i$.
We can write down the recursive equation
\begin{align}
e^\mu&= \frac{1+v_2 Z +v_3 Z^2+ \cdots}{Z} := \frac{f(Z)}{Z} :=F(Z) \;.
\end{align}
This recursion can also be represented graphically, for the CDT case it is represented in the lower line of Figure \ref{fig:DimersBP} disregarding the colouring.
Generically the weights $v_i$ are non-negative, in which case $F(Z)$ has a minimum at $Z_0$. 
The non-analytic behaviour at this point is then
\begin{align}
Z(\mu) - Z(\mu_0) &\sim (\mu - \mu_0)^{\frac{1}{2}} , & e^{\mu_0} &= F(Z_0) \;.
\end{align}
Loosening the requirement that the weights be positive, we can construct solutions to
\begin{align}
\pd{^{k} F(Z)}{Z^{k}}&=0\;,
\end{align}
for $k=1,\dots,m-1$ and $m>2$. 
In this case, the non-analytic behaviour of the partition function is
\begin{align}
Z(\mu)-Z(\mu_0)&\sim (\mu -\mu_0)^{\frac{1}{m}} \;.
\end{align} 
We thus introduce multicritical points into our system.
These multicritical points lead to a new continuum limit, and for matrix models it has been argued that they correspond to conformal matter coupled to the system \cite{Kazakov:1986hu}.

We will construct a multicritical model of CDT, by placing hard dimers with weight $\xi$.
If we chose $\xi$ to be negative, we introduce negative weights for the decorated links.
The partition function including hard dimers is
\begin{align}
Z(\mu,\xi) &= \sum_{BP}\prod_i v_i \prod_l e^{-\mu} \sum_{HD(BP)}\xi^{|HD(BP)|} \;.
\end{align}
The last sum is over all possible hard dimer configurations on a BP, $|HD(BP)|$ denotes the number of hard dimers contained in a configuration.
Introducing the hard dimers modifies the recursive equations. 
We now need to keep track of BPs whose root is a hard dimer, with partition function $\tZ$, and those whose root is not a dimer, with partition function $Z$, separately.
For the CDT model, where $v_i=1$ 
\begin{align}
e^\mu &= \frac{1}{Z} \left( \frac{1}{1-Z}+ \frac{\tZ}{(1-Z)^2}\right) & e^\mu \tZ&= \xi \frac{1}{1-Z} \;.
\end{align}
These recursive relations are illustrated in Figure \ref{fig:DimersBP}.
For $\xi_c=-1/12$ this model has a $m=3$ multicritical point with the critical exponents
\begin{align}\label{eq:MCexponentsCDT}
d_h&=\frac{3}{2} &\sigma &=\frac{1}{2} & \gamma &= \frac{1}{3} \;.
\end{align}
Our model is a restricted model of dimers coupled to CDT.
Not all possible hard dimer configurations are included, since the BPs do not include spacelike edges or the leftmost forward pointing timelike edges of any vertex.
It does, however, seem likely that including the omitted configurations would not lead to a qualitative change in these results.
To examine this, we obtained the same results for unrestricted hard dimers on generalised CDTs as the scaling limit of multicritical matrix models.

\subsection{Multicritical matrix models}
\begin{figure}
\subfigure[\label{fig:MMprop}]{\includegraphics[width=0.45\textwidth]{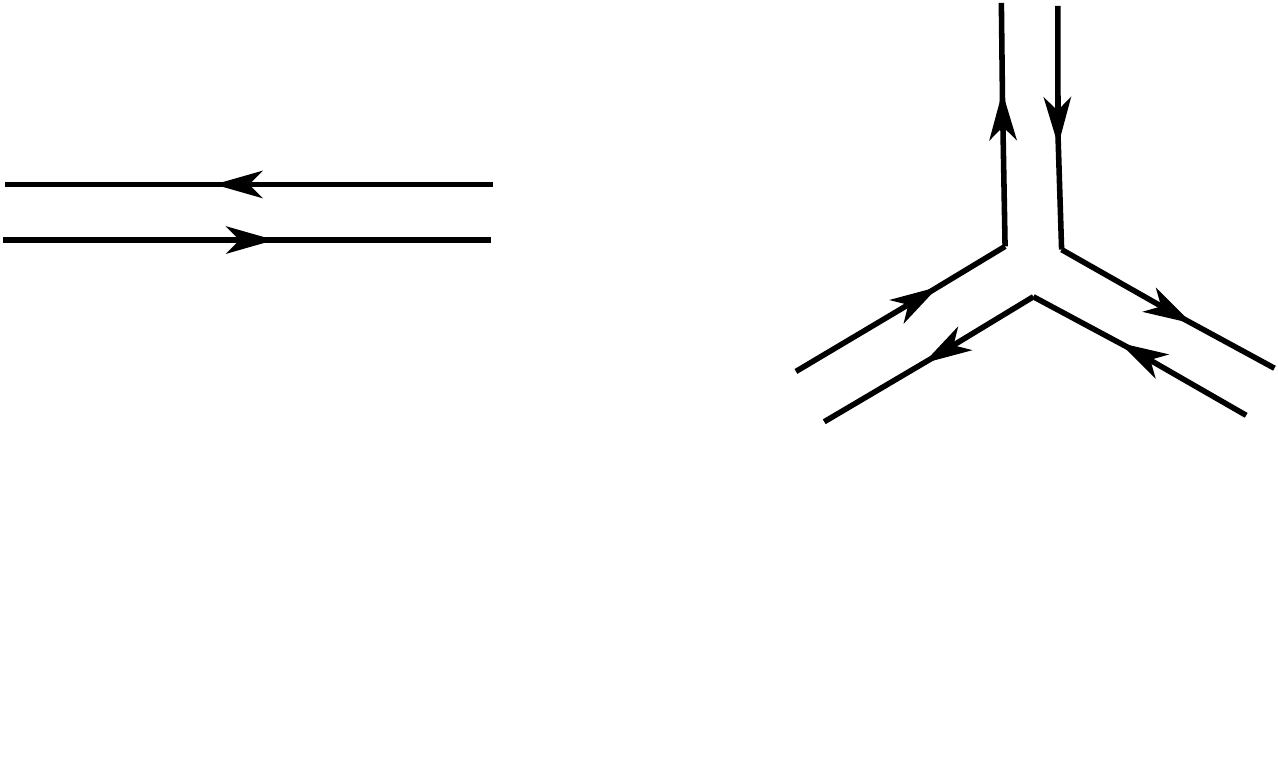}} \hspace{40pt}
\subfigure[\label{fig:MMexplain}]{\includegraphics[width=0.4\textwidth]{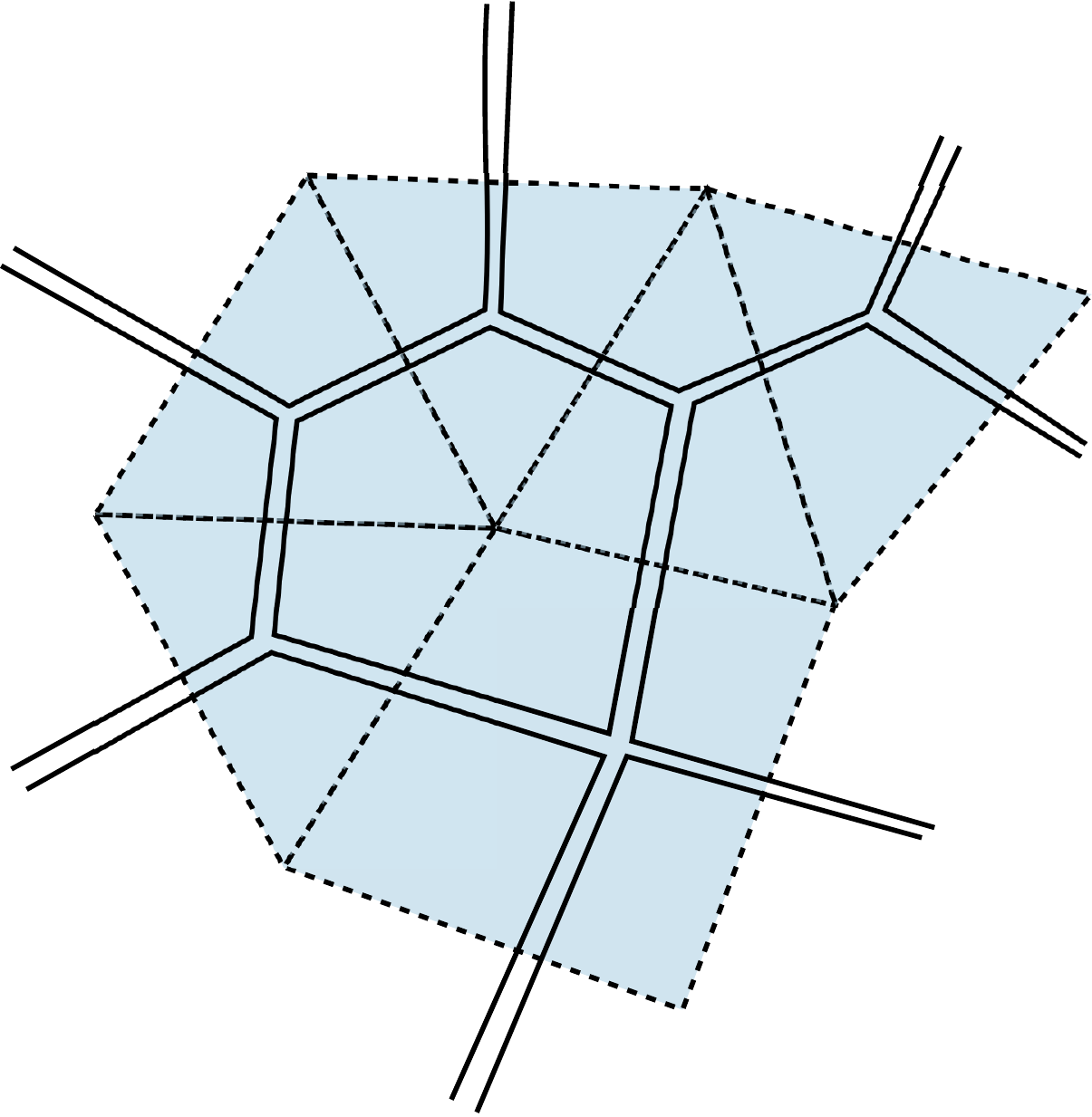}}
\caption{The left figure shows the propagator and a $\phi^{3}$ term in the matrix model, higher order interactions follow the same pattern. The right figure shows how propagators and interactions combine to determine a surface. The propagators are drawn as double lines, and the surface of the dual graph is shaded light blue.}
\end{figure}
The idea behind matrix models is that the partition function of a matrix field theory characterises a $2$d surface\footnote{We will here provide an extremely basic overview, for a real introduction into matrix models see \cite{DiFrancesco:1993nw}.}.
We can write the partition function for a $N\times N$ matrix $\phi$ as
\begin{align}\label{eq:Zmatrix}
\mc{Z}(\lambda_i,g_s)=\int \md \phi \; e^{-\frac{N}{g_s}V(\phi,\lambda_i)}\;,
\end{align}
where the $\lambda_i$ are arbitrary coupling constants.
The potential $V(\phi,\lambda_i)$ has to be a scalar, so it is made up of contractions and traces of the matrix.

The Feynman expansion of this uses the propagator and interactions as in Figure \ref{fig:MMprop}. 
In the limit of high loop order, it then contains graphs like Figure \ref{fig:MMexplain}. 
The solid double lines are propagators of the Feynman graph, while the dotted lines show the dual simplicial complex.

This construction uses matrices because the matrix indices translate to double lines and thus make it possible to define an oriented simplicial complex, which is necessary to describe a Riemannian surface.
A $\phi^3 $ term encodes triangles, a $\phi^4$ term squares and so forth. 
In the partition function \eqref{eq:Zmatrix}, all possible triangulations built out of the elementary blocks included in the potential are included.
As in calculating partition functions in QFT, we take the logarithm $\log{\mc{Z}}$ to remove all disconnected diagrams.
We can then take the large $N$ limit, to restrict the topology to that of a sphere, and for the continuum limit we take the couplings $\lambda_i \to \lambda_{i,c}$.

It has been shown that the scaling limit of the potential
\begin{align}
V(\phi)=\frac{1}{2}\phi^2-\lambda \phi -\frac{\lambda}{3}\phi^3 \;,
\end{align}
leads to a generalised CDT \cite{Ambjorn:2008gk}.
The linear part of the potential is introduced to make taking the scaling limit easier.
If we assume this potential in \eqref{eq:Zmatrix} above, we can solve  the partition function by expanding the exponential in powers of $\lambda$.
The result is an ensemble of random surfaces, generated by gluing together triangles and `tadpoles' corresponding to the linear term.
To calculate the sum over connected surfaces of spherical topology, we take the logarithm of $\mc{Z}$ and the large $N$ limit.
\begin{figure}
\centering
\includegraphics[width=0.8\textwidth]{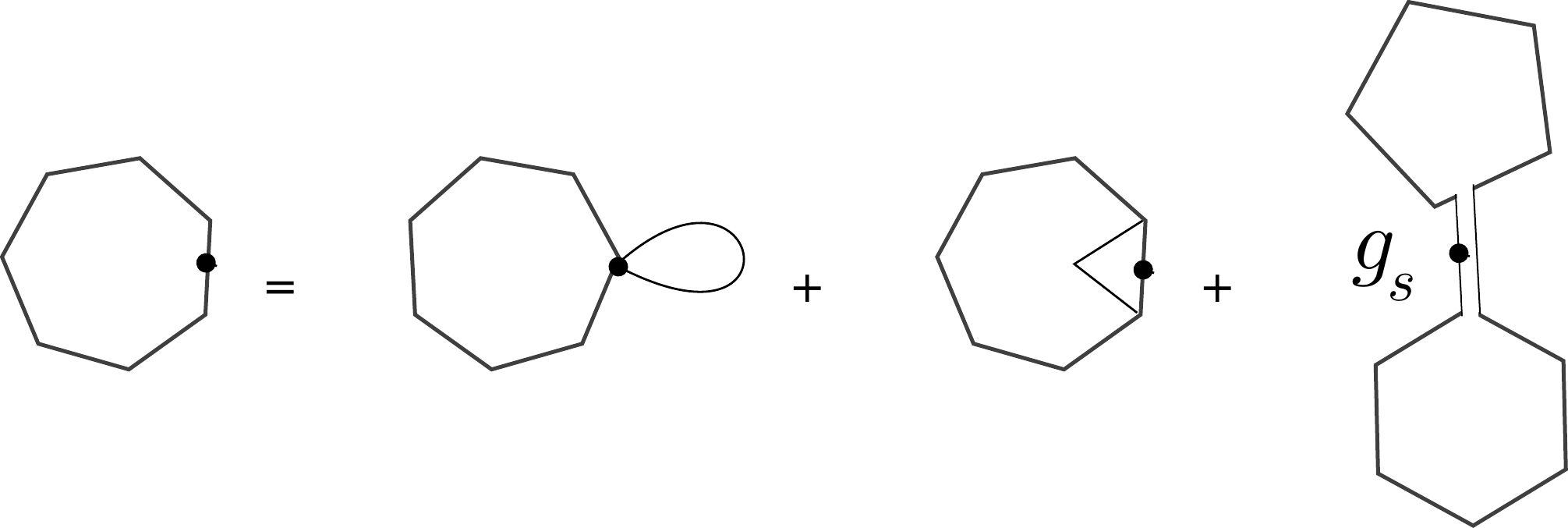}
\caption{\label{fig:loop}This figure shows the graphic representation of the loop equation. This figure is taken from \cite{Ambjorn:2012zx}.}
\end{figure}
The disk-amplitude that describes planar triangulations with a boundary is
\begin{align}
W(x)=\frac{1}{N} \left\langle \mathrm{tr}\frac{1}{x-\phi} \right\rangle \;,
\end{align}
where the expectation value is taken with respect to the partition function \eqref{eq:Zmatrix}.
In the large $N$ limit, $W(X)$ satisfies the loop equation
\begin{align}
g_s W(x)^2 &= V'(x) -Q(x) &  Q(x)&=c_1 x +c_0 \;,
\end{align}
here $V'$ denotes the derivative with respect to $x$. 
This equation can be represented graphically as in Figure \ref{fig:loop}.
Like above, $x$ is the fugacity of a boundary cosmological constant $x=e^{\mu_b}$.
We can solve the loop equation
\begin{align}
W(x)= \frac{1}{2 g_s}\left(V'(x)-\sqrt{V'(x)^2-4 g_s Q(x)}\right) 
\end{align}
and determine the constants $c_0, c_1$ by requiring that $W(x)$ has a single cut on the real axis, and $W(x) \to \frac{1}{x} + \mc{O}(\frac{1}{x^2})$ as $x \to \infty$.
We can rewrite it as
\begin{align}
W(x)=\frac{1}{2 g_s}\left(V'(x)+ \lambda(x-c) \sqrt{(x-b)(x-a)}\right)
\end{align}
with $a,b,c$ functions of the coupling constants $\lambda, g_s$.
This expression has several continuum limits.
One possible choice is to take $c(\lambda) \to b(\lambda)$ for a fixed $g_s$.
This limit can be associated with Euclidian quantum gravity \cite{jansbook}.

Of  interest to us is the scaling limit that recovers CDT.
In this limit, in addition to $c(\lambda)\to b(\lambda)$ we scale $g_s \to 0$, more specifically
\begin{align}
g_s&= G_s \epsilon^3 \;,
\end{align}
where we introduced the small parameter $\epsilon$.
To lowest order, the critical couplings are
\begin{align}
\lambda_c(g_s)&=\frac{1}{2} -\frac{3}{4} G_s^{\frac{2}{3}} \epsilon^2 & x_c(g_s)&= 1+G_s^{\frac{1}{3}} \epsilon \;.
\end{align}
The cosmological constant and the boundary cosmological constant are 
\begin{align}
\lambda&=\lambda_c(g_s) - \epsilon^2 \Lambda & x&=x_c(g_s)+\epsilon X \;.
\end{align}
Under this rescaling, the disc function itself undergoes multiplicative renormalisation
\begin{align}
W_{\text{cont}}(X)&= \lim_{\epsilon \to 0} \epsilon W(x)\\
&=\frac{1}{2 G_s} \left(\Lambda_\CDT -\frac{1}{2} X_\CDT^2 + (X_\CDT-H)\sqrt{(X_\CDT+H)^2-\frac{4G_s}{H}} \right)
\end{align}
with
\begin{align}
\Lambda_{\CDT} &= \Lambda+\frac{3}{2} G_s^{\frac{2}{3}} & X_\CDT &= X +G_s^{\frac{1}{3}} & 2 \Lambda_\CDT H- H^3 &= 2G_s \;.
\end{align}
$G_s$ can be interpreted as the coupling constant governing the branching off of baby universes. 
The limit $G_s\to0$ serves to suppress the branching off.
Taking this limit recovers the CDT disc function 
\begin{align}
W_{\text{cont}}(X) \to \frac{1}{X+\sqrt{2 \Lambda}} \;.
\end{align}
To generate a multicritcal matrix model, we use the potential
\begin{align}
V(\phi)=\frac{1}{2} \phi^2 -\lambda \phi -\frac{\lambda}{3} \phi^3 -\frac{\lambda^3 \xi}{2}\phi^4\;.
\end{align}
The surfaces this potential generates are glued together from triangles and squares.
We can identify the dimers as the diagonals of the squares.
The fugacity of a link with a dimer is $\lambda \xi$.
Using this rule, the dual triangulations are decorated with hard dimers.

To determine the multicritical behaviour of this model, we can solve the loop equation as above.
The new potential leads to a polynomial $Q(x)$ of second order, but leaves the rest of the loop equation unchanged.
The constants in $Q(x)$ are fixed by requiring that $W(x)$ fall off like $\frac{1}{x}$ for $x \to \infty$.
At the multicritical point $\lambda_c(g_s), \xi_c(g_s)$, the disc function is\footnote{This corresponds to equation (35) in  \cite{Ambjorn:2012zx}, unfortunately there is a typographic mistake in equation (35), the equation should have read as above.}
\begin{align}
W(x)=\frac{1}{2 g_s}\left(V'(x)- 2\lambda_c^2(g_s)\xi_c(g_s)\left(x-b_c(g_s)\right)^2  \sqrt{(x-c_c(g_s))(x-a_c(g_s))}\right)\;.
\end{align}
To obtain a limit as above, we take
\begin{align}\label{eq:gsscaling}
g_s=G_s \epsilon^4 \;.
\end{align}
The scaling is changed from that for pure CDT, since the multicritical point is characterised by
\begin{align}
V'(\lambda_c,\xi_c,x_c)=V''(\lambda_c,\xi_c,x_c)=V'''(\lambda_c,\xi_c,x_c)=0 \;.
\end{align}
From this we find that the critical values for $g_s=0$ are
\begin{align}
x_*=\frac{1}{\lambda_*}=-\frac{1}{\xi_*}=\sqrt{3}\;.
\end{align}
We can then calculate
\begin{align}
\lambda_c(g_s)&=\lambda_* \left(1-\frac{\sqrt{5}}{9}G_s^\frac{1}{2} \epsilon^2 -\frac{16 \sqrt{5}}{27} G_s^{\frac{3}{4}} \epsilon^3\right) + \mc{O}(\epsilon^4) \\
\xi_c(g_s)&=\xi_* \left(1-\frac{\sqrt{5}}{9}G_s^\frac{1}{2} \epsilon^2 +\frac{16 \sqrt{5}}{27} G_s^{\frac{3}{4}} \epsilon^3\right) + \mc{O}(\epsilon^4)\\
x_c(g_s)&=b_c(g_s)=x_* \left(1+ \frac{2}{35^{\frac{1}{4}}} G_s^{\frac{1}{4}} \epsilon \right) + \mc{O}(\epsilon^2) \;,
\end{align}
using the scaling of $g_s$ in \eqref{eq:gsscaling}.
Since the theory has two couplings $\lambda, \xi$, our solution has two parameters.
We can parameterise the perturbation around the continuum  constants $\tilde \Lambda, \Lambda$
\begin{align}
\lambda&= \lambda_* +\tilde \Lambda \epsilon^2 -\Lambda \epsilon^3 & \xi &= \xi_* - \frac{1}{2}\tilde \Lambda \epsilon^2 \;.
\end{align}
It is possible to restrict the solution to one parameter by choosing a convenient background $\tilde \Lambda =0$ and redefining
\begin{align}
\Lambda_\CDT &=\Lambda +\frac{32 \sqrt{3} 5^{\frac{1}{4}}}{81} G_s^\frac{1}{4} & X_\CDT&= X+\frac{2}{\sqrt{3}5^{\frac{1}{4}}} G_s^{\frac{1}{4}} \;.
\end{align}
Using this, we calculate $W(x)$ in the CDT limit $G_s \to 0$
\begin{align}
W(x)=\frac{1}{\epsilon} \frac{1}{X_\CDT + \Lambda_\CDT^\frac{1}{3}} \;.
\end{align} 
One can then calculate the same critical exponents as in \eqref{eq:MCexponentsCDT} above.

Introducing hard dimers couples a matter system to CDT. 
We can generalise the hard dimer model on CDT, to obtain higher order multicritical points by introducing more types of hard dimers (see  \cite{Glaser:2012ej}).
A slightly different model of dimers coupled to CDT was obtained in \cite{Atkin:2012yt}.
Interestingly, although the models seem very similar they do lead to different critical exponents \cite{Ambjorn:2014voa}.

In the context of matrix models of random geometries, the hard dimers with negative weight are well understood.
The high temperature expansion of the Ising model is equivalent to hard dimers with a positive weight in a magnetic field.
Analytic continuation of the field then leads to imaginary field strengths and dimers with negative weight.
This is related to the Lee-Yang edge singularity  \cite{Kurtze:1979zz}.

In simulations, however, the need for negative dimer weights is hard to interpret.
Normally weights would be understood as (unnormalised) probabilities that can then be used to calculate partition functions. 
Negative weights cannot be categorised in such a way, which makes it hard to implement them in computer simulations.
Thus we do not know how to examine the theory further or extend the analysis to higher dimensions, for which we do not have analytic results.

\section{\label{sec:HL}Connection to foliated continuum theories}

CDT is as a discretisation of the path integral for gravity. 
While the existence of a second order phase transition hints that a continuum theory exists, it does not tell us what this continuum theory is.
The original hope was that the continuum limit of CDT would lead to quantum Einstein gravity, a quantum theory of gravity at its asymptotically safe fixed point.
However, as mentioned above when discussing the phase diagram and the spectral dimension, there are similarities to Ho\v{r}ava-Lifshitz (HL) gravity.
These were first presented in \cite{Ambjorn:2010hu}.

Ho\v{r}ava-Lifshitz gravity is a proposed UV completion of general relativity \cite{Horava:2009uw}, c.f.  \cite{Sotiriou:2010wn} for a good review.
General relativity is not power counting renormalisable.
It is an effective theory, that is only valid for a limited domain of energies and has to be superseded by a new theory at high energies.
It could be made power counting renormalisable by including higher order curvature terms.
These higher order derivatives in time lead to ghost degrees of freedom and thus turn the theory non-unitary.
HL gravity is inspired by the anisotropic scaling of space and time that is observed in many condensed matter systems.

The important insight is, that if time scales as $t \to a^z t$ while space scales as $x \to a x$ this can lead to a power counting renormalisable theory.
With this scaling the mass dimension of space is $[x]=-1$, and that of time is $[t]=-z$ .
From this, we can calculate the mass dimension of $\kappa=\frac{1}{16 \pi G_N}$ as $[\kappa]=d-z$, in $d+1$-dimensional space-times.
Thus for $z \geq d$ the mass dimension of the coupling becomes less than or equal to zero, rendering the theory power counting renormalisable.

This scaling breaks diffeomorphism symmetry  to the smaller group of foliation preserving diffeomorphisms $t \to \tau(t)$ and $x \to \xi(x,t)$.
This allows for the inclusion of new terms in the action.

Using the ADM parameterisation of the metric \cite{Arnowitt:1962hi}, the action for HL gravity in $d+1$ dimensions is \cite{Donnelly:2011df}\footnote{HL gravity is a Lorentzian theory. We will Wick rotate it later to compare to CDT but for the time being we are back in Lorentzian space-time.}
\begin{align}\label{eq:actiongeneral}
	I= \kappa \int \md t \int \md x^{d}\sqrt{g} N \left( K_{ij} K^{ij} - \lambda K^{2} -V(g_{ij},a_{i}) \right) \;.
\end{align}
Here $a_{i}= \nabla_{i} \log{N} $ these terms only arise in non-projectable Ho\v{r}ava–Lifshitz theory. 
In projectable Ho\v{r}ava–Lifshitz theory, one requires that $N=N(t)$.
This restriction is not fundamental, but ensures that the gauge freedom of the smaller symmetry group suffices to set $N=1$ as in general relativity. 
It also reduces the number of terms that need to be included in the potential.
The extrinsic curvature $K_{ij}$ is defined as
\begin{align}
K_{ij}=\frac{1}{2 N} \left(\partial_{t} g_{ij} - \nabla_i N_j - \nabla_j N_i \right)\;.
\end{align}
This action is only fixed up to boundary terms.
The potential is of the form
\begin{align}
V(g_{ij},a_{i})= - \xi R^{(d)} -\alpha a_{i} a^{i} -2 \Lambda + \dots \;,
\end{align}
where $R^{(d)}$ indicates that this is the Ricci scalar of the spatial metric $g$, and the $\dots$ stand for all terms up to scaling dimension $2z$.
To be a description of reality the low energy limit of this theory should match with general relativity.
Thus $\lambda$ should scale to one and all couplings in the potential, apart from the cosmological constant $\Lambda$, should go to zero.
There are variants of this theory including a variety of terms in the potential; 
for example not all non-projectable models include the $a_i$ terms, and many models use the condition of detailed balance, introduced in  \cite{Horava:2009uw}, to reduce the number of independent couplings.
The requirement of foliation preservation excludes all terms that are of higher order in $\partial_t$.
These are the terms that give rise to ghost degrees of freedom.
Thus the requirement of foliation preservation also preserves unitarity.

There are several indications that CDT and HL gravity might be connected.
Both theories describe a foliated version of general relativity that is anisotropic in space and time.

As mentioned above, the measured spectral dimension of CDT agrees with the calculated spectral dimension in HL gravity \cite{Horava:2009if}.
A further hint comes from the similar phase diagrams.
As expounded above, CDT contains three phases. 
In this, it is similar to an effective Lifshitz theory, which can for example describe a magnetic system.
Say we take a $d$-dimensional system and split these $d$ indices into two classes $\alpha=m+1, \dots,d$ and $\beta=1,\dots,m$, allowing us to take into account anisotropic scaling. 
The free energy of the theory can be written in terms of an order parameter $\phi(x)$
\begin{align}
F(x)= a_2 \phi(x)^2 + a_4 \phi(x)^4 + \dots + c_2 \left(\partial_\alpha \phi\right)^2 +d_2 \left(\partial_\beta \phi\right)^2 + e_2 \left(\partial_\beta^2 \phi\right)^2 \;.
\end{align}
The phase structure of this system depends on the coupling constants.
In an isotropic system $m=0$, and all directions are $\alpha$ directions.
If we take the couplings to depend on a parameter, call it temperature $T$, then the phase transition occurs at $a_2(T)=0$. 
Assuming that $a_4>0$, the phases are $\phi=0$ for $a_2>0$ and $|\phi|>0$ for $a_2<0$.
In a magnetic system, the phase with $\phi=0$ is paramagnetic, while $|\phi|>0$ is ferromagnetic. 
Including anisotropy, $m>0$, another phase transition arises when $d_2$ passes through zero.
For $d_2<0$, the order parameter $\phi$ oscillates in the $\beta$ directions.
This helical phase can be reached from both the para and ferromagnetic phases, depending on the sign of $a_2$.
The triple point where these three phases meet is the so-called Lifshitz point.

These three phases can be mapped to the phases of CDT.
Although no formal order parameter exists in CDT, we loosely equate it with the geometry.

Phase $A$, its geometry fluctuating in time, corresponds to the helical phase of a Lifshitz theory.
In phase $B$, all geometry is compressed into one slice.
The Hausdorff dimension in this phase is very high, so even though all simplices cluster together the true `extent' of the geometry is zero.
We can then equate this phase with the paramagnetic phase $\phi =0$.
The macroscopic, ordered phase $C$ is associated with the ferromagnetic phase of a Lifshitz theory.

Thus foliation, spectral dimension, and the phase diagram point to a connection between CDT and HL gravity.

To examine this further, we quantise projectable $2$d HL gravity and obtain its Hamiltonian, to compare to the CDT Hamiltonians (\ref{eq:CDTham1},\ref{eq:CDTham2},\ref{eq:CDTham3})\footnote{This section summarises work published in \cite{Ambjorn:2013joa} which is reprinted in Appendix \ref{app:HL}}.
We parameterise the metric as
\begin{align}
\md s^2 = - N(t)^2 \md t^2 + \left( \g(x,t) \, \md x + N^{(1)}(x,t) \md t\right)\left( \g(x,t) \, \md x + N^{(1)}(x,t) \md t\right) \;.
\end{align}
This leads to the action
\begin{align}
I= \int d t \; d x \; N \g \left( (1-\lambda) K^{2} - 2 \Lambda \right) 	
\end{align}	
with
\begin{align}
 K= \frac{1}{N} \left( \frac{1}{\g} \partial_{0} \g - \frac{1}{\g^2} \partial_{1}N_{1} + \frac{N_{1}}{\g^3} \partial_{1} \g \right) 
\end{align}
the external curvature. 
In projectable $2$d HL gravity, the only term in the potential is the cosmological constant $\Lambda$.
To quantise this theory, we calculate its Hamiltonian.
The only non-trivial momentum is
\begin{align}
\pd{\mc{L}}{(\partial_t\g)}:=\pg=2 (1-\lambda) K \;.
\end{align}
Using this we can write the Hamiltonian
\begin{align}
	H&= \int \md x \left[ N(t)\mc{H}+ N^{(1)}(x,t) \mc{H}_{1} \right]\;,
\end{align}
where we introduced the Hamiltonian constraint $\mc{H}$ and the momentum constraint $\mc{H}_1$
\begin{align}
	\mc{H}&= \g \frac{(\pg )^{2}}{4(1-\lambda)}+2 \Lambda &
	\mc{H}_{1}&=\frac{- \partial_{x} \pg}{\g}	\;.
\end{align}
The momentum constraint $\mc{H}_1=0$ tells us that $\pg=\pg(t)$.
Restricted to this constraint surfaces, $H$ can be expressed as
\begin{align}
H&= N(t) \left( L(t) \frac{\pg(t)^2}{4(1-\lambda)} + 2\Lambda L(t) \right) & \text{with}&& L(t)&= \int \md x\; \g(x,t) \;.
\end{align}
This introduces the loop length $L(t)$, which is invariant under foliation preserving diffeomorphisms.
$N(t)$ can be regarded as the Lagrange multiplier of the Hamiltonian. 
This fixes $\pg(t)$ to be
\begin{align}
\pg(t)^2=8(1-\lambda) \Lambda \;.
\end{align}
This requirement fixes the constraint surface to be a surface of constant extrinsic curvature.
Real solutions to this only exist for $(1-\lambda) \Lambda >0$.
For cases where \mbox{$(1-\lambda) \Lambda <0$} we can still force a genuinely dynamic evolution by treating the cosmological constant $\Lambda$ as a Lagrange multiplier.
An example of this treatment is $2$d Liouville gravity.

To simplify the notation, we can rescale 
\begin{align}
N(t) &\to \frac{N(t)}{4(1-\lambda)}&\text{and}&&\Lambda &\to 2(1-\lambda) N(t)
\end{align}
leading to
\begin{align}
H&= N(t) \left( L(t) \pg(t)^2 + \Lambda L(t) \right) \;.
\end{align}
In projectable HL gravity, the proper time
\begin{align}
\tau (t):= \int_0^t \md t' N(t') 
\end{align}
is a physical observable. 
Since the loop length is invariant under time redefinition, we can gauge fix to proper time $\tau$ and make $N(t)$ drop out of the action.
The classical evolution of the loop length is entirely determined by the action
\begin{align}
S_E=\int \md t \left( \frac{\dot L^2}{4 L(t)} +\Lambda L(t) \right) \;.
\end{align}
Here we Wick rotated to the Euclidian action, to make the connection to CDT in the next steps more obvious.
We can calculate the propagator for the loop length as
\begin{align} 
G(L_1,L_2;T)= \int \mc{D} L(t) e^{-S_E[L(t)]} \;.
\end{align}
This is equivalent to the propagator in equation \eqref{eq:therightpropagator}.
We can expand this for small times $\epsilon$
\begin{align}
G(L_1,L_2;\epsilon)= \bra{L_2} 1-\epsilon \hat H +\mc{O}(\epsilon^{\frac{3}{2}}) \ket{L_1}
\end{align}
and thus obtain the Hamiltonian operator
\begin{align}
\hat H = - \pd{}{L}L \pd{}{L} +\Lambda L \;.
\end{align}
An alternative way to obtain this result, is to gauge fix the Hamiltonian to $H=L \Pi^2 + \Lambda L$  and impose the canonical commutation relations
\begin{align}
\{ L,\Pi\}&=1& \to&  &[ \hat L,\hat \Pi]&=i &\qquad \text{and} \qquad  \hat \Pi&= -i \frac{d}{d L} \;.
\end{align}
Using this approach, we encounter the operator ordering problem
\begin{align}
\left(L\Pi^2\right)_0&= - \frac{d}{d L} L\frac{d}{d L} &\left(L\Pi^2\right)_{-1}&= -  L\frac{d^2}{d L^2} &\left(L\Pi^2\right)_1 & = - \frac{d^2}{d L^2} L \;.
\end{align}
Comparing these orderings to equations (\ref{eq:CDTham1},\ref{eq:CDTham2},\ref{eq:CDTham3}) in section \ref{sec:CDTintro}, we see that they do exactly correspond to the CDT Hamiltonians for the distinct measures.

We have thus identified the continuum theory of $2$d CDT, it is projectable Ho\v{r}ava Lifshitz gravity with $\lambda<1$ and $\Lambda>0$.
This knowledge will open up new opportunities of inquiry.

Since we have no analytic solutions for CDT in higher dimensions, we have few expressions to compare to higher dimensional HL gravity.
One possible path to strengthen this connection would be to examine more and new observables that can be compared between the theories.
If one were able to extract analytic results from higher dimensional CDT, it might be possible to verify the connection there in the same manner as we did for $2$ dimensions.

CDT in $2$ dimensions can be generalised to allow for a controlled branching off of baby universes \cite{Loll:2005dr}. 
This theory might be a candidate to connect to non-projectable HL gravity in $2$d, but this has not been successful so far.

In $2$ dimensions, projectable HL gravity is self consistent, however in higher dimensions projectable HL gravity is plagued by instabilities \cite{Sotiriou:2009} and massive scalars \cite{Charmousis:2009tc}.
It is unclear how these pathologies will influence this programme of research.

One of the reasons the connection between HL gravity and CDT was conjectured, was that both theories do have a time foliation.
Recent work in CDT introduced a reformulation that obtains the same results without requiring a strict time foliation \cite{Jordan:2013iaa,Jordan:2013awa}.

One possibility would be, that this foliation-free CDT is equivalent to Einstein Aether theory which does not have a strict foliation but contains a timelike vector field.
Einstein Aether theory has been shown to be equivalent to HL gravity \cite{PhysRevD.81.101502} if the vector field can be written as the divergence of a potential.

\section{\label{sec:EDT}Higher curvature terms in EDT}
There are two main differences between CDT and EDT.
One is that CDT only includes geometries with a time foliation, which makes it possible to ascertain a unique time direction and thus a Wick rotation.
The other difference is the parameter $\alpha$.
This parameter quantifies the anisotropy between timelike and spacelike edges. 
Its introduction is important for the Wick rotation, since it allows us to understand the Wick rotation as a parameter Wick rotation as illustrated in  \cite{Sorkin:2009ka}.

The research into EDT stalled, because its phase diagram did not contain a higher order transition. 
In  \cite{Laiho:2011zz}, it is suggested  that EDT might not have a higher order phase transition, because it only contains one dynamical coupling.
It is then an interesting idea to introduce a new coupling that could give rise to a new phase and maybe a higher order phase transition.

In  \cite{Laiho:2011zz}, the partition function of EDT is generalised to include terms that can be interpreted as couplings to higher order terms in the Ricci scalar $R$
\begin{align}
\mc{Z}_{E}=\sum_{T\in \mc{T}} \frac{1}{C(T)}\left[ \prod_{t=1}^{N_2} o_t^{\beta} \right] e^{-\mc{S}_E}\;.
\end{align}
The sum is over labelled triangulations $\mc{T}$, taking into account the combinatorial factor $C(T)$. 
The new coupling $\beta$ is introduced as part of a measure factor, and the product in this measure factor runs over all $N_2$ triangles of each configuration.
The order of a triangle $t$ is $o_t$, defined as the number of four-simplices the triangle belongs to. 
The Regge action $\mc{S}_E$  for dynamical triangulations is
\begin{align}
\mc{S}_E = -\kappa_2 N_2 + \kappa_4 N_4 \;,
\end{align}
as above.
One can interpret the measure factor $o_t^\beta$ as a coupling to higher order corrections in the curvature.
The deficit angle in four dimensions depends on the number of four-simplices meeting at a given triangle.
Changing the order of triangles then changes the curvature.
Writing
\begin{align}
o_t^\beta&= e^{\beta \log{o_t}}
\end{align}
and formally expanding the $\log{}$ as a power series in $o_t$, $\beta$ is a coupling to all powers of the curvature.
After the first claims in  \cite{Laiho:2011zz} and  \cite{Coumbe:2012qr}, we decided to investigate this new idea\footnote{The results presented in this section are from  \cite{Ambjorn:2013eha} which is reprinted in Appendix \ref{app:EDT}}.
We use MC simulations to map a region in the $(\kappa_2,\beta)$ parameter plane for geometries consisting of $N_4=160k$ four-simplices.
To examine the geometries, we chose to record six observables, the average number of triangles $N_2$, the average number of vertices $N_0$, and the susceptibilities of these two
\begin{align}
\chi(N_0)&= \frac{\av{N_0^2} -\av{N_0}^2}{N_4} & \chi(N_2)&= \frac{\av{N_2^2} -\av{N_2}^2}{N_4} .
\end{align}
The last observables are the radius volume profile $V(r)$, and the average radius $\av{r}$.
The graph geodesic distance between two four simplices is the number of steps that one takes between them on the shortest route, the set of all possible routes being the graph that connects the centres of neighbouring four-simplices. 
For a given simplex $i_0$, the volume $V(r,i_0)$ is given by the number of simplices at geodesic distance $r$ from it.
Using this, we define the radius volume profile $V(r)$ and the average radius $\av{r}$
\begin{align}
V(r)&= \av{\frac{1}{N_4}\sum_{i_0} V(r,i_0)}_{\text{conf}} &\av{r}&= \frac{1}{N_4} \sum_r r \cdot V(r)\;.
\end{align}

\begin{figure}
\includegraphics[width=0.5\textwidth]{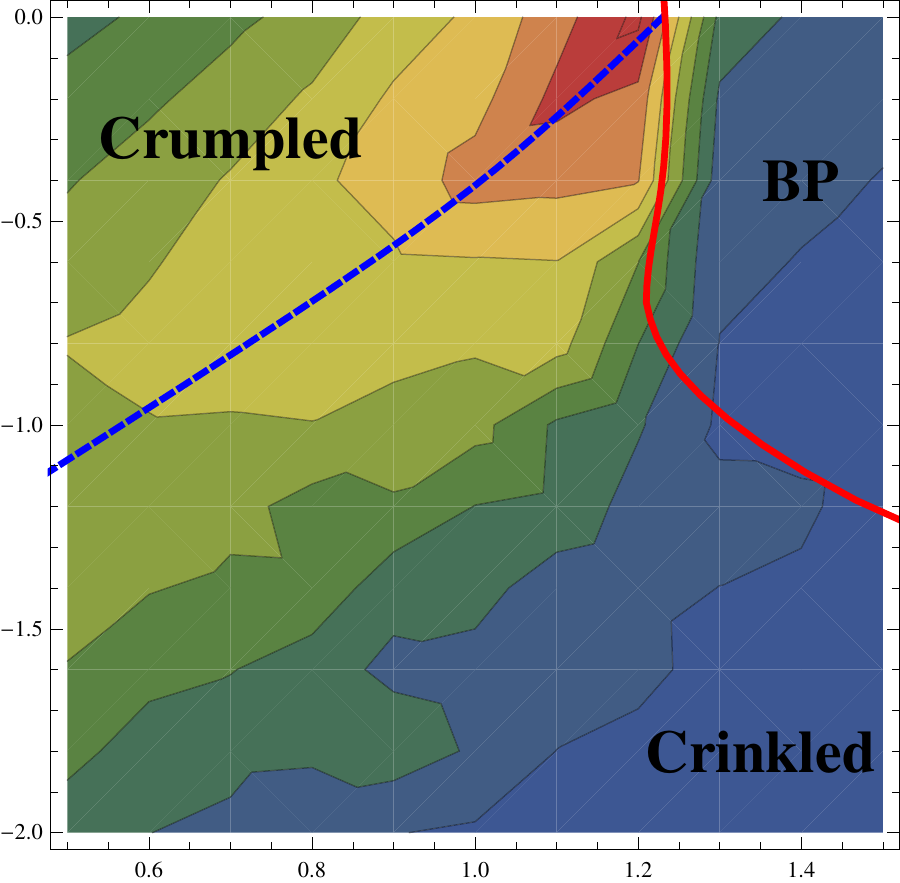}\hfill
\includegraphics[width=0.5\textwidth]{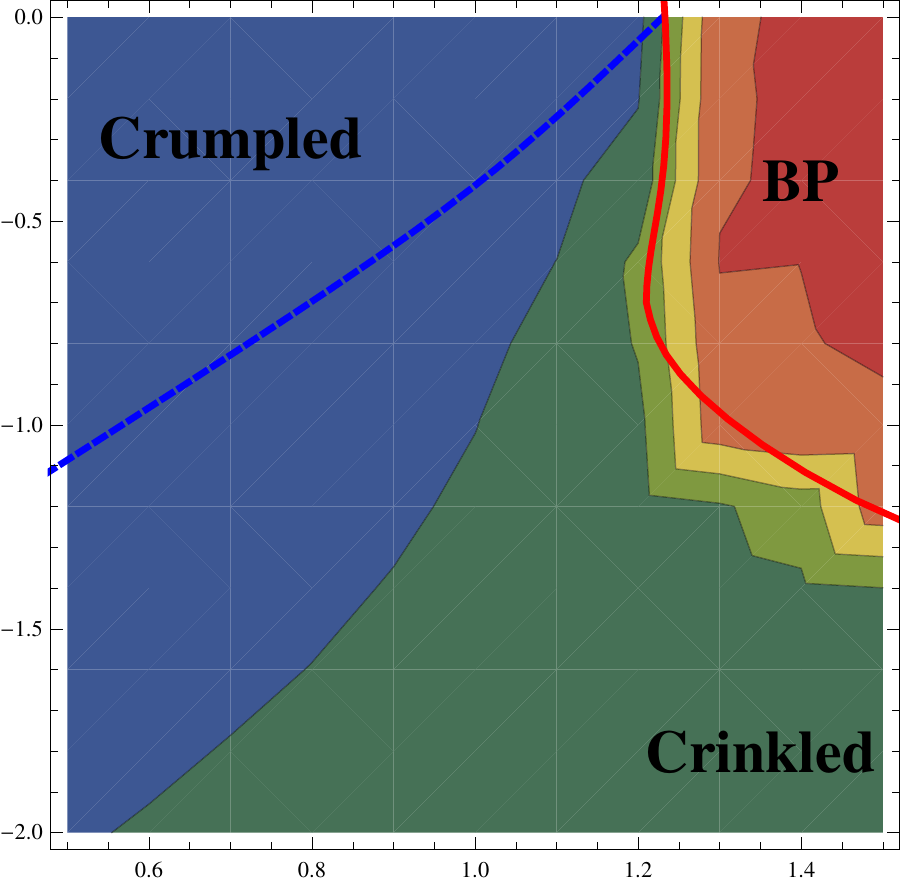}
\caption{\label{fig:EDT_phases}The left hand figure shows the variance of $N_0$, a peak in this variance would be a possible sign of a phase transition. The peak at the phase transition between the crumpled and the branched polymer phase at $\beta=0$ extends out into the $(\kappa_0,\beta)$ plane, but becomes lower with smaller $\beta$, so there is no phase transition. We mark the trajectory of the peak with a blue, dashed line. On the right hand side we plot the average radius of the geometry. Here we can clearly delineate the branched polymers from the crumpled and crinkled phases, the phase transition is marked in red. The difference between crumpled and crinkled phase, however, is very small, and the change progresses smoothly.  These figures are taken from  \cite{Ambjorn:2013eha}.}
\end{figure}
We examine these observables for signs of a phase transition. 
After careful consideration, we find that there is only one phase transition, stretching out from the transition between crumpled geometries and branched polymers at $\beta=0$.
In Figure \ref{fig:EDT_phases}, we show plots of $\chi(N_0)$ and $\av{r}$ in the $(\kappa_2, \beta)$ plane.
To understand the geometries better, we analyse the structure of baby universes. 
A baby universe is a part of the triangulation that is only connected to the remainder of the triangulation by a minimal neck.
A minimal neck consists of five four-simplices that are connected to each other such that they encase an `empty' four-simplex, a simplex that is not part of the triangulation.
Cutting the geometry at such a minimal neck leads to two disconnected parts, which each have a boundary consisting of these five simplices.
The study of minimal baby universes is used in  \cite{Ambjorn:1993vz} and  \cite{Ambjorn:1993sy} to examine two and four-dimensional dynamical triangulations.

\begin{figure}
\includegraphics[width=0.3\textwidth]{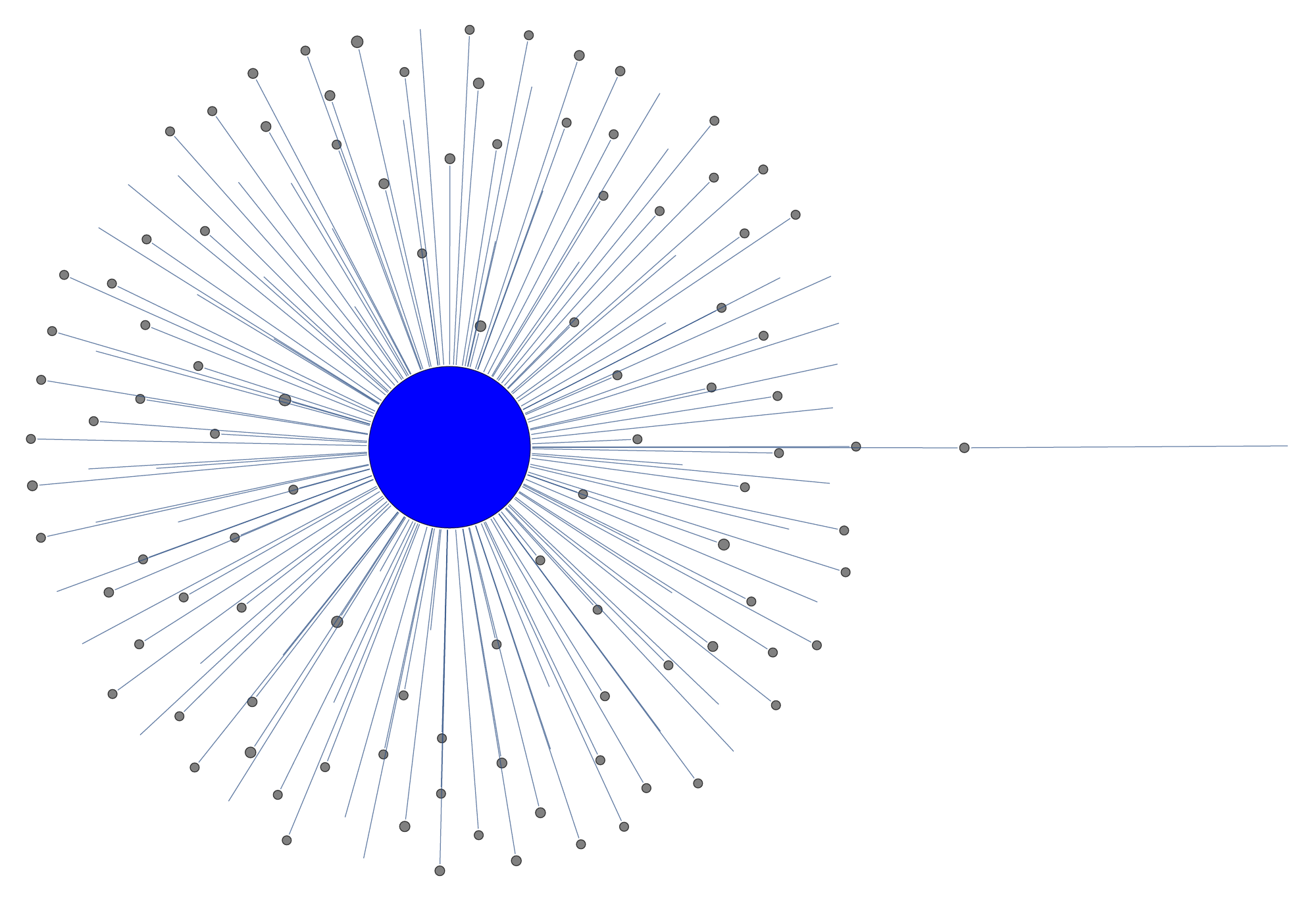}\hspace{15pt}
\includegraphics[width=0.3\textwidth]{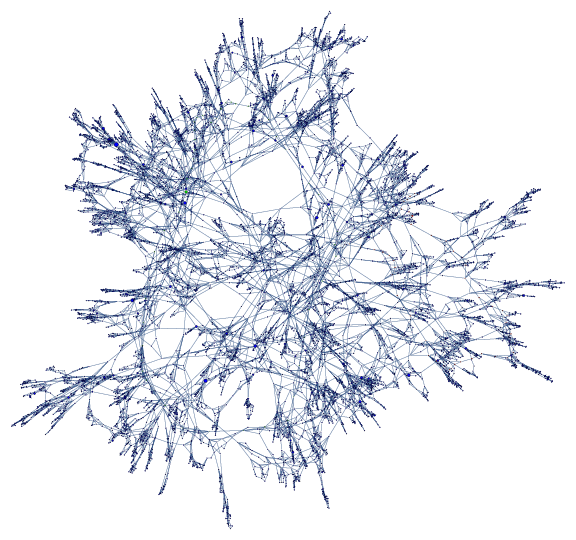}\hspace{15pt}
\includegraphics[width=0.3\textwidth]{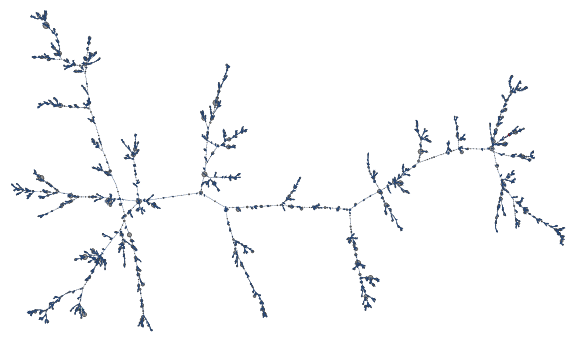}
\caption{\label{fig:EDT_quali} These figures show the baby universe structure of the geometries in the three phases. 
Every circle denotes a baby universe, and every line denotes a connection between these baby universes. 
The outer left picture shows the crumpled phase, in which there are almost no baby universes. 
The geometry is concentrated in one universe, the big blue blob at the centre, from which very few and small baby universes branch out. 
In the crinkled phase, the middle picture, the baby universes form a dense network. 
This is because the additional coupling gives rise to triangles of high order, at which several baby universes start. 
The last plot on the right shows the branched polymer phase. This is a treelike structure, where baby universes grow from other baby universes. 
These figures are taken from  \cite{Ambjorn:2013eha}.}
\end{figure}

While there is a qualitative difference between the geometries in the crumpled and crinkled regions of phase space, the change from one to the other happens gradually in a crossover type behaviour.
We thus identify three regions that are qualitatively different.
\begin{itemize}
\item {\bf Branched Polymer:} Geometries in this phase are elongated $r \propto N_4^{1/2}$. 
They consist of many baby universes and contain many minimal necks, leading to a treelike structure as shown in Figure \ref{fig:EDT_quali}. 
The probability to find a baby universe of size $V$ is $P(V) \propto V^{\gamma-2}(N_4-V)^{\gamma-2}$, where $\gamma=1/2$ is the string susceptibility exponent. 
The Hausdorff dimension in this phase is $d_h=2$ and the spectral dimension $D_s=4/3$.
\item {\bf Crumpled:} In this phase, the geometries are collapsed. 
The average radius $\av{r}$ grows slower than $N_4^\alpha$ for any $\alpha>0$. 
The geometry is collapsed around two singular vertices and one singular link, the vertices having order $o_v \propto N_4$ and the link having order $o_l \propto N_4^{2/3}$. 
There are very few baby universes, and those that exist are very small. 
Thus the susceptibility exponent $\gamma$ cannot be measured (formally it could be taken to be $\gamma= -\infty$). 
The Hausdorff dimension in this phase is $d_h=\infty$, as is the spectral dimension $D_s$.
\item {\bf Crinkled:} The `new' phase that arises through the coupling $\beta$. 
Its properties are an interpolation between the branched polymer and the crumpled phase, but seem to approach the crumpled phase with increasing volume. 
$\av{r}$ is larger than in the crumpled phase, but still grows slowly with $N_4$. 
The new coupling causes triangles of high order $o_t \propto N_4^{0.16}$, these triangles of high order do not arise in the other two phases.
Although there are many baby universes their size is still small, thus it is again unfeasible to measure $\gamma$. 
These minimal baby universes build a `treelike' structure, but are connected to loops through the triangles of high order (c.f. Figure \ref{fig:EDT_quali}). 
The Hausdorff dimension $d_h$ is large and most likely infinite, and the same is true for the spectra dimension $D_s$.
\end{itemize}
To examine how the behaviour of the phases scales with the volume of the simulation, we examine a path in the $(\kappa_2 ,\beta)$ plane that transverses all three phases.
It thus has to cross a hypothetic phase transition between the crumpled and the crinkled phase. 
The path follows a hook from $(\kappa_2=0.5,\beta=0)$, in the crumpled phase, to $(\kappa_2=2.0,\beta=-2.0)$, in the crinkled phase, and ends at $(\kappa_2=2.0,\beta=-1.0)$, in the branched polymer phase. 
We run simulations along this path at volumes of $N_4=40k,80k$ and $160k$. 

We do not detect any signal of a phase transition between the crumpled and the crinkled phase along this line. 
More detail on this investigation is provided in  \cite{Ambjorn:2013eha} (reprinted in Appendix \ref{app:EDT}).

Since there is no phase transition between the crumpled and the crinkled phase, the only chance for a continuum limit is if the phase transition between the crinkled and the branched polymer phase is of higher order.
This question is investigated carefully in  \cite{Rindlisbacher:2013gka}, and they determine that the transition is of first order.

We can thus conclude that while the new coupling leads to a qualitatively new region, it does not lead to new phase transitions or to a viable continuum limit for EDT.

\section{The road forward in CDT}
Currently, causal dynamical triangulations is the only fully non-perturbative theory of quantum gravity, that has shown that it can reproduce genuinely classical large scale behaviour.
Discretising space-time through $d$-simplices does not have to lead to a $d$-dimensional space-time, as is exemplified by the theory of dynamical triangulations.
Yet, CDT recovers an extended phase that has four-dimensional behaviour.

The three phases of CDT can be understood as contrasting effects dominating the path integral.
In phase $A$, where the coupling $\kappa_0$, corresponding to the inverse Newton's constant, is large, the geometry fluctuates strongly.
The distribution at one time slice does not influence the distribution in the next.
This can be interpreted as a decoupling of gravity. 
In phase $B$, $\kappa_0$ is small leading to a collapse of geometry.
The interesting phase $C$ then presumably arises as a result of entropic contributions, its existence requires the anisotropy between time and space.

Simulations show that the phase transition between phases $B$ and $C$ is of second order and thus should allow for a continuum limit.

In section \ref{sec:dimers}, we summarised the article  \cite{Ambjorn:2012zx}.
Coupling dimers to CDT allows for a simple model of matter that leads to a new universality class of the theory.
While the analytic work in $2$ dimensions is simple, it is unclear how to extend this model to higher dimensions.
The fact that the weight $\xi$ is negative renders simulations impracticable and needs to be examined further.
One proposal is that for each triangulation $T$ the sum over all possible decorations with dimers could lead to an overall positive weight that can then be used in simulations.

In the article \cite{Ambjorn:2013joa}, which is summarised in section \ref{sec:HL}, we show that in $2$d CDT and projectable HL gravity have the same Hamiltonian.
It is not clear, how this will carry over to higher dimensions, since HL gravity in higher dimension requires the explicit introduction of higher order curvature terms.
In CDT, such curvature terms are not added explicitly.
Constructing such higher order curvature invariants from simplices is a non-trivial task (c.f.  \cite{Anderson:2011bj}).
However, the sum over configurations that calculates the path integral in CDT picks up entropic contributions.
These contributions are quite important in the continuum phase $C$ and introduce higher order curvature terms in the effective action.

Section \ref{sec:EDT} examined the idea that EDT with an additional coupling might lead to a new, higher order phase transition, which we clearly disprove.
However, EDT is still an active area of research with analytic methods, and proponents of the tensor track believe that repeated scalings of the theory might lead to a continuum limit \cite{PROP:PROP201300032}.

Recently a modification  of CDT that circumvents the strict foliation has been proposed \cite{Jordan:2013iaa,Jordan:2013awa}.
Instead, a local causality condition leads to a space-time that still has a local time direction but more freedom.
While first results of simulations in $2$ and $3$ dimensions are impressive, the simulations need to overcome technical difficulties in $4$ dimensions.
The new freedom also makes it harder to solve the theory analytical.
For example, the transfer matrix method, which was demonstrated above, cannot be used.

As a non-perturbative lattice theory of quantum gravity, CDT can be useful to test asymptotic safety ideas; discrete theories can be understood very well through Wilsonian ideas.

Another interesting result in CDT is the study of renormalisation group flows in CDT itself.
In absence of an analytic expression, these flows are studied as lines of constant physics within the parameter space of the theory \cite{Ambjorn:2014gsa}.
These simulations hint at a flow towards the triple point, in which anisotropic scaling between space and time arises naturally.

\chapter{Causal set theory}
\section{Introduction to causal set theory}
Causal set theory (CST) is a minimalist approach to quantum gravity.
Like CDT, it discretises the path integral of gravity.
In CDT, the discretisation is a short distance cut-off, a non-physical regularisation to be taken to zero at the end of the calculation.
Causal set theory, on the other hand, postulates that space-time is fundamentally discrete. 
A continuum limit is neither intended nor necessary.

Causality, here understood as the Lorentzian structure of space-time, is an interesting, yet not always easy to implement concept.
Riemannian metrics are, in many ways, easier to handle than Lorentzian ones.
It starts with their topology being conceptually easier.
For a Riemannian manifold $(\mc{N},d)$, we can define open balls of radius $\epsilon$ as $B_{\epsilon}(x)=\{y \in \mc{N}|d(x,y)<\epsilon\}$.
This definition uses that $d(x,y)\geq 0$, i.e. the metric is positive definite.
Every open ball is thus bounded.
The metric topology on $\mc{N}$ is then determined by the set of all open balls.

For a Lorentzian manifold $(\mc{M},g)$, the metric $g$ is not positive definite but has signature $(-1,1,1,1)$.
We can define a manifold topology, by using any Riemannian metric $g_{R}$ defined on $\mc{M}$.
However, these open balls are not natural objects of our Lorentzian space-time.

A completely Lorentzian topology is given through the interval topology, called the Alexandrov topology by physicists.
The basic open sets in the interval topology are given as $I(x,y)=\{z \in \mc{M}|$for $z$ causally in between $x$ and $y\}$, that is all points that lie on timelike curves between $x$ and $y$.
For a strongly causal space-time, the interval topology is identical to the manifold topology.

Another interesting topology is the path topology $\mc{P}$.
It was introduced by Hawking, King and McCarthy  \cite{Hawking:1976fe} and is defined as the finest topology that induces the Euclidian topology on arbitrary timelike curves.
This definition can be recast in a simple, physical form; a set is open if every observer times it as open, independent of the velocity or acceleration of the observer.
$\mc{P}$ is Hausdorff, connected, and locally connected, but neither normal, nor locally compact.
A basis for this topology can be defined as $B_\epsilon(p) \cap (C(p) \cup \{p\})$, with $p$ a point in $\mc{M}$ and $C(p)$ all points in the chronological future or past of $p$.
One can think of these as the interior of `cut off' light cones.
Based on this work, David Malament proves \cite{malament:1399} that a bijection between space-times that preserves future directed continuous timelike curves is a smooth conformal isometry.
In other words, if the causal structure of two space-times is identical, these space-times are the same up to conformal transformations.

The causal structure of a space-time describes the causal relationship between its points.
Two points $x,y$ can be either timelike $g(x,y)<0$,  lightlike $g(x,y)=0$, or spacelike $g(x,y)>0$ to each other.
If $x$ is to the past of $y$ and $g(x,y)<0$, then $x$ is in the chronological past of $y$, and an observer could travel between the two points.
If $x$ is to the past of $y$ and $g(x,y)=0$, then $x$ is on the past light-cone of $y$, and light can travel between the points.
If $g(x,y)>0$, then $x$ and $y$ are spacelike to each other, and nothing can travel between them.
These relations encode the causal structure of a space-time, and together with a time orientation they induce a partial order on all points of the space-time.
The causal structure can thus be described as a partially ordered set.

\begin{figure}
\center
\includegraphics[width=0.45\textwidth]{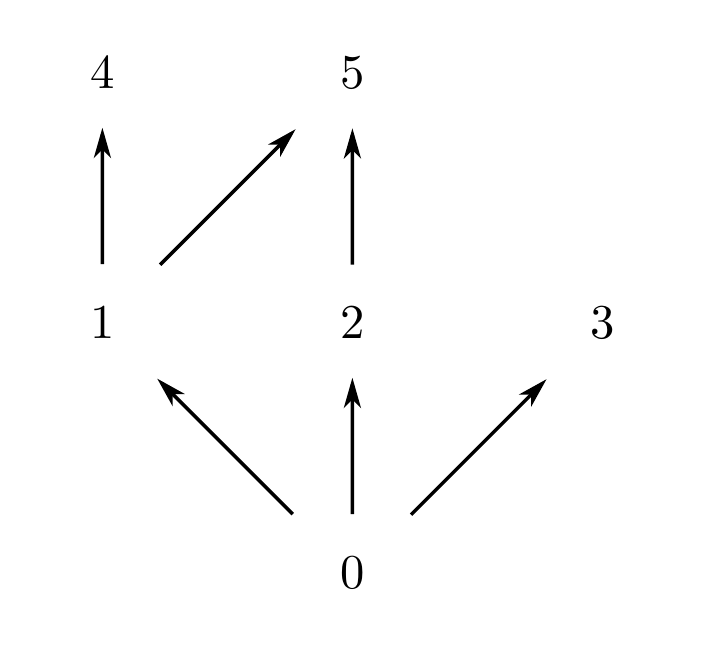}
\caption{\label{fig:causetIllu2}This is the Hasse diagram of a simple partial order. Element $0$ is linked to $1,2,3$ which we write as $0<1,0<2,0<3$ and also related to $4,5$ written as $0\p 4,0\p5$.}
\end{figure}

Formally a partially ordered set $\mc{C}$ with $x,y,z \in \mc{C}$ and the order relation $\peq$ is
\begin{itemize}
	\item {\bf reflexive} for all $x \in \mc{C}$ $x \peq x$
	\item {\bf transitive} for all $x,y,z \in \mc{C}$ and $x \peq y$ and $y \peq z$ then $x \peq z$
	\item {\bf antisymmetric} if $x,y \in \mc{C}$ and $x \peq y \peq x$, then $x =y$.
\end{itemize}
If $x\p y$, we say $x$ is in the causal past of $y$.
We say the points $x$ and $y$ are related, if either $x \p y$ or $y \p x$.
A partial order can be depicted through a so-called Hasse diagram, as illustrated in Figure \ref{fig:causetIllu2}.

If the volume is discretised with a fundamental volume $V_f$ associated to every discrete event, the partial order can also encode the volume information.
Thus a locally finite partially ordered set can describe space-time \cite{Bombelli:1987aa}.

To define local finiteness, we use the inclusive Alexandrov interval $I_A(x,y)$, defined as all points that lie causally between $x$ and $y$, for $x \p y$  
\begin{align}
I_A(x,y)=\{z \in \mc{C}| x \p z \p y \}\; .
\end{align}
Then $|I_A(x,y)|$ is the number of points in this interval, and a partially ordered set $\mc{C}$ is locally finite if $| I_A(x,y)| < \infty $ for all $x,y \in \mc{C}$.
The condition of local finiteness together with the fundamental volume $V_f$ associated with every event, allows us to fix the conformal degree of freedom, while the partial order is  equivalent to the causal order of events.
We call the relation $x\p y$ a link, if there are no elements causally in between $x$ and $y$, we then write $x \pl y$.
A series of events $x_1,x_2,\dots,x_n$ that are pairwise linked, i.e. $x_1<x_2< \dots <x_n$, we call a path.
A chain is a set of causal set element $e_1,e_2,\dots,e_j$, where $e_i \prec e_j$ for $i<j$, but not $e_i \pl e_{i+1}$ for all $i$.
Conversely, we call a series of events $x_1,x_2,\dots,x_n$ that contains no related elements an anti-chain.
For a maximal anti-chain $A$ there are no elements in the causal set that can be added to $A$ while keeping $A$ an antichain.

There are a few observables that can be reconstructed from the causal set quite easily.
The simplest one is the proper time distance $\tau_{xy}$ between two events $x,y$.
One can define $\tau_{xy}$ as the longest path connecting $x$ and $y$  \cite{Myrheim:1978ce}.

If our causal set encodes a space-time, one of the most fundamental things we want to know about the space-time is its dimension.
We then need a way to reconstruct the space-time dimension from the causal set.
Maybe the best-known example is the so-called Myrheim-Meyer dimension $d_m$.
For an interval $I$ that is embeddable into a region of volume $V$ of $\mathbb{M}^{d}$ with a density of points $\rho$, the expected number of $k$ chains is
\begin{align}
\av{S_k}=\frac{(\rho V)^k \G{\frac{d+1}{2}}\G{d+1}\G{d+2}^{(k-1)}}{2^{k-1}k \G{k(\frac{d+1}{2})}\G{(k+1)(\frac{d+1}{2})} }\; .
\end{align}
To measure the Myrheim-Meyer dimension $d_m$ of a causal set, we count the number of relations between two elements $S_2$ and calculate $d_m$ from this \cite{meyer_dimension_1988}.

Another simple estimator is the midpoint scaling dimension $d_s$ (as defined for example in \cite{Sorkin:2003bx}). 
It estimates the dimension of a causal set through the growth in volume if an interval is doubled in radius.
One defines the midpoint $m$ as the point $m$ in the interval $I(p,q)$ for which $N'=min(|I(p,m)|,|I(m,q)|)$ is maximal.
The midpoint scaling dimension $d_s$ is then given as $d_s=\log_2 \left(N/N'\right)$.

\begin{figure}
\vspace{20pt}
\center
\includegraphics[width=0.4\textwidth]{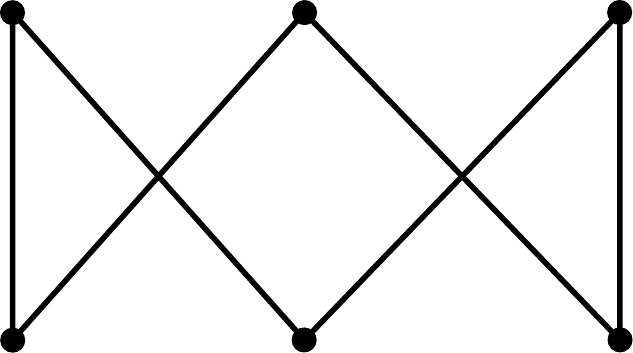}\vspace{25pt}
\caption{\label{fig:crownset}A crown poset. This poset cannot be embedded into $2$d Minkowski space, the relation between the two points in the middle would always be amiss.}
\end{figure}

A third dimension estimator is the embedding dimension.
While the first two estimators depend on average quantities over intervals of the causal set, the embedding dimension is a more local concept.
For any $\mathbb{M}^d$, there exist causal sets that cannot be embedded into it.
A simple example is the set consisting of two unrelated points. 
This cannot be embedded into $0+1$ dimensions.
In higher dimensions, one can define the pixy posets.
The simplest pixy poset is the one that cannot be embedded into $1+1$d.
It consists of six elements in two layers of three, we illustrate its Hasse diagram in Figure \ref{fig:crownset}.
There are generalisations of this to higher dimensions \cite{crownsets}.
To assign an embedding dimension to a causal set, we examine which pixy posets it contains as subsets. 
If we identify pixy posets of dimension up to $d_e$ but not higher, then the space-time has embedding dimension $d_e$.
This estimator will of course do quite badly if the causal set does not have uniform dimension of if the dimension of the space-time is different at different scales,  for example because it describes a space-time that contains regions of different dimension.

We can thus recover volume, dimension and timelike distances. 
What about spacelike distances?
A maximal antichain is the closest causal set analogue to a spacelike hyper surface.
But to establish where on this hypersurface a point is and how far two points are from each other, we need to use relations.
To reconstruct spatial information, we need at least some elements to the future and/or past of the maximal anti-chain.
Which elements to include has to be carefully chosen depending on the information we want to reconstruct \cite{Rideout:2009zh}.

In CDT, the spacelike hypersurfaces are fundamental ingredients; a foliation of the space-time is introduced by the discretisation. 
Spacelike hypersurfaces are also fundamental in approaches that are based on a Hamiltonian formalism.
This makes clear how fundamentally different causal set theory is.

\subsection{Die Hauptvermutung}

An embedding of a causal set is an injective map $\phi : \mc{C} \to \mc{M}$.
To every element of the causal set, it associates a point in the manifold so that the causal relations in the manifold and the partial order relations in the causal set are in agreement.
Such an embedding is said to be faithful, if on average the volume of a region is proportional to the number of points contained in it.
This should be true for any Lorentz invariant distribution of points, and in $2$d, there exists a class of so-called Lorentzian lattices for which it is \cite{Saravani:2014gza}. 
However, in higher dimensions Lorentzian lattices do not provide a faithful embedding.
Thus in practice, the distribution of points in a causal set is taken to be a Poisson distribution.
In a Poisson distribution the probability to find $m$ points in the volume $V$ if the fundamental volume is $V_f=1/\rho_f$ is
\begin{align}
P(m,V,\rho_f)= \frac{\left(\rho_f V\right)^m}{m!} e^{-\rho_f V}\;.
\end{align}
An important assumption in causal set theory is that a causal set $\mc{C}$ cannot be faithfully embedded into two macroscopically distinguishable space-times.
This is the so-called Hauptvermutung, which is proven in the infinite density limit \cite{Bombelli1989226}, but is still undecided for finite density.

A strong indication for the Hauptvermutung are the results in \cite{Major:2006hv}.
There, a mapping between discrete and continuum topologies for causal sets that faithfully embed into globally hyperbolic space-times is constructed.
To define the discrete topology on the causal set, they use thickened anti-chains.
A thickening of an anti-chain $A$ is defined as
\begin{align}
T_n(A) \equiv \bigg\{ p \bigg| |( \{p\} \sqcup \mathrm{Past}(p))\cap({A}\sqcup \mathrm{Fut}(A))  | \leq n \bigg\} \;,
\end{align}
with $p$ being a point to the future of $A$.
Then every element in $T_n(A)$ has a finite past in $T_n(A)$, as long as $n$ is finite.
We can define a set of maximal points in $T_n(A)$, as points which have no future links in $T_n(A)$.
The shadow of such a maximal point $p_i$ in $T_n(A)$ onto the anti-chain $A$ is called $A_i$
\begin{align}
A_i = \bigg\{ p \bigg| \mathrm{Past}(p_i) \cap A\bigg\} \;.
\end{align}
These shadows cover the total anti-chain completely and can be used to construct a nerve simplicial complex that carries the topology.
The nerve simplicial complex consists of a vertex for every $A_i$ and a $n$-simplex for every non-empty overlap between $n$ shadows.
In  \cite{Major:2006hv}, it was shown that this simplicial complex leads to the correct continuum homology for a wide range of thickenings $n$, if $A$ is  suitably chosen.

One way to test whether a causal set faithfully embeds into a space-time, is to construct an explicit embedding.
Currently no construction for exact embeddings is known, but an algorithm that yields an approximate embedding is described in  \cite{Henson:2006dk}.
To embed $N$ points $e_0, e_1 , \dots, e_{N-1}$ into an interval of flat space, we assume that all points lie between $e_0$ and $e_1$.
For simplicity, let us assume that the causal set embeds into $2$d Minkowski space.
We can immediately assign the coordinates $(u_0,v_0)=(0,0)$ and $(u_1,v_1)=(\sqrt{N},\sqrt{N})$, where we chose to work in light-cone coordinates $u,v$.
To embed the remaining points, we need to assign coordinates to them.
To assign coordinates to a point $e_i$, we first calculate its distance from the top and bottom points $g(e_0,e_i)$ and $g(e_1,e_i)$ using equation \eqref{eq:Mvolume} above.
Simple geometry then leads us to
\begin{align}
X&= \frac{1}{2} \left(g(e_0,e_1)^2+ g(e_0,e_i)^2-g(e_1,e_i)^2 \right)				\\
u_i&=\frac{1}{2v_1}\left( X \pm \sqrt{X^2 -g(e_0,e_1)^2 g(e_0,e_i)^2}\right)			\\
v_i&= \frac{g(e_0,e_i)^2}{2 u_i}\;.
\end{align}
Fluctuations in the distance measure could lead to imaginary coordinates, in which case the imaginary part is simply ignored.
We still have a left right symmetry through the $\pm$.
This is fixed by choosing one sign to place the element $e_2$ in relation to $e_0, e_1$. 
After this, we need to place all elements related to $e_2$ consistent with this choice.
For an element to the past of $e_2$ we calculate its position using $e_0,e_1$ and then $e_0, e_{2}$(or conversely $e_2,e_1$ for an element to the future of $e_2$).
Each calculation leads to two possible positions for the point, so four positions in total, two of which should approximately match. 
We then choose one of the approximately matching two as the position of the point. 
Not all elements will be related to $e_2$.
We can then pick an unrelated element $e_j$, chose one of the two coordinate assignments for it, and place all points related to it accordingly. 
In any set, we will only need to arbitrarily fix a small number of elements.
The construction can be generalised to higher dimension by comparing to additional points to fix the higher dimensional spherical symmetry.

We can estimate the quality of this embedding, by comparing the original causal order to the causal order in Minkowski space induced by the embedding. 
The rough scheme outlined above leads to a rate of improper relations of just $\mc{R}=1.7 \%$.
This can be improved by including a relaxation phase, in which the elements move around slightly to improve the relation matching, into the algorithm.

\begin{figure}
\center
\includegraphics[width=0.65\textwidth]{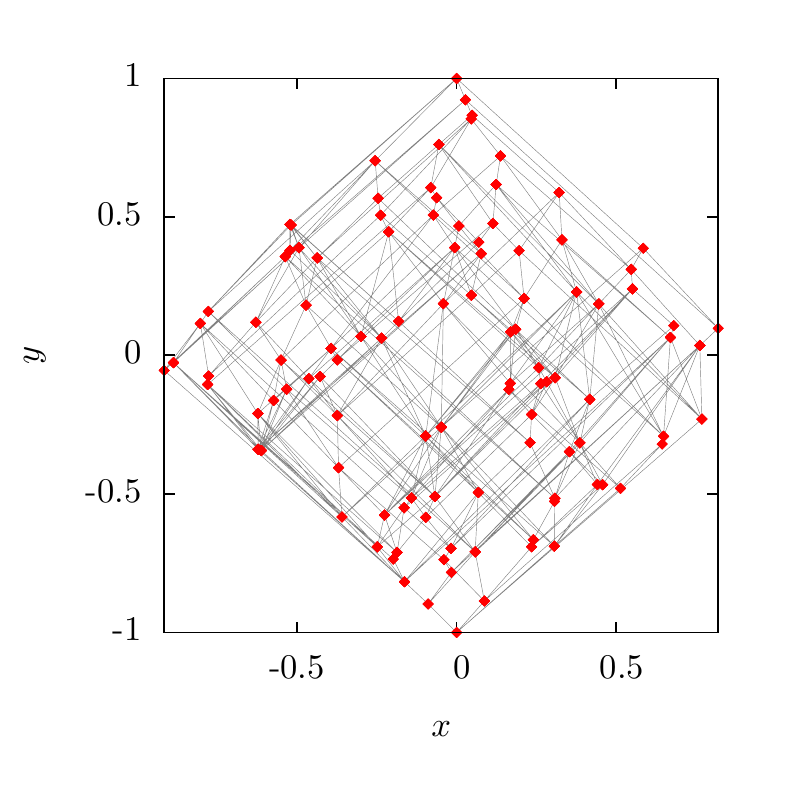}
\caption{\label{fig:causetIllu}A $100$ element sprinkling into flat $2$d space. The grey lines mark links between two elements.}
\end{figure}

The inverse process, generating a causal set from a known space-time  $(\mc{M},g)$, is called a sprinkling.
The sprinkling process is well understood and often used to generate causal sets as test-beds for new methods.
To generate a sprinkling, we chose points in the manifold to include in the causal set at random, according to a Poisson distribution.
After thus picking the elements of our causal set, we can calculate the partial order between them using the manifold and its metric.
Once this is done, all information that $\mc{M}$ encodes above the discreteness scale should be contained in the partial order.
Figure \ref{fig:causetIllu} illustrates a $100$ element causal set sprinkled into Minkowski space.

One common problem for approaches to gravity that introduce a fundamental discrete structure is Lorentz invariance.
The bounds on Lorentz violations are extremely tight \cite{LV}, so any theory that allows Lorentz violations needs to suppress them.
The elements of the causal set are Poisson distributed in the manifold.
This distribution only depends on the space-time volume $V$.
The discrete structure of causal sets is thus random in a way that makes it impossible to specify a frame and respects Lorentz invariance.

This Lorentzian randomness introduces non-locality into the causal set.
Consider a point $p$ and its nearest neighbours. 
In a Riemannian space-time, these nearest neighbours are contained in a small open ball $B_\epsilon(P)$ with $\epsilon$ proportional to the discreteness scale. 
The nearest neighbour is the element that is closest to the point $p$.
In a Lorentzian space-time, points that have metric distance zero from $p$ lie on its light-cone.
The light-cone is a non-compact object, thus a nearest neighbour in a Lorentzian setting is not actually `near' in any intuitive way.
If we discretise an infinite space-time randomly, there will always be infinitely many elements lying arbitrarily close to the lightcone.
In a hypothetical infinite causal set that would embed into Minkowski space-time, every element would be of infinite valency.
The infinite valency is necessary to preserve Lorentz invariance, since the invariant subspaces under Lorentz boosts are non-compact.
Thus insisting on Lorentz invariance and discreteness leads to non-local effects.

It can be proven that this infinite valency and randomness preserve Lorentz symmetry \cite{Bombelli:2006nm}.

The Poissonian distribution of causal set elements is instrumental in the prediction of the cosmological constant \cite{Sorkin:1990bj}.
The argument goes as follows:
If space-time is discrete with $V_p=l_p^3 t_p$, the universe consists of $N=10^{246}$ elements.
Assuming unimodular gravity, the cosmological constant $\Lambda$ is the conjugate of the size of the universe $N$.
The Poissonian distribution of space-time leads to fluctuations in $N$ of the order $N^{1/2}$.
Then, assuming a mechanism that sets the cosmological constant to zero, there would still be a residual fluctuation around it, of order $N^{-1/2}\sim 10^{-123}$.
The observed value of $\Lambda \sim 10^{-122}$ is of this order of magnitude.
Of course, this reasoning also predicts a change in the cosmological constant over time \cite{Ahmed:2002mj}. 
This could, for example, be tested through the study of gamma ray bursts.

\subsection{\label{subsec:pathintegral}The path integral as a sum over partial orders}
To define a quantum theory of gravity from causal sets, we use the path integral.
Instead of Wick rotating before discretising, as done in CDT, in CST the path integral is just discretised.
It then becomes
\begin{align}
\mc{Z}_{\CS} &= \sum_{\mc{C}\in \mathrm{C}} e^{\frac{i}{\hbar} \mc{S}_{\CS}(\mc{C})}\;,
\end{align}
where instead of integrating over geometries we now sum over a class of causal sets $\mathrm{C}$.
The weight is given by a causal set action $S_{\CS}(\mc{C})$ that we will need to specify.
One proposal for the causal set action, will be presented in section \ref{sec:factor}.
In addition to choosing an action, we will need to decide which class of causal sets  $\mathrm{C}$ to sum over.
There are several contrasting approaches to this.

The simplest proposal is to just sum over the class of all partial orders.
This has two problems. 
The first is that it is a very large class.
As stated above, assuming $V_f=V_{pl}$, the universe contains $N=10^{246}$ elements.
Partial orders have been studied in detail from a mathematical point of view.
For a $N$ element set, there are $P_N \sim 2^{N^2/4+\mc{O}(N)}$ possible partial orders \cite{MR0369090}.
This makes the proposition of summing over all possible orders an ambitious undertaking.

The other problem is, that not every partial order corresponds to a space-time.
In fact, there exist partial orders of size $\sim 100$ that do not faithfully embed into any Minkowski space \cite{0264-9381-15-11-009}.
The sum over all possible partial orders then needs to suppress all non-manifoldlike orders through the action.

\begin{figure}
\center
\includegraphics[width=0.8\textwidth]{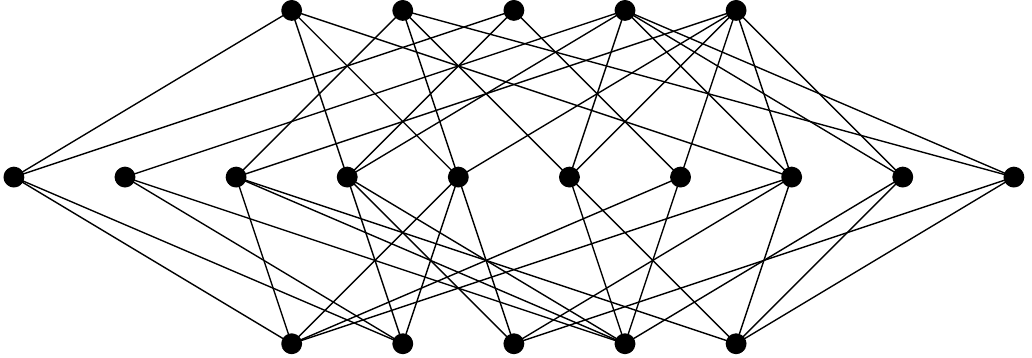}
\caption{\label{fig:KRorder}Illustration of a 20 element KR order. The top and bottom layer each include $5$ elements. Each of these elements is connected to about half of the $10$ elements in the middle layer.}
\end{figure}

In \cite{MR0369090} it is shown, that the number of so-called Kleitman-Rothschild (KR) orders is asymptotically equal to the number of all partial orders.
A KR order is a partial order consisting of three layers $L_1,L_2,L_3$.
The layers $L_1,L_3$ each contain $N/4+o(N)$ elements, with the remaining elements in the layer $L_2$.
Then each element in the layer $L_1$ is linked to roughly half of the elements in layer $L_2$, and each element in $L_2$ is linked to roughly half of the elements in layer $L_3$, this is illustrated in Figure \ref{fig:KRorder}.
Thus the action needs to suppress these orders stronger than $N^2/4$, to have a chance at recovering space-time.
If the KR orders were the only pathological class of partial orders, we might consider adding an explicit counter term to the action to suppress them.
However, after the KR orders there is a next class of KR-like orders that contain $2$ layers. 
These $2$-layer orders grow slower than the KR orders but faster than all other orders.
After these $2$-layer orders there are $4$-layer orders, and this continues so that there are countably infinitely many classes of pathological partial orders \cite{Dhar:1980}.

Another proposal for the class of causal sets to sum over, is all sets originating in a growth dynamics.
We could then determine their weights by the probability for them to grow according to this dynamics.
This type of dynamics routinely assigns very small probabilities to KR orders and can thus serve to suppress them.
It can also circumvent the need for a causal set action, since the weight of a causal set is assigned according to the growth dynamics.

\begin{figure}\center
\includegraphics[width=0.8\textwidth]{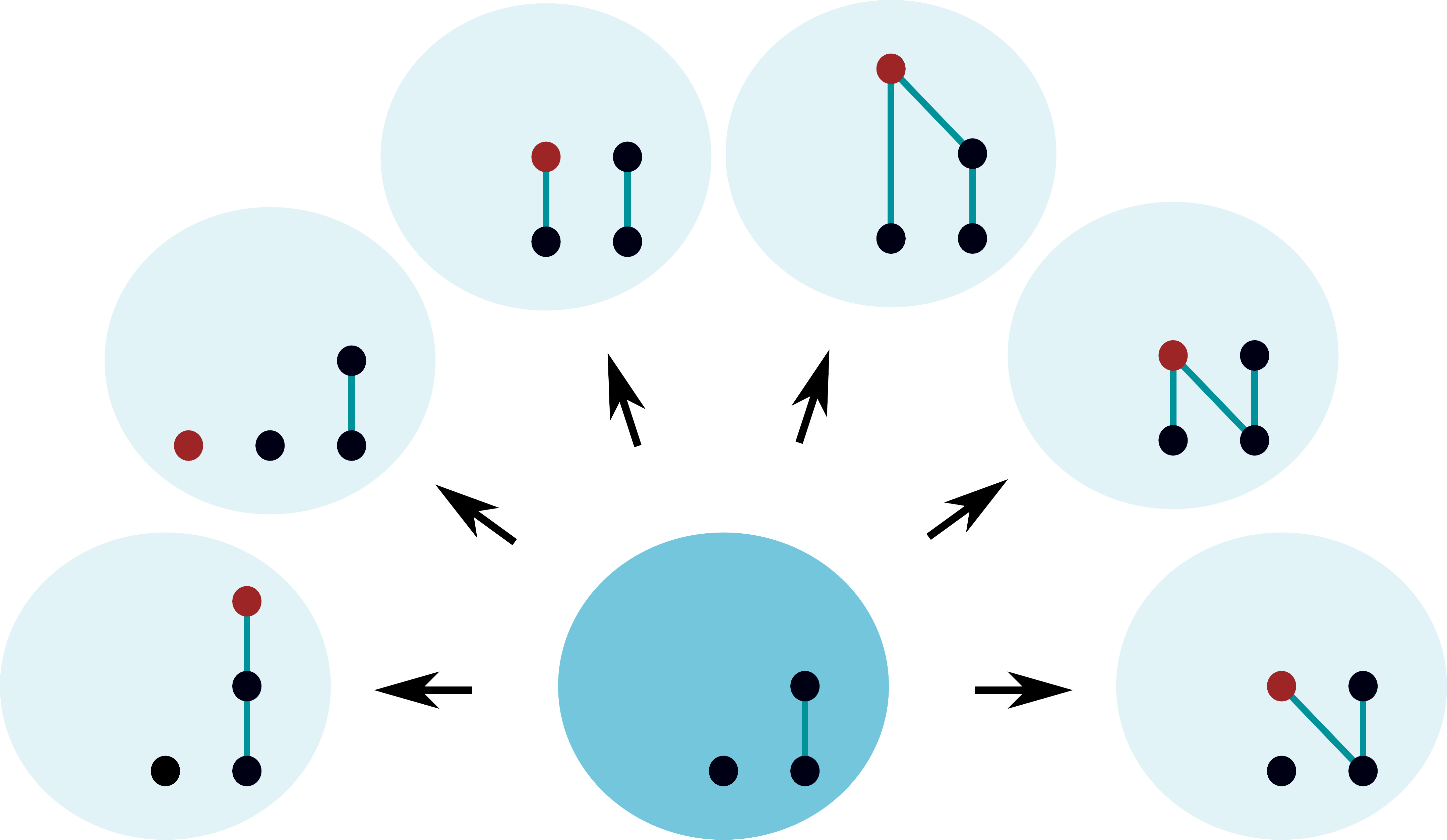}
\caption{\label{fig:seqgrow}Illustration of a sequential growth dynamic for a causal set. To the original set, the darker on in the middle, a new point, marked in red, is added in all possible ways.}
\end{figure}

Most growth dynamics for causal sets are sequential, starting from one first element they add further elements one by one.
Each element added chooses a set of the elements already in existence to be in its past, as is illustrated in Figure \ref{fig:seqgrow}.
If this choice is random, we need to take the transitive closure, adding the relations that are implied by transitivity.
In classical sequential growth dynamics, the probability for the $(k+1)$-th point to take a precursor set $S$ of the preexisting $k$ elements into its past is
\begin{align}
P(S)=\frac{t_{|S|}}{\sum_{i=0}^k \binom{k}{i}t_i} \;.
\end{align}
The $t_i$ are a sequence of non-negative coupling constants.
We then need to add the transitive closure of $S$ to the past of the element $k$.
Growing a set step by step like this, we can label each point by when it was added to the causal set.
This is a natural labelling.
In a natural labelling, if two points have labels $k,l$, then if $k<l$ $k$ cannot be to the future of $l$.
The sequential growth of a causal set gives rise to a type of bouncing cosmology, in which the coupling $t_i$ undergoes a renormalisation after every bounce \cite{Sorkin:1998hi}.

The best examined growth dynamics for causal sets, at the moment, is transitive percolation.
In transitive percolations, the coupling in the $n$-th step is $t_n=t^n$.
One can show, that transitive percolation is an attractive fixed point under the renormalisation that is induced by bouncing cosmologies \cite{Martin:2000js}.
There are indications that transitive percolation can reproduce some properties of de Sitter space-times.
In  \cite{Ahmed:2009qm} they measured the volume of an Alexandrov interval as a function of its height, measured in proper time, and fit it to the same curve for a de Sitter space-time.
We will further examine these percolated causal sets using a new test of manifoldlikeness in section \ref{sec:local}.

The most modest proposal is to restrict the class of causal sets to causal sets of a fixed dimension.
Most proponents of CST do prefer the more ambitious proposals as models of reality.
The fact that causal sets could lead to a dynamically emergent dimensionality of space-time, is seen as a desirable quality.
Additionally, it is not trivial to restrict the dimensionality of a causal set.
Nevertheless, the fixed dimension proposal has merit as a test bed.
In $2$ dimensions the class of $2$d orders is very simple to implement and describes causal sets that are on average equivalent to $2$d Minkowski space.
The set of $N$-element $2$d orders $\Omega_{2d}$ is defined as follows:
	Let $S = (1, . . . , N )$ and $U = (u_1 , u_2 , . . . ,u_N ), V = (v_1 , v_2 , . . ., v_N )$, with $u_i , v_i \in S$. 
	$U$ and $V$ are then total orders w.r.t. the natural ordering $<$ in $S$.
	An $N$-element 2D order is the intersection $C = U \cap V$ of two total $N$-element orders $U$ and $V$ , i.e., $e_i \prec e_j$ in $C$ iff $u_i < u_j$ and $v_i < v_j$.
Physically speaking, we can think of $c_i \in C$ as the lightcone coordinates $(u_i,v_i)$ of a point in $2$d Minkowski space.
This model has been examined in simulations using the Benincasa-Dowker action, more detail on which will be presented in section \ref{sec:factor}.
The simulations expose two distinct phases, delimited by a clear phase transition \cite{Surya:2011du}.

\subsection{Particle dynamics on causal sets}
Another interesting problem is how to couple matter to a causal set.
There are a number of solutions to this.
A very simple phenomenological model are  swerves.
Swerves are an attempt at modelling how the discreteness of the causal set might influence particle motion.
The premise is simple; a particle hops from causal set element to causal set element, moving only along links.
This movement introduces a characteristic diffusion in energy and momentum  \cite{Philpott:2008vd}.

The swerve model is a classical model.
A particle moves along one trajectory and tries to follow the geodesic as closely as possible.
In causal sets we want the discreteness scale to be small, so certainly the movement of a particle on this scale should be a quantum theory.

We can quantise the movement of a particle on a given causal set, using the path integral.
In a first step, this corresponds to a quantum particle moving on a classical background geometry.
For a full quantum theory of gravity coupled to matter, we need to also sum over all geometries.

The idea of a quantum particle moving on a given causal set is investigated in  \cite{Johnston:2008za}.
The retarded propagator for a scalar particle on a classical, $d$-dimensional space-time is 
\begin{align}
\tilde K_m^{(d)}(p)&= -\frac{1}{(p_0+i\epsilon)^2-\vec p^2-m^2} \;.
\end{align}
On a causal set, we can calculate this as a sum over all trajectories the particle could take.
There are two possible classes of trajectories, chains and paths. 

To each trajectory of length $n$, a weight $a^n b^{n-1}$ is associated, the parameter $a$ is associated with each move from one element to the next and $b$ is associated with each stop at an element.

They then show that, for a causal set sprinkled into $1+1$d, summing over chains leads to
\begin{align}
K_{\mc{C}}(p)&= - \frac{ 2a}{(p_0+i\epsilon)- p_1^2 +2 a b \rho}\;,
\end{align}
where $\rho$ is the density of the causal set $\mc{C}$. 
This expression agrees with the continuum propagator, if we take $a= 1/2$ and $b= - m^2/\rho $.
On a causal set sprinkled into $3+1$d, on the other hand, it is the sum over paths that leads to the propagator
\begin{align}
K_{\mc{C}}(p)&=- \frac{ 2 \pi \sqrt{6} \sqrt{\rho} a}{(p_0+i\epsilon)- p_1^2 + \frac{2 \pi \sqrt{6}  a b}{\sqrt{\rho}}}\;,
\end{align}
which agrees with the continuum propagator for $a= 1/(2 \pi) \sqrt{\rho/6}$ and $ b=- m^2/\rho$.

This definition of a propagator can also be extended to define a distinguished ground state for quantum fields on curved space-times \cite{Afshordi:2012jf}.
Another tool to examine propagation of a field on a causal set is the causal set d'Alembertian, which will be introduced in the next section.

\section{\label{sec:factor}The causal set d'Alembertian operator}

We can define a scalar field $\phi$ on a causal set by associating values $\phi(e)$ with each element $e$.
To define the dynamics of such a field, we need a derivative operator.
In most discretisations, one uses a sum over nearest neighbours and next-to-nearest neighbours to define the d'Alembertian operator.
But, as alluded to above, each causal set element has very many nearest neighbours, in the limit of an infinite causal set it has infinitely many.

\begin{figure}
\center
\includegraphics[width=0.6\textwidth]{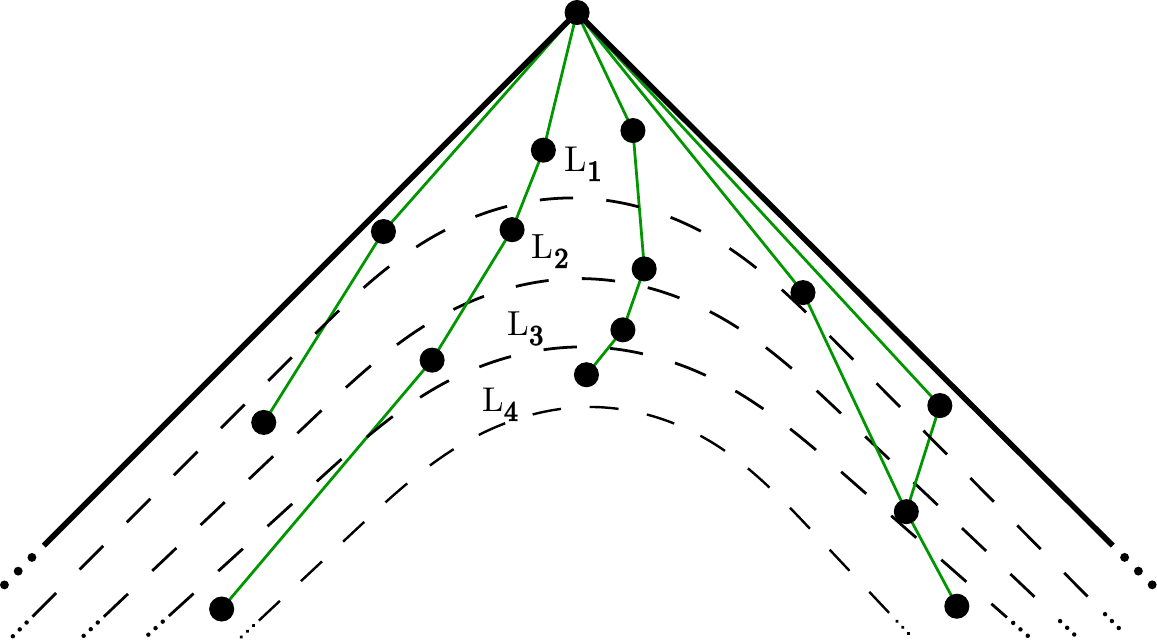}
\caption{\label{fig:layercake} This picture illustrates the layers of a causal set. Of course they are not always this orderly, and the dashed lines are drawn as eye-guides.}
\end{figure}

A solution to this conundrum is proposed by Rafael Sorkin in \cite{Sorkin:2007qi}.
For a two dimensional causal set, he proposed the operator
\begin{align}
B^{(2)}\phi(x):=\frac{1}{l^2}\Big[- 2\phi(x)
+4\Big(\sum_{y \in L_1(x)}\!\!\!\phi(y)-
2\!\!\!\sum_{y\in L_2(x)}\!\!\!\phi(y)+\sum_{y\in L_3(x)}\!\!\!\phi(y)\Big)\Big]\;, 
\label{eq:Bop2}
\end{align}
where $l=V_f^{1/2}$ is the fundamental discreteness scale.
The three sums run over layers of the causal set. 
The first layer $L_1$ are all nearest neighbours to the past of $x$, the second layer $L_2$ all elements with one intervening element and so forth, as illustrated in Figure \ref{fig:layercake}.
For functions that change slowly compared to the discreteness scale, the average of this sum over layers over many causal sets, indeed reproduces the d'Alembertian.
For each causal set in the average, the result fluctuates around the continuum value and these fluctuations grow with the number of points included.
Thus the infinite density limit of this operator is not well defined.
Causal set theory does not attempt a continuum limit, so this is not a fundamental problem.
Still, the discreteness scale of the universe is small and would presumably induce large fluctuations, so a mechanism to dampen them is necessary.

Sorkin proposes to smear the layers over an intermediate non-locality scale $\zeta$.
To achieve this, he introduces the function
\begin{align}
f(n,\epsilon)&=(1-\epsilon)^n \left( 1-\frac{2 \epsilon n}{1-\epsilon} + \frac{\epsilon^2 n (n-1)}{2 (1-\epsilon)^2}     \right) \;,
\end{align}
where $\epsilon=(l/\zeta)^2$. 
The intermediate non-locality scale $\zeta$ can run from $l$ to $\infty$.
The new d'Alembertian is then
\begin{align}
B^{(2)}\phi(x):=\frac{\epsilon}{l^2}\Big[- 2\phi(x) + \sum_{y\prec x} f(I_A(x,y),\epsilon) \; \phi(y) \Big]\,. 
\end{align}
The sum runs over all elements $y$ to the past of $x$, and the weight that each element contributes is determined by the function $f$.
For $\zeta=1$ we recover equation \eqref{eq:Bop2} above. 
Larger $\zeta$ smear the operator over several layers and serve to dampen the fluctuations.

The proposal generalises to $4$ and higher dimensions  \cite{Dowker:2013vba}.

For general $d$, the operator is defined as
\begin{align}
B^{(d)}\phi(x)=\frac{1}{l^2}\left( \alpha_d \phi(x) +\beta_d \sum\limits_{i=1}^{n}  C^{(d)}_i \sum\limits_{y \in L_{i}}\phi(y) \right) \, .
\end{align}
This is the expression without an intermediate non-locality scale.
For the expression including the intermediate non-locality scale, we can introduce a generalisation $f^{(d)}$ of the function $f$.

The sum runs over the first $n$ layers to the past of $x$.
Despite each layer being infinite, the overall sum remains finite due to cancellations between the layers.
To obtain these cancellations, requires a dimension dependent minimal number of layers $n_{d}$.
In  \cite{Glaser:2013xha}, we show  that $n_d=\dtwo +2$ for even $d$ and $\frac{d-1}{2}+2$ for odd $d$, is the minimal number required for the cancellation.
This number of layers is first introduced in \cite{Dowker:2013vba}, where the $C^{(d)}_i$ and the factors $\alpha_d$ and $\beta_d$ are determined explicitly for $d=2, \dots , 7$.

The $C^{(d)}_i$ can be generated using an operator $\Od{d}$
\begin{align}\label{eq:defCi}
\Od{d} \int\limits_{J^-(x)} \!\!\!\! \mathrm{d}V_y \exp(-l^{-d} V_d(x,y)) = \sum_{i =1}^{n_d} C^{(d)}_i  \!\!\!\! \int\limits_{J^-(x)} \!\!\!\!\mathrm{d}V_y \frac{(l^{-d} V_d(x,y))^{i-1}}{(i-1)!} \exp(-l^{-d} V_d(x,y)) \;.
\end{align} 
We here use $V_d(x,y)$ to denote the Alexandrov interval for general space-times.
In  \cite{Dowker:2013vba}, the operator for even dimensions $d=2n$, is determined to be
\begin{align}\label{eq:op}
\Od{2n}=\frac{(H+2)(H+4) \dots (H+2n+2)}{2^{n+1}(n+1)!}\, ,
\end{align}
where $H=-l\pd{}{l}$.
In odd dimensions, $d = 2n +1$, the operator is given as $\Od{2n+1} = \Od{2n}$. 
Using this operator, we can write the average over sprinklings for $B^{(d)}\phi(x)$
\begin{align}\label{eq:avB}
\bar{B}^{(d)} \phi(x) = \alpha_d{l^{-2}}  \phi(x)+ {\beta_d}{l^{-(d+2)}}  
\Od{d} \!\! \! \int\limits_{J^-(x)} \!\! \mathrm{d}V_y \phi(y) e^{-V_d(x,y)l^{-d}}\,. 
\end{align}
The lightcone coordinates are defined as $u=\frac{1}{\sqrt{2}}(t-r)$ and $ v=\frac{1}{\sqrt{2}}(t+r)$.
Using this, we can write down integral equations for  the constants $\alpha_d$ and $\beta_d$
\begin{align}\label{eq:defbeta}
\frac{1}{\beta_{d}}&= \lim_{l \rightarrow 0} \frac{S_{d-2} }{2(d-1)l^{d+2}} \Od{d} \intl{-\infty}{0} \md u \intl{u}{0} \md v  \left( \frac{v-u}{\sqrt{2}} \right)^{d} e^{-l^{-d} V_{A_d}(u,v) } \\
\frac{\alpha_d }{\beta_d}&= - \lim_{l \rightarrow 0} 
\frac{ S_{d-2}}{l^{d}}  \Od{d} \intl{-\infty}{0} \mathrm{d}u \intl{u}{0}  \mathrm{d}v 
\left(\frac{v-u}{\sqrt{2}}\right)^{d-2} 
e^{-l^{-d} V_{A_d}(u,v)} \;. \label{eq:defalpha}
\end{align}
Here $V_{A_d}(u,v)$ is the Alexandrov volume between the origin and the point of integration in Minkowski space, expressed in light-cone coordinates.

In  \cite{Benincasa:2010ac}, this d'Alembertian operator is used to calculate the Ricci scalar of a weakly curved manifold and to propose the first causal set action.
To define the Ricci scalar, they observe that
\begin{align}
\lim_{l \rightarrow 0} 
{\bar{B}}^{(d)}\phi(x) &= \Box^{(d)} \phi(x) + a_d R(x)\phi(x) \;.
\end{align}
For a sprinkling into a weakly curved manifold this can be shown using the Riemann normal coordinate expansion.
Expanding the integral in \eqref{eq:avB} in Riemann normal coordinates we find
\begin{align}
& \beta l^{-(d+2)} S_{d-2} \Od{d}\int_{\mathcal{N}} \!\! \mathrm{d} t\, \mathrm{d}r \, r^{d-2} 
  \nonumber \\
&\bigspace \times \bigg[ \bigg( R_{00} t^2 + \frac{1}{d-1}(R_{00}+ R) r^2\bigg)
             \bigg(-\frac{1}{6} - \frac{1}{24(d+1)}l\frac{\partial}{\partial l}\bigg) e^{-l^{-d}V_{A\, 0}}
	\nonumber	\\ 				
             &\bigspaced  + R(t^2 - r^2)\frac{1}{24(d+1)(d+2)} l  \frac{\partial}{\partial l} e^{-l^{-d}V_{A_d}} \bigg]\,. \label{eq:rncexp}
\end{align} 
These integrals determine the factor $a_d$.

Applying the operator $B^{(d)}$ to the constant $a_d^{-1}$ then yields the Ricci scalar.
For the action, we sum this expression over all causal set elements, leading to
\begin{align}
\frac{1}{\hbar}\mathcal{S}^{(d)}(\mc{C}) = \frac{1}{\hbar} \sum_{x \in \mc{C}} B^{(d)}\frac{1}{a_{d}}(x)=-\left(\frac{l}{l_p}\right)^{{d-2}}\bigg[\alpha_d N + \beta_d \sum_{i = 1}^{n_d} C_i ^{(d)} N_i \bigg]\;,
\end{align}
where $N$ is the size of the causal set, $N_i$ the number of intervals of size $|I_A(x,y)|=i-1$ in the entire causal set, and the factor of $\hbar$ is absorbed into the Planck length $l_p$.

In $2$d,  this action has been investigated for flat regions of Minkowski space-time, a cylinder space-time, and the topology changing trousers  \cite{Benincasa:2010as}.
This action is also used in Monte Carlo simulations  \cite{Surya:2011du}.
The factor $a_d$ has been conjectured to be $-\frac{1}{2}$ for all dimensions.

It is then obvious that closed form expressions for equation \eqref{eq:defCi},\eqref{eq:defbeta}, and \eqref{eq:defalpha} and a proof for the conjectured factor $a_d=-1/2$ are of interest.
Here we will sketch how to calculate $\beta_d$ and then show how to determine the $C_i^{(d)}$ using a generating function\footnote{This section summarises calculations first presented in  \cite{Glaser:2013xha} which is reprinted in Appendix \ref{app:Factor}.}. 
For concreteness we show the calculation in even dimensions and then comment on the changes necessary for the odd dimensions afterwards.
The integral defining $\beta_d$ is
\begin{align}
\frac{1}{\beta_{d}}&= \lim_{l \rightarrow 0} \frac{S_{d-2} }{2(d-1)l^{d+2}} \Od{d} \intl{-\infty}{0} \md u \intl{u}{0} \md v  \left( \frac{v-u}{\sqrt{2}} \right)^{d} e^{-l^{-d} V_{A_d}(u,v) }\;.
\end{align}
Expanding this as a power series in $u,v$ we calculate
\begin{align}
&\intl{-L}{0} \md u \intl{u}{0} \md v  \left( \frac{v-u}{\sqrt{2}} \right)^{d} e^{-l^{-d}  c_d (uv)^{\dtwo  }} \\
&=\suml{k=0}{d}\binom{d}{k} \frac{(-1)^k}{2^{\dtwo  }} \suml{n=0}{\infty} \frac{\left(-l^{-d}c_d \right)^n}{n!} \frac{L^{d(n+1)+2}}{(\dtwo  n+k+1)(d(n+1)+2)}\;. \label{eq:intbeta1}
\end{align}
The constant $c_d$ is implicitly defined by $V_{A_d}=c_d (uv)^\dtwo$.
We can then apply the operator $\Od{\deve}$ to $l^{-dn}$
\begin{align}
\Od{\deve}l^{-d n}= \frac{1}{2^{\frac{d}{2}+1} (\frac{d}{2}+1)!} \prod\limits_{i=1}^{\frac{d}{2}+1} (2 i -l \pd{}{l}) l^{-dn} = l^{-dn} \prod\limits_{i=1}^{\frac{d}{2}+1} \frac{\Poch{\twod i+1}{n}}{\Poch{\twod  i}{n}} \;. \label{eq:evenopp}
\end{align}
Inserting this into the sum in \eqref{eq:intbeta1}, we can express it as a generalised hypergeometric function
\begin{align}
&\frac{\Od{\deve}}{l^{d+2}} \intl{-L}{0} \md u \intl{u}{0} \md v  \left( \frac{v-u}{\sqrt{2}} \right)^{d} e^{-l^{-d} c_d (uv)^{\dtwo  }}  \\
=&\suml{k=0}{d}\binom{d}{k} \frac{(-1)^k }{2^{\dtwo} (k+1)(d+2) c^{1+\twod}} z^{1+\twod}   \mFm{\dtwo+1}{\twod(k+1),\twod +1,\dots, 2 }{\twod(k+1)+1,\twod , \dots, 1 }{-z} \;.\label{eq:sumeve}
\end{align}
In the above equation,  we introduced  $z= c_d (\frac{L}{l})^d$ to simplify the notation.
In the upper row of the argument of the hypergeometric function, we shortened the $\frac{d}{2}$ terms in the sequence $\twod j +1$ with $j=1,\dots,\frac{d}{2}$ to $\twod+1,\dots,2$ (in the lower row $\twod j$ with $j=1,\dots,\frac{d}{2}$ is shortened to $\twod,\dots,1$). 

This is a nice closed form expression. 
The limit $l \to 0$ is now replaced by the limit $z \to \infty$.
This calculation needs the following identities
\begin{align}
&\lim_{z\rightarrow \infty} e^{z} \mFm{q}{a_{1},\dots, a_{q}}{a_{1}-1,\dots, a_{q}-1}{-z}  = \frac{(-z)^{q}}{\prod\limits_{j=1}^{q} (a_{j}-1)} +O\left( (-z)^{q-1}\right)  \label{eq:limit0} \\
&	\lim_{z\rightarrow \infty} z^{a_{0}} \mFm{q+1}{a_{0},a_{1}, \dots,a_{q}}{a_{0}+1,a_{1}-1, \dots,a_{q}-1}{-z} = \G{a_{0}+1}  \prod\limits_{j=1}^{q}\frac{a_{j}- a_{0} -1}{a_{j}-1} \;, \label{eq:limit}
\end{align}
which are derived in Appendix B of  \cite{Glaser:2013xha}.

Under the limit $z\to \infty$, the terms in \eqref{eq:sumeve} split into three categories.
Terms with $k<\dtwo$ do not contribute to the sum.
Their first argument in the upper row of the hypergeometric, simplifies against one of the terms in the lower row.
The function then becomes of the form \eqref{eq:limit0} and is thus exponentially suppressed.
Terms with $k \geq \dtwo$ contribute a term $z^{-\twod(k+1)}$, as the function is of the form \eqref{eq:limit}.
This contribution falls of faster than $z^{1+\twod}$ for $k>\dtwo$, thus terms with $k>\dtwo$ do not contribute.
Thus only $k=\dtwo$ leads to a finite contribution
\begin{align}
&\lim_{z\to \infty} \binom{d}{\dtwo} \frac{(-1)^\dtwo}{2^{\dtwo} (\dtwo+1)(d+2) c^{1+\twod}} z^{1+\twod} \mFm{\dtwo+1}{\twod+1, \twod +1,\dots, 2 }{\twod+2,\twod , \dots, 1 }{-z}\\
&=\binom{d}{\dtwo} \frac{1}{2^{\dtwo-1}d (d+2) c^{1+\twod}} \G{\frac{2+d}{d}} \;.
\end{align}
This can be inserted into \eqref{eq:defbeta}
\begin{align}
	\beta_\deve &=\frac{2 \; \G{\frac{d}{2}+2} \G{\frac{d}{2}+1}}{\G{\frac{2}{d}} \G{d}} \; c_d^{\twod} \label{eq:solbetaeven}\;.
\end{align}
For odd dimension, the calculation follows the same pattern with a slight twist.
Taking the limit $z \to \infty$ requires a few more steps of simplification, the real difference, however, is the result of the limit.
In odd dimensions, several terms survive the limit, leading to a sum over $k$.
This sum can be solved and a simple expression remains.
The detailed calculation is explained in  \cite{Glaser:2013xha} reprinted in Appendix \ref{app:Factor}.

In a similar manner, we can calculate \eqref{eq:defalpha}. We thus prove that
\begin{align}
\beta_d&= \begin{cases} 
\frac{2 \; \G{\frac{d}{2}+2} \G{\frac{d}{2}+1}}{\G{\frac{2}{d}} \G{d}} \; c_d^{\frac{2}{d}} & \text{for even }d\\
 \frac{d+1}{2^{d-1}\G{\frac{2}{d}+1}} \; c_d^\frac{2}{d} & \text{for odd }d \\ 
 \end{cases}\\
	\intertext{and}
	\alpha_d&= \begin{cases}
\frac{- 2 c_d^{\frac{2}{d}}}{\G{\frac{d+2}{d}}} & \text{for even }d \\
	\frac{- c_d^{\frac{2}{d}}}{\G{\frac{d+2}{d}}} & \text{for odd }d \;.\\ \end{cases}
\end{align}
Using the same calculational techniques we also solve the integrals in \eqref{eq:rncexp} and prove the conjecture that $a_d=-1/2$ for all dimensions.

To completely know the d'Alembertian in any dimension, we would also like easier expressions for the $C_i^{(d)}$ defined in \eqref{eq:defCi}.
Applying the operator to $exp(-l^{-d}V)$ we calculate
\begin{align}\label{eq:Ohyper}
\Od{\deve} e^{-l^{-d}V}		&= \mFm{\dtwo  +1}{\twod +1, \frac{4}{d}+1,\dots,2, \twod +2}{\twod , \frac{4}{d},\dots,1, \twod +1}{-l^{-d}V} \; .
\end{align}
We can compare this to  \eqref{eq:sumeve} and notice that the $\dtwo$ arguments in the middle  of the hypergeometric functions are generated by the operator $\Od{\deve}$.
In fact, the last arguments arise there as well, but are cancelled by other arguments.

These terms are instrumental in cancelling out contributions to the integral that would lead to an infinite $l \to 0 $ (equivalently $z \to \infty$) limit.
The limit is always taken on a hypergeometric function
\begin{align}
z^a \mFm{p+1}{b_k,b_1+1, \dots, b_p+1}{b_k+1,b_1, \dots , b_p}{-z}
\end{align}
with a rising sequence of $b_i$.
For every $b_k <a$ which would give rise to a term $z^{a-b_1}$ blowing up as $z \to \infty$ the operator generates a matching argument to cancel it.
Thus the operator generates exactly the minimal number of layers $n_d$ necessary for a finite result.

The hypergeometric function \eqref{eq:Ohyper} simplifies to a power series in $l^{-d}V$ multiplied by $e^{-l^{-d}V}$ that defines the $C_i^{(d)}$.
To simplify the notation, we again introduce $z=l^{-d}V$ we can then obtain a generating function for the $C_i^{(d)}$ 
\begin{align}\label{eq:defGC}
G_\deve &=e^{z} \mFm{\dtwo  +1}{\twod +1, \frac{4}{d}+1,\dots, \twod +2}{\twod , \frac{4}{d},\dots, \twod +1}{-z}\;.
\end{align}
This defines the $C_i^{(d)}$ as
\begin{align}
C_i^{(d)}=\left( \pd{}{z} \right)^{i-1} G_d \; \bigg|_{z=0} \;,
\end{align}
which is calculated in Appendix C.1 (C.2 for the odd result) of  \cite{Glaser:2013xha}, and leads to
\begin{align}
C_i^{(\deve)}=\sum_{k=0}^{i-1} \binom{i-1}{k} (-1)^k \, \frac{\G{\dtwo(k+1)+2}}{\G{\dtwo+2}{\G{1+\frac{d k}{2}}}}\\
C_i^{(\dodd)}=\sum_{k=0}^{i-1} \binom{i-1}{k} (-1)^k \, \frac{\G{\dtwo(k+1)+\frac{3}{2}}}{\G{\frac{d+3}{2}}{\G{1+\frac{d k}{2}}}} \;.
\end{align}
For a given $i$, these finite sums can be evaluated for general $d$.
They are well suited to be used in computer programs to implement the action in any dimension.

We have thus been able to determine simple expressions for all constants that are necessary to define both the d'Alembertian operator and the Benincasa-Dowker action on a causal set.
It was also possible to show that $n_d=\dtwo +2$ for even $d$ and $\frac{d-1}{2}+2$ for odd $d$ is the minimal number of layers needed to achieve a finite result, in $d$ dimensions, and to prove that the prefactor of the Ricci scalar is $- 1/2$ for all dimensions.

Knowing these analytic expression will be useful for computer simulations and investigations of the causal set action.

The non-locality that arises as a byproduct of the strictly Lorentzian perspective in causal sets could also lead to  interesting continuum phenomenology.
A continuum version of the d'Alembertian might lead to new predictions.

In  \cite{Aslanbeigi:2014zva}, the d'Alembertian was generalised to a class of operators.
To do so, they derive a set of consistency equations that the constants in the d'Alembertian have to fulfil.
These equations can be solved for any $n\geq n_d$, thus defining a class of operators in any dimension.
Using these consistency equations and the integral expression for the new generalised operators they examine their Fourier transforms.
They observe that in $2$d the evolution defined by this operator is stable, while in $4$d it seems to be unstable.
This work leads to a proposal for a Lorentz invariant cut-off in quantum field theory.

\section{\label{sec:local}Recovering locality}
Defining locality in a causal set is difficult, yet a definition of locality would be very useful.
Many constructions to recover continuum quantities depend on such a definition.
Measuring the midpoint scaling dimension or the Myrheim-Meyer dimension we assume a region that is approximated by flat space.
Both known definitions for the curvature of a causal set  \cite{Benincasa:2010ac,Roy:2012uz} depend on the measurement being taken in a small local region.
The construction of an embedding and a homology mentioned above also depend on a local region \cite{Major:2006hv,Henson:2006dk}.

It would thus be an important step to discover a characteristic of causal sets that allows us to distinguish between local intervals, those small w.r.t. the curvature scale, and non-local intervals, those large w.r.t. the curvature scale. 
To render the qualitative idea of the curvature scale more precise, we define the scale of flatness.
The scale of flatness of an ``approximately flat'' space-time region in which Riemann normal coordinates are valid is a dimensionless scale $\varsigma \gg 1$ and $\varsigma^{-1} \sim R \tau^2$.
Here $R$ can be any component of the Riemann tensor in Riemann normal coordinates and $\tau$ the proper time between any events within the region. 
For a flat space-time $\varsigma \to \infty$, as appropriate for a scale of flatness.

\begin{figure}
\center
\includegraphics[width=0.5\textwidth]{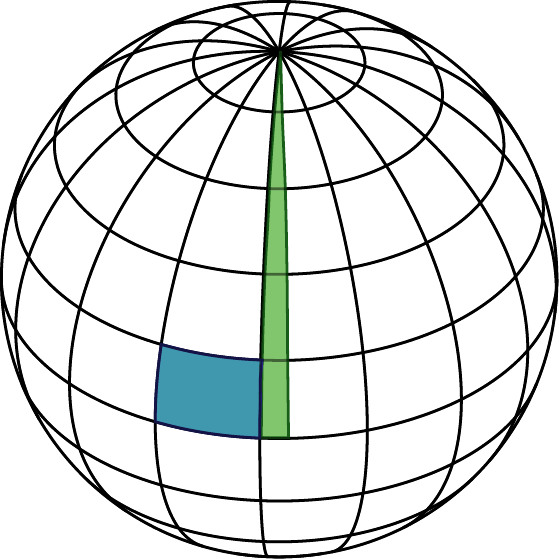}
\caption{\label{fig:sphere}The blue and the green interval have approximately the same size. Yet, the green clearly probes more of the curvature than the blue.}
\end{figure}

One simple illustration of the question whether a region is big or small w.r.t. to the scale of flatness is shown in Figure \ref{fig:sphere}.
An interval of fixed size on a sphere can be local and approximately flat or not flat at all, depending on how it is placed on the sphere.
The same is true for curved Lorentzian manifolds.
It is thus necessary to establish new tools to determine if an interval in a causal set probes the curvature or if it is local.

\subsection{The abundance of intervals}
In simulating the $2$d orders in \cite{Surya:2011du}, a new observable is employed.
\begin{figure}
\center
\includegraphics[width=0.5\textwidth]{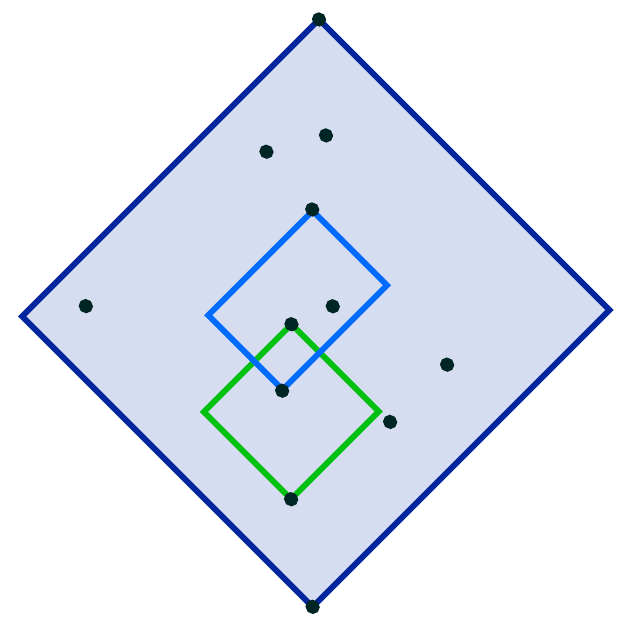}
\caption{\label{fig:intervalillu}This figure shows a large interval of size $N=10$, marked in dark blue. Within this large interval there are several smaller intervals of different size. We have marked an interval of size $m=1$ in green and one of size $m=2$ in blue. }
\end{figure}
A large interval contains smaller sub intervals.
These are the Alexandrov neighbourhoods between two points within the larger interval (c.f. Figure \ref{fig:intervalillu}).
We can then count how many intervals of size $m=0$, $m=1$ and so forth the large interval contains.
The curve that arises when we plot the number of intervals vs the size of the intervals we call the interval abundance.

For $2$d orders that seem good approximations of a $2$d manifold, the interval abundance curve shows a smooth falloff, while the behaviour is markedly different for obviously non-manifoldlike sets.
We thus investigate whether a similar relation holds in higher dimensions, and if we can predict the exact curve\footnote{
In this section we summarise \cite{Glaser:2013pca}, which is reprinted in Appendix \ref{app:Local}.}.
For a causal set sprinkled into the $d$-dimensional Minkowski interval $V$ with density $\rho$, the average number of intervals of size $m$ is
\begin{align}
\Nmd(\rho, V) =&  \rho^{2} \intl{\diamond}{} \md V_y \intl{\diamond_{y}}{} \md V_{x}
        \frac{ (\rho V_{0\,d}(x,y))^{m}}{m!} e^{-\rho V_{0\,d}(x,y)} \nonumber \\ 
        = & 
	\frac{\left( -\rho\right)^{m+2}}{m!} \pd{^{m}}{\rho^{m}} \intl{\diamond}{} \md V_{x}
        \intl{\diamond_{y}}{} \md V_y\; e^{-\rho V_{0\,d}(x,y)}  \nonumber \\
        =&\frac{ (-\rho)^{m+2}}{m!} \pd{^{m}}{\rho^{m}}\rho^{-2} \Nzerod(\rho,V) \;.
\end{align}
The domain of integration $\diamond$ is the entire interval $V$, while the domain of integration $\diamond_y$ is the region causally between the point $y$ and the upper tip of the interval $V$.
Thus if we can calculate the expected number of links $\Nmdv{0}{d}(\rho,V)$ the number of larger intervals can be calculated by differentiation.
The calculation is exhibited in detail in \cite{Glaser:2013pca}, the result is:  
\begin{align}
\Nmd(\rho, V)=&\frac{(\rho V)^{m+2}}{(m+2)!} \frac{\G{d}^{2}}{\Poch{\frac{d}{2} (m + 1) + 1}{ d - 1}} \frac{1}{\Poch{\frac{d}{2} m + 1}{d - 1}} \nonumber \\
	& \mFm{d}{1+m,\frac{2 }
	{d}+m,\frac{4}{d}+m, \dots ,\frac{2 (d-1)}{d}+m)}
	{3+m,\frac{2 }{d}+m+2,\frac{4}{d}+m+2, \dots ,\frac{2 (d-1)}{d}+m+2}
	{- \rho V} \;,
\label{eq:flatclosedform} 
\end{align}
where $_{p}F_{q}(\{a_1, \ldots, a_p \}, \{b_1, \ldots, b_q\}|-z)$ is a generalised hypergeometric function and $(a)_n=\G{a+n}/\G{a}$ is the Pochhammer symbol.
This expression is convergent, since generalised hypergeometric functions converge for all $z$ values if $p \leq q$.  
The details of obtaining this form for the $\Nmd$ are explained in appendix A of  \cite{Glaser:2013pca}.
\begin{figure}
\center
\includegraphics[width=0.6\textwidth]{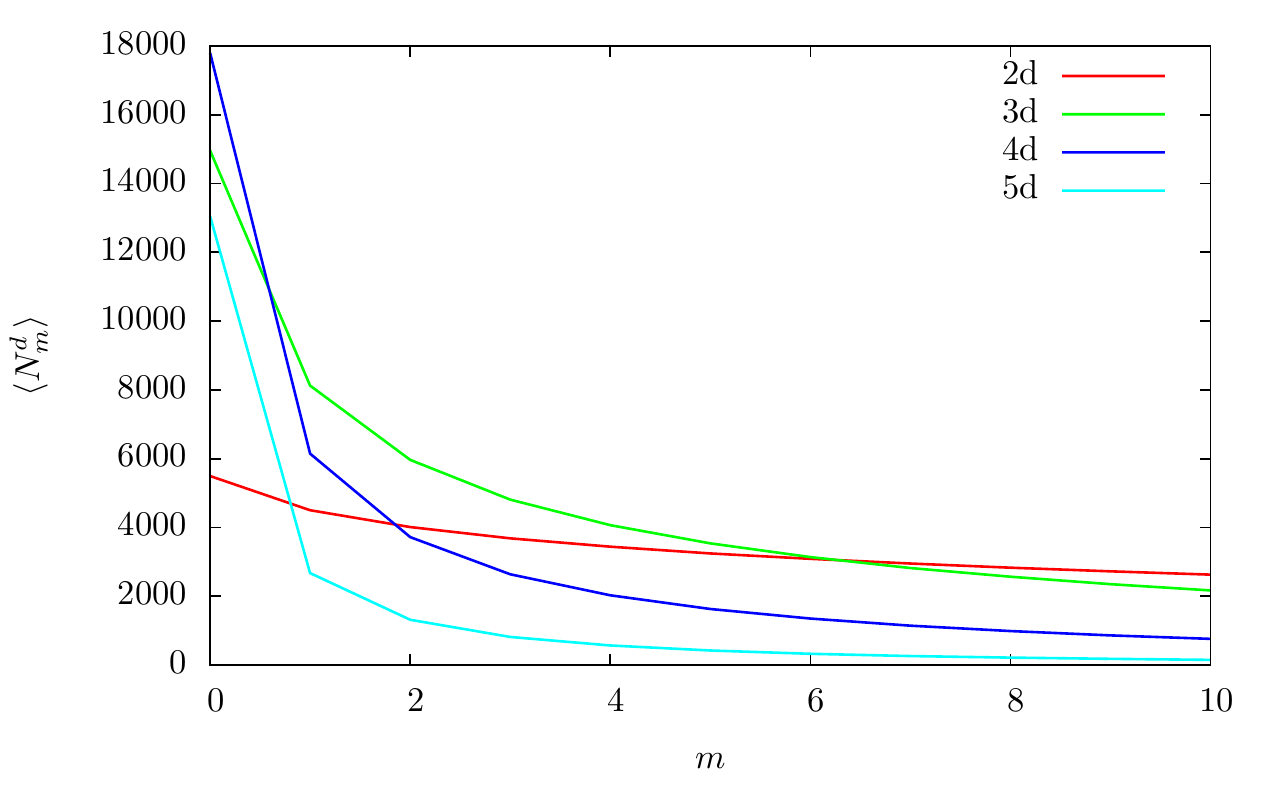}
\caption{\label{fig:m-intervals}Plot of the analytic function for the interval abundance for intervals of size $N=1000$ for $d=2,\dots,5$. This figure is taken from  \cite{Glaser:2013pca}.}
\end{figure}
In Figure \ref{fig:m-intervals}, one can see that the curves are very characteristic of the space-time dimension.
The interval abundance can thus be used as a holistic dimension estimator.
It has the advantage of only giving rise to integer dimension values, and the amount of overlap between the measured and the predicted curves could be used as a quality measure.

We rewrite this expression by replacing $\rho V \to N$.
This can be seen as changing the perspective from the manifold into which the causal set is sprinkled to the causal set itself.
The large $N$ limit can then be either the large density, continuum limit $\rho \to \infty$, at fixed size $V$, or the large size $V \to \infty$ limit, at fixed density $\rho$.
We can take this limit on equation \eqref{eq:flatclosedform} and find
\begin{align}
 \Nmd(N) =\frac{N^{2-\frac{2}{d}}}{m!} \G{\frac{2}{d}+m} \frac{\G{d} }{ \left(\frac{d}{2} -1\right) \Poch{\frac{d}{2}+1}{d-2}}+ \begin{cases}
 \mathcal{O}(N ) & \text{ for } d=3\\
 \mathcal{O}(N \log{N}) & \text{ for } d=4\\
 \mathcal{O}(N^{2-\frac{4}{d}}) &\text{ for } d>4 
 \end{cases}
\end{align}
for $d>2$ and
\begin{align}
\Nmdv{m}{2}(N)= N \log{N} + \mathcal{O}(N) 
\end{align}
for $d=2$. A remarkable property of this is that the  behaviour in $N$ is independent of the size $m$ of the interval in question.
\begin{figure}
\subfigure[\label{fig:contlimita}]{\includegraphics[width=0.49\textwidth]{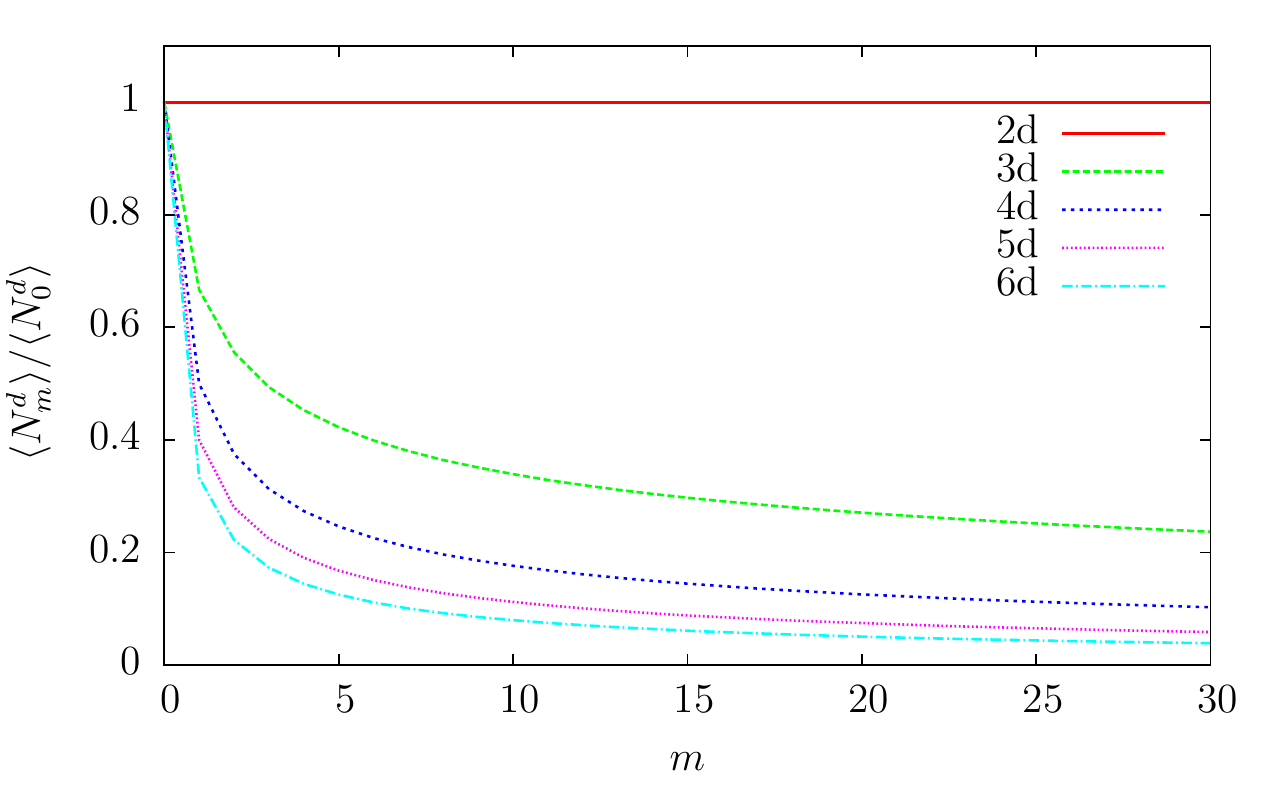}}
\subfigure[\label{fig:contlimitb}]{\includegraphics[width=0.49\textwidth]{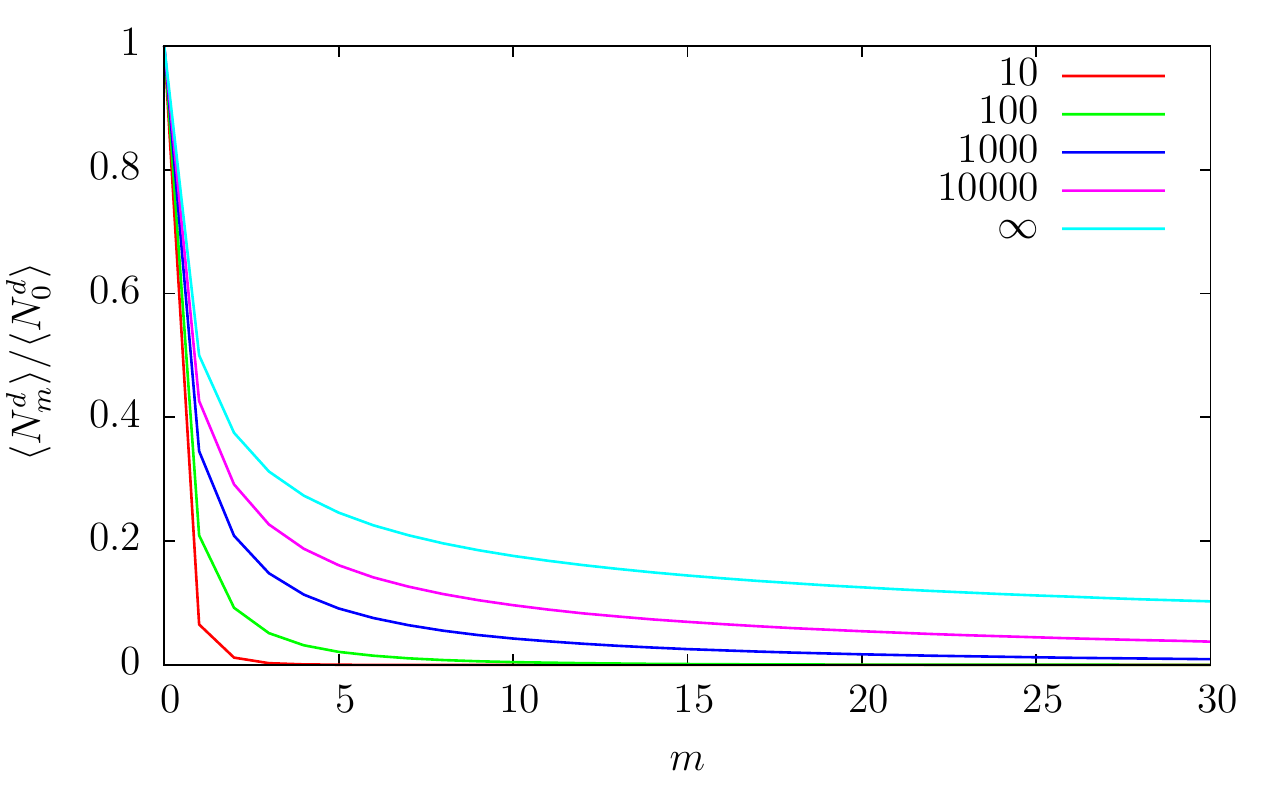}}
\caption{\label{fig:contlimit}The left figure shows how the ratio of abundances looks in different dimensions. Interestingly, in $2$d the abundance is independent of the size $m$ of the interval. On the right the convergence towards the continuum limit is tested. We plotted the ratio of abundances for $N=10,100,1000,10000$ elements and the continuum limit for $d=4$. This figure is taken from  \cite{Glaser:2013pca}.}
\end{figure}
We can define the ratio of abundances by dividing with $\Nzerod(N)$
\begin{align} \label{eq:limitdeftwo}
\mc{S}_m^d \equiv \lim_{N \rightarrow \infty} \frac{\Nmd(N)}{\Nzerod(N)}=\frac{
  \G{\frac{2}{d}+m} }{\G{\frac{2}{d}} \G{m+1}}\; . 
\end{align}
The number of $m$-intervals in an interval of size $N$ thus grows as $N^{2-\frac{2}{d}}$.
Unfortunately the convergence to this limit is very slow and thus not feasible to test with simulations (c.f. Figure \ref{fig:contlimitb}).

\subsection{Defining locality from the interval abundance}
We set out to establish $\Nmd$ as a measure of locality on a causal set that faithfully embeds into a manifold $\mc{M}$.
This implies its other uses as a dimension estimator and a measure of manifoldlikeness.

We calculated that for a Poisson sprinkling the average $\Nmd$ over an Alexandrov interval shows a characteristic behaviour in $m$.
Since the distribution of points is Poissonian, we expect that for large enough $N$ a typical realisation will be close to the average realisation, thus 
\begin{align}\label{eq:localtrack}
N_m (\mc{C}) \sim \Nmd(N \pm \sqrt{N})
\end{align}
for all $m$. 
We have confirmed this through simulations, the results of which are summarised in the next subsection.
We can use this to define locality. 
We call a (sub) Alexandrov-interval in a causal set causal set $\mc{C}$ local if it satisfies equation \eqref{eq:localtrack}.

For a causal set $ \mc{\tilde C}$ that embeds into an arbitrary interval $I_A(p,q)$ in a curved space-time the distribution $N_m( \mc{\tilde C})$ will diverge from $\Nmd$.
We thus call this causal set non-local.
If, however, the discreteness scale is significantly smaller than the scale of flatness, we expect to be able to locate Alexandrov intervals $I(p',q')$ of sufficient size $N'$ that lie entirely in $I_a(p,q)$ and can be identified as local\footnote{The following definition, claim and conjecture are quotes from  \cite{Glaser:2013pca}.}.

\begin{definition} 
  We will say that a $N$-element causal set $C$ is {\bf strongly $d$-rigid} if $\exists$ a
  $d$ for which $\NmCC \sim \Nmd(N \pm \sqrt N)$.  If $C$ possesses an $N'$-element sub causal
  set $C'$ which is strongly $d$-rigid, then $C$ is said to be {\bf weakly $d$-rigid}
  with respect to $C'$.
\end{definition} 

For a causal set $\mc{C}$ with $N>>1$ elements, strong $d$-rigidity is a necessary condition to faithfully embed into a flat $d$-dimensional Alexandrov interval.
A causal set $\mc{\tilde C}$ that embeds into a $d$-dimensional curved space-time should fulfil the condition of weak $d$-rigidity.
We would even expect it to fulfil something slightly stronger, every point of it should be in some weakly $d$-rigid subset.
A more in depth analysis of this case is still pending.

This leads us to the following claim:

\begin{claim} \label{claimone} Let $C$ be a $N$-element causal set that faithfully embeds into an
  Alexandrov interval $I_A(p,q)$ in a $d$-dimensional space-time such that the discreteness
    scale is much smaller than the scale of flatness everywhere. Then there exists a sub causal set
       $C' \subset C$ of cardinality $N'>>1$ such that $C'$ is strongly $d$-rigid. Moreover,
     if $I_A(p,q)$ is an Alexandrov interval in $d$-dimensional Minkowski space-time then for large
  enough $N$, $C$ is itself strongly $d$-rigid.
 \end{claim}

To make the arguments, we require arbitrarily large $N$ to suppress fluctuations.
The simulations in the next section show that these conditions work extremely well, even for causal sets of size $N=100$.

Since these requirements seem to work very well, we can ask whether they are strong enough to define manifoldlikeness.
In the general case this is clearly not sufficient, as a curved space-time would have to include a large number of strongly $d$-rigid sub causal sets.

We can argue that many non-manifoldlike partial orders will not fulfil even weak $d$-rigidity.
For example, the KR orders that dominate the space of all causal sets will not fulfil it. 
We will examine this in more detail in the next section.
This is very promising, but we have no prove that there exists no non-manifoldlike causal sets that fulfil weak $d$-rigidity.

Strong $d$-rigidity might suffice to establish that an Alexandrov interval embeds into $d$-dimensional Minkowski space.
This is also supported by the observation that the ratio $\Nmd/\Nmdv{0}{d}$ is scale invariant in the large $N$ limit.
This mimics the scale invariance of flat space-time. 
We thus conjecture:
\begin{conjecture} \label{conjecture1}
  If the interval abundances $\NmCC$ for a $N$-element causal set $C$ are such that that   $\NmCC
  \sim \Nmd(N\pm \sqrt N)$ for some $d$ in the large $N $ limit, then $C$
  faithfully embeds into an Alexandrov interval in $d$-dimensional Minkowski space-time.
\end{conjecture}

\subsection{Examining locality in sprinkled causal sets}

The analytic expressions for the interval abundance, that we derived above, can be tested on sprinkled causal sets.
We use the Cactus Code framework and the causal set thorns written by David Rideout to do so  \cite{dr-cactus,cactus2}.
Here, we will present a selection of the results obtained, additional cases are contained in \cite{Glaser:2013pca}.

\begin{figure}
\center
\includegraphics[width=\textwidth]{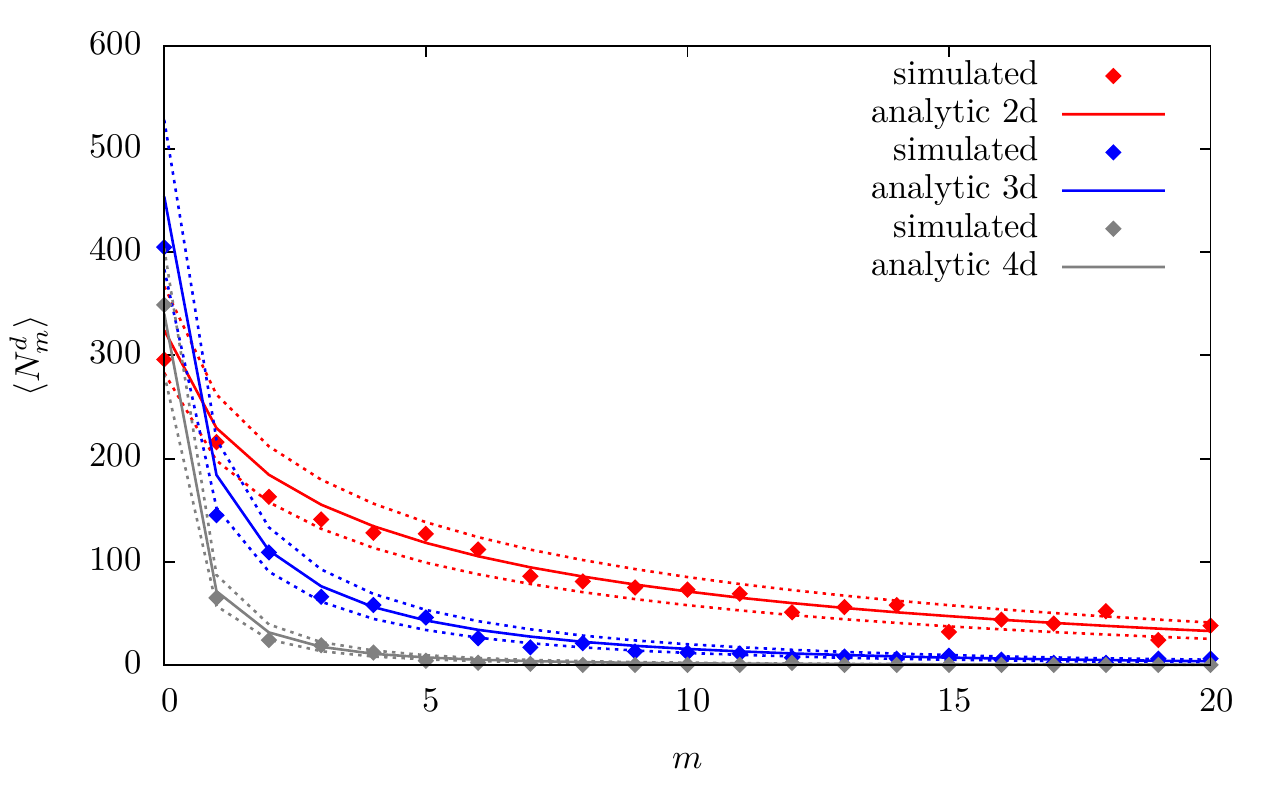}
\caption{\label{fig:singletogether}Comparing a single sprinkled causal set of size $100$ to the analytic curves for $N=100$ (solid lines) and $N=90,110$ (dashed lines) for $d=2,3,4$. This figure is taken from  \cite{Glaser:2013pca}.}
\end{figure}

The first check to make is that our analytic curves are reproduced for flat space.
For this purpose, we sprinkle one causal set of size $N=100$ each in $2,3$ and $4$ dimensions and compare it to the analytic curve.
These single instances clearly obey strong $d$-rigidity for their respective dimension, see Figure \ref{fig:singletogether}.
\subsubsection{Curved space}
\begin{figure}
\centering
\includegraphics{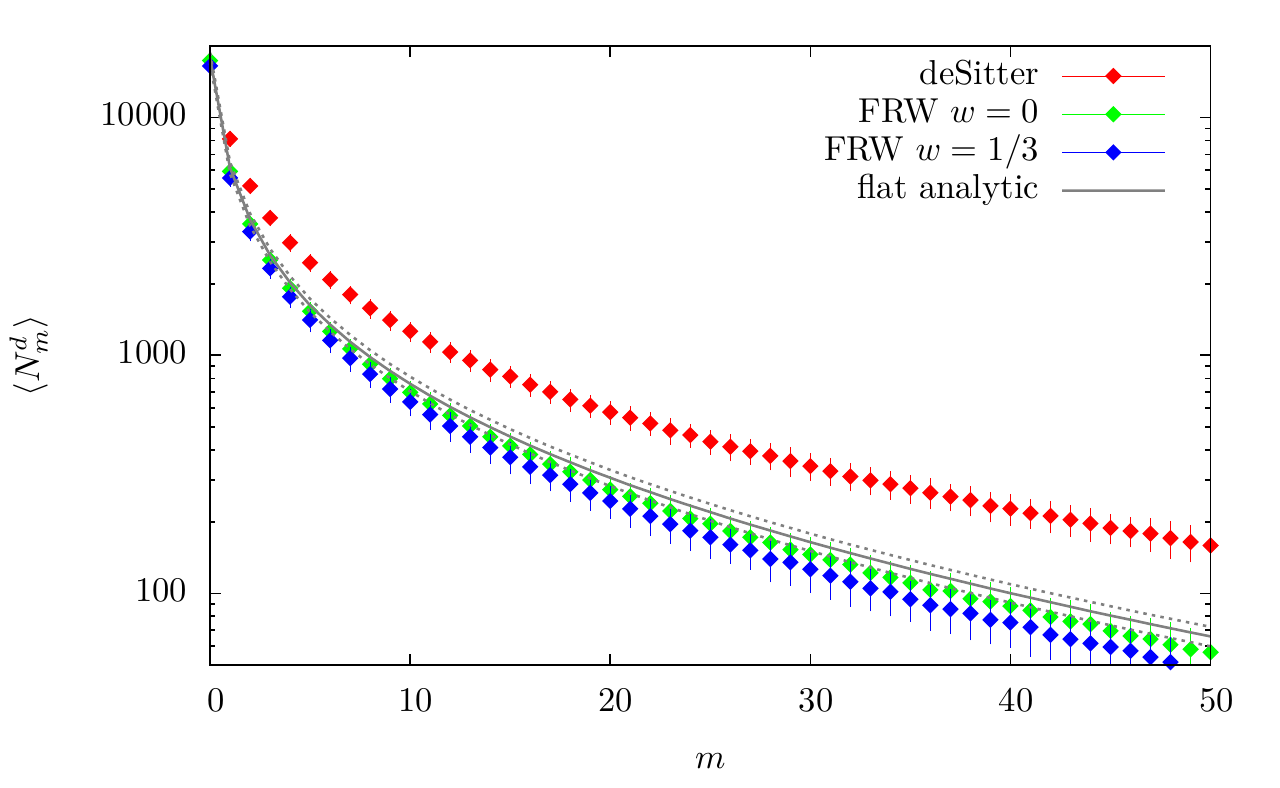}
 \caption{\label{fig:FRW_whole} $\langle \NmCC \rangle$ for   $N=1\,000$ element causal sets obtained from
          sprinkling $100$ times into $4d$ FRW space-times which are cosmological constant, matter, or radiation dominated. This figure is taken from  \cite{Glaser:2013pca}.}
\end{figure}

To test how the interval abundance behaves in curved space, we examined causal sets sprinkled into $4$d, $k=0$ FRW space-time
\begin{align}
\md s^{2} =	-\md t^2 + {a(t)}^2 \left( \sum_{i=1}^4 (dx^i)^2 \right)\;, 
\end{align}
where 
\begin{align}
	a(t)=&a_{0} \; t^{q}&  \text{with} &&	q=& \frac{2}{3(1+w)}\;,
\end{align}
and the equation of state is $p= w \rho$. 
We choose to simulate matter, radiation, and cosmological constant dominated space-times, corresponding to $w=0,1/3$, and the de Sitter case $w=-1$.
The de Sitter radius is taken as $\ell =1.3$.
To examine how much $\NmCC$ deviates from the flat space expectation $\Nmd$, we examine $100$ realisations of $\av{N}=1\,000$ element causal sets.
Figure \ref{fig:FRW_whole} shows the significant deviations this case leads to.

This  shows that a large interval that probes curvature is not strongly $d$-rigid.
Another interesting question, is if we can show in simulations that such an interval is weakly $d$-rigid.

\begin{figure}
\vspace{40pt}
\subfigure[FRW with $w=0$]{\includegraphics[width=0.6\textwidth]{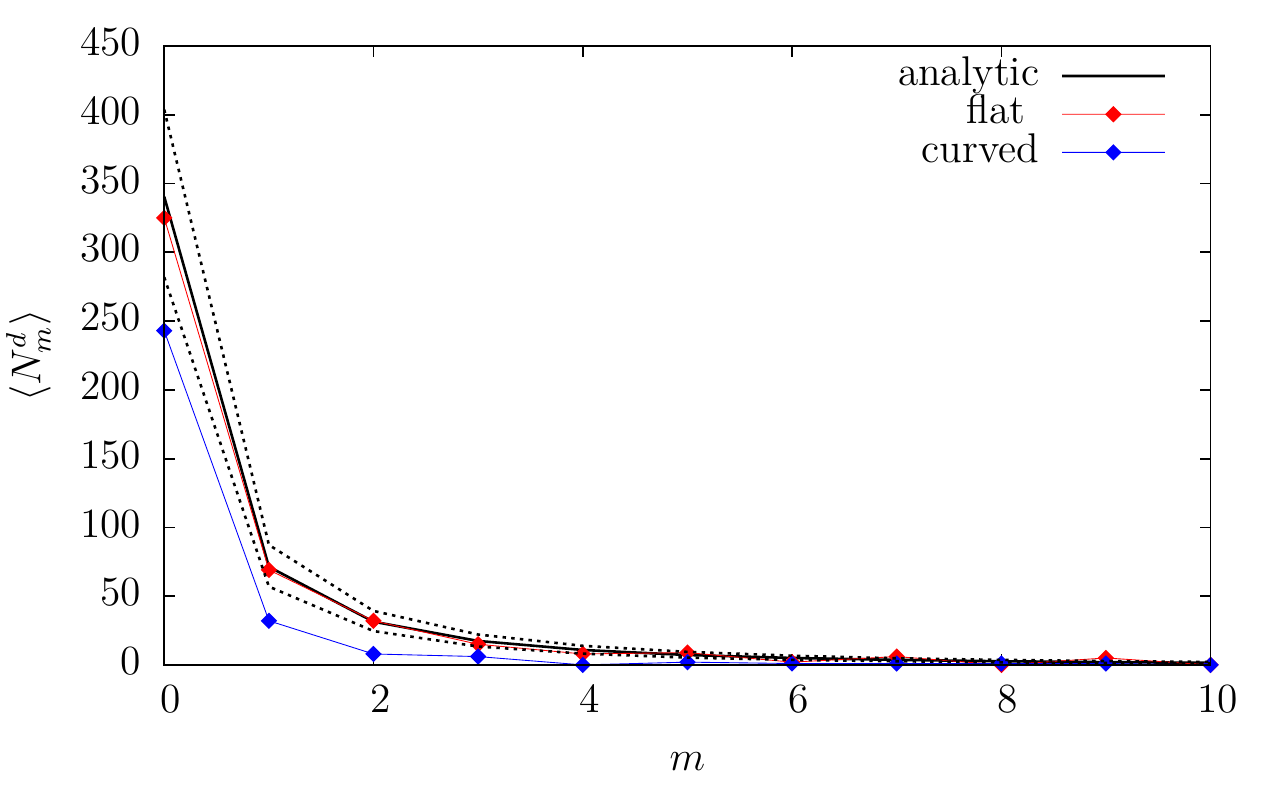}  \hspace{13.5pt} \includegraphics[width=0.35\textwidth]{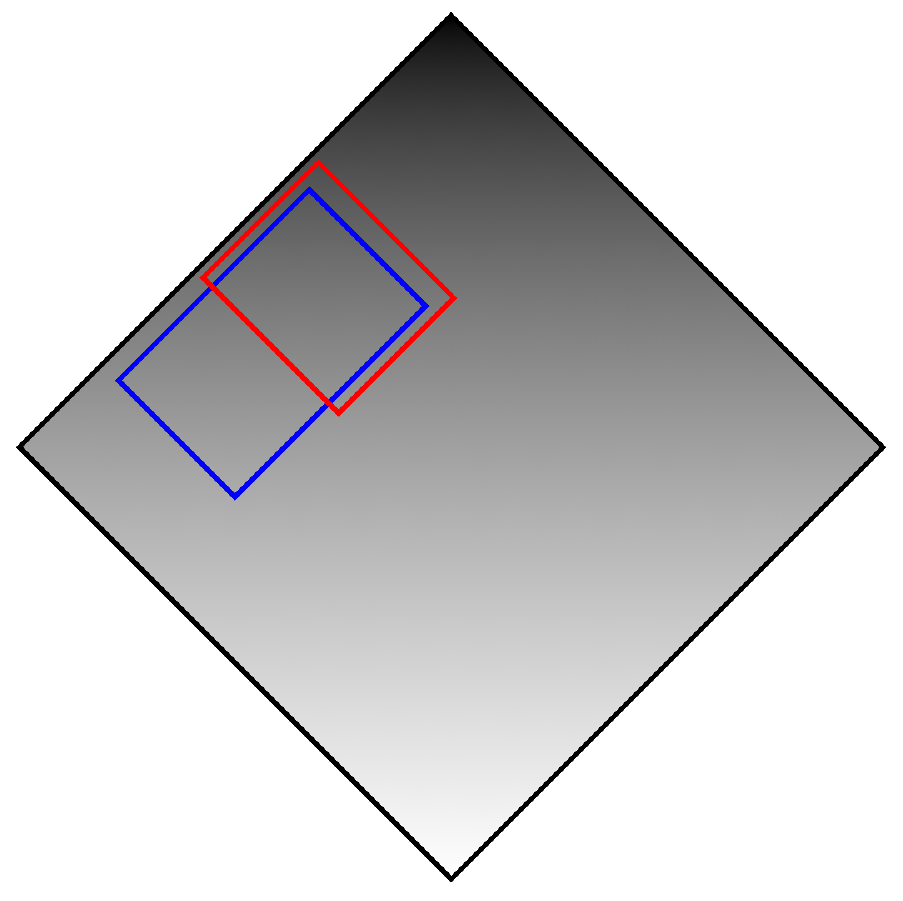}}
 \vspace{40pt}
\subfigure[FRW with $w=\frac{1}{3}$]{\includegraphics[width=0.6\textwidth]{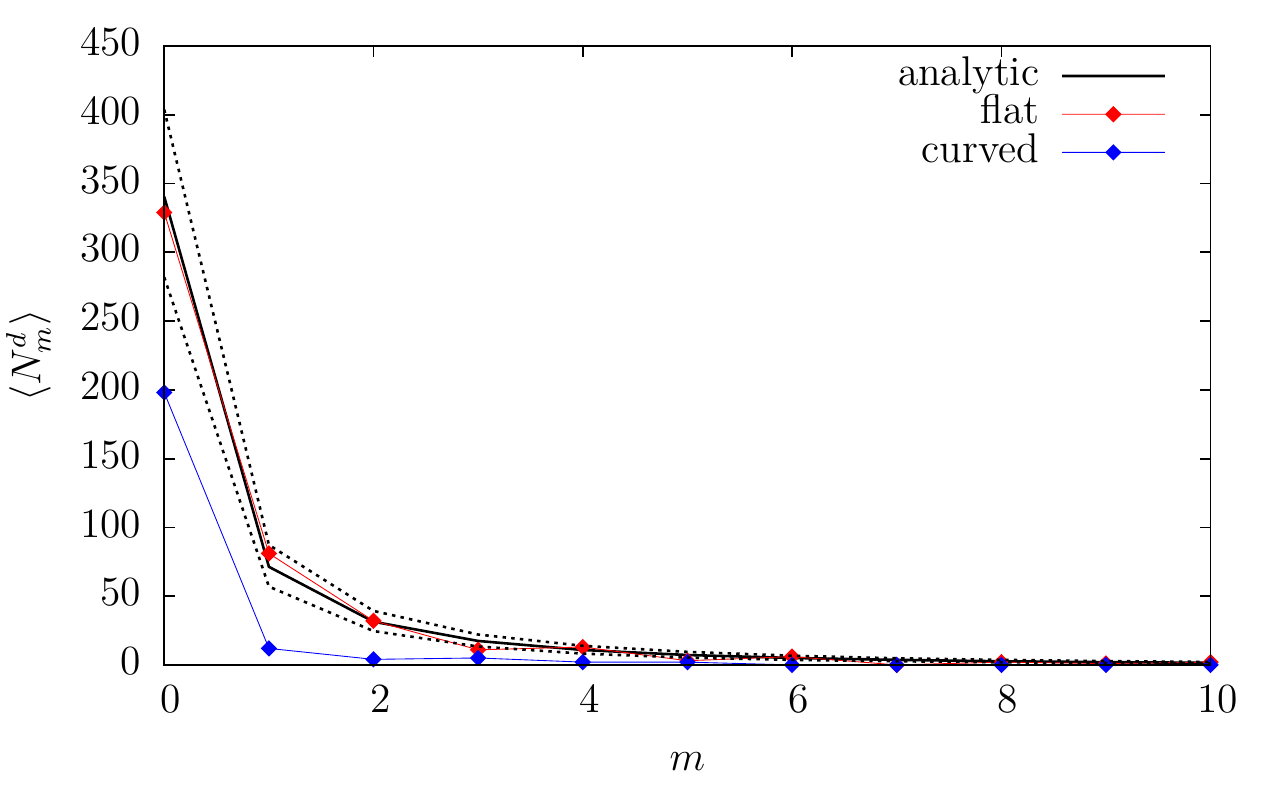} \hspace{13.5pt} \includegraphics[width=0.35\textwidth]{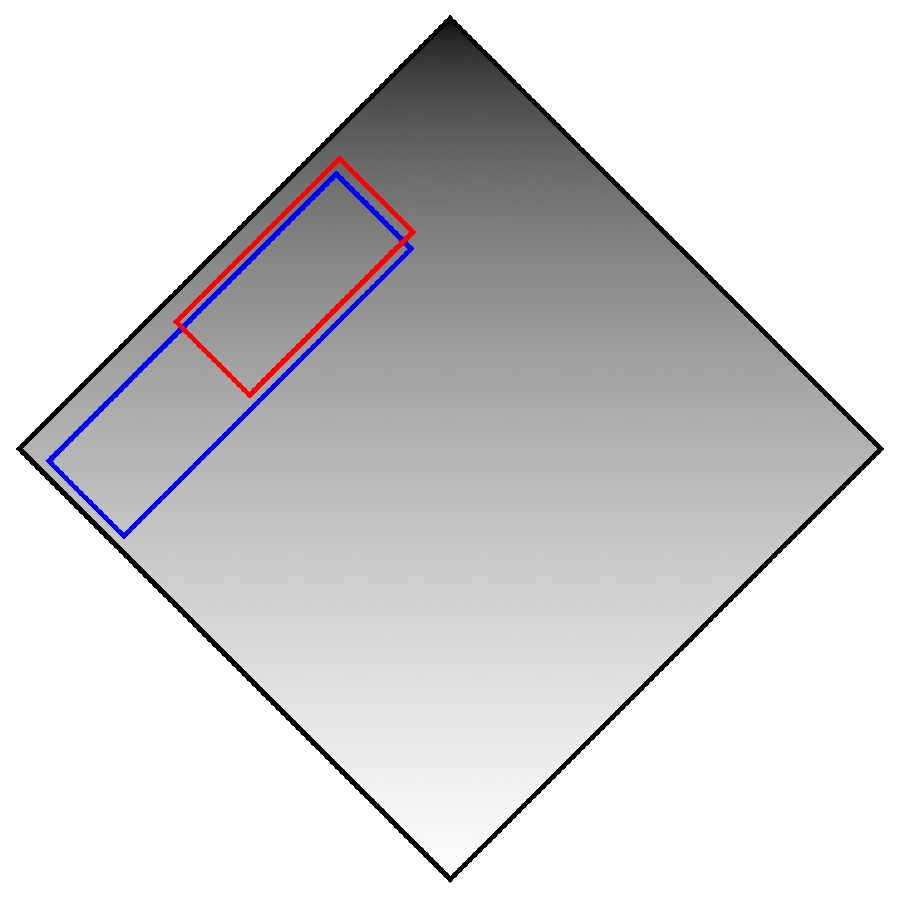}}
\caption{\label{fig:FRW_subsets} Single realisations of small interval causal sets $C'$ contained in a
  $N=10\,000$ element causal set $C$ obtained from a sprinkling into $4d$ FRW space-times which are matter or radiation dominated. 
  The sketches on the right hand side show which intervals are local and which non-local, while  the shading indicates
  the scale factor of the universe. This figure is taken from \cite{Glaser:2013pca}.
  } 
\end{figure}
To answer this, we simulate a single realisation of a $10\,000$ element causal set, for each of the three cases.
We then examine a random sample of $100$ element sub intervals for strong $d$-rigidity.
For the matter and radiation dominated FRW universes $w=0,1/3$, we are able to identify intervals whose abundance follows $\Nmd$ and those whose does not.
Examining the position of the points defining these intervals, we can even recognise them as `more' or `less' local in a qualitative sense.
This is shown in Figure \ref{fig:FRW_subsets}.

The same type of test on a $10\,000$ element causal set sprinkled into a flat interval will not show such differences.
Neither does the example of de Sitter space.
Since de Sitter space-time is maximally symmetric, the characteristics of any interval should only depend on its size.
\begin{figure}
\subfigure[$100$ element de Sitter intervals]{\includegraphics[width=0.45 \textwidth]{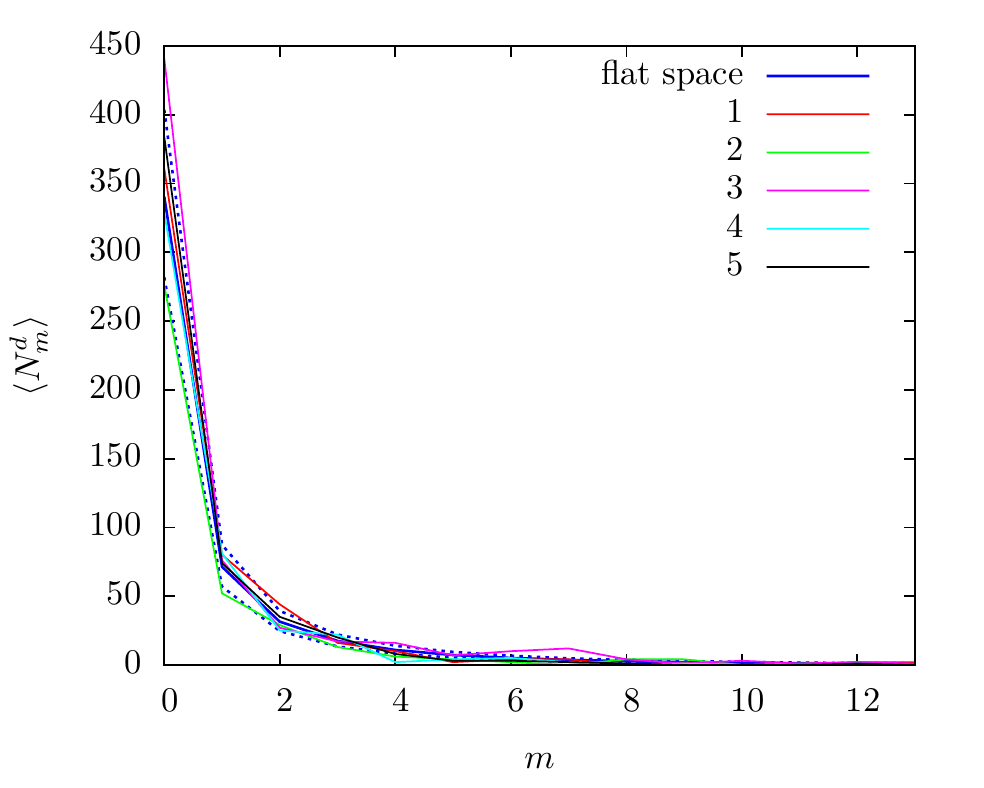}}
\subfigure[$2000$ element de Sitter intervals]{\includegraphics[width=0.45 \textwidth]{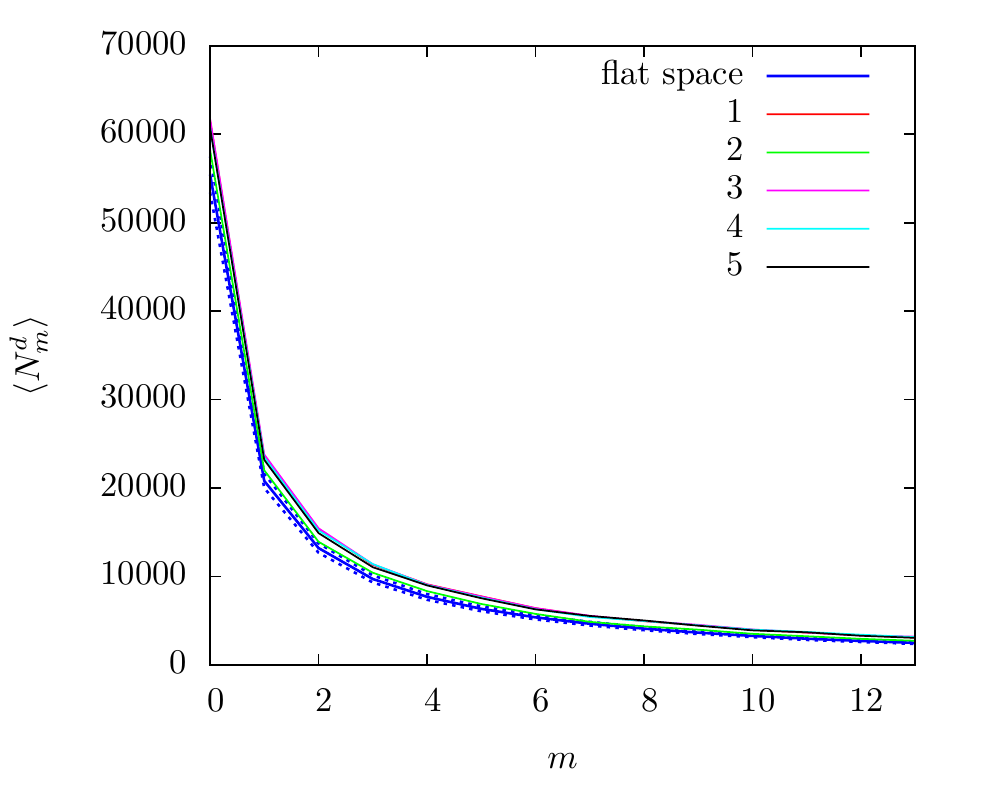}}
\caption{\label{fig:deSitter} 5 single realisations of small interval causal sets $C'$ contained in a $N=10\,000$ element causal set $C$ obtained from a sprinkling into $4d$ de Sitter space-time. This figure is taken from  \cite{Glaser:2013pca}.}
\end{figure}
To confirm this, we examine randomly chosen $100$ and $2\,000$ element sub intervals out of a $10\,000$ element de Sitter sprinkling.
The results are shown in Figure \ref{fig:deSitter}.
The curves $\NmCC$ for the $100$ element intervals are all very close to each other and to the curve $\Nmd$, as expected for relatively small intervals in a maximally symmetric space-time.
For the $2\,000$ element intervals, the curves $\NmCC$ clearly differ from $\Nmd$, but are still very similar to each other.
The interval abundance thus also captures this characteristic of de Sitter space.

\subsubsection{Transitive percolation}
In subsection \ref{subsec:pathintegral}, when talking about the sum over causal sets in the path integral, we introduced transitive percolations.
Transitive percolations are one of the easiest model to grow a causal set, and the causal sets they create show characteristics of de Sitter space  \cite{Ahmed:2009qm,Rideout:1999ub}.

The characteristic of de Sitter space that is examined is the relation between the proper time and volume of an interval.
For this, a fit to the relation between the volume and the proper time extent  for some intervals in the causal set is made.
This fit is performed on those intervals that, for a given proper time, have the largest volume.
For many values of the couplings $p=t/(t+1)$ and sizes of causal sets, this fits the relation for de Sitter space.
The free parameters in this fit are the de Sitter radius $\ell$ and a proportionality between the length of the longest path and the proper time.

We can examine if this similarity to de Sitter space-time extends to the interval abundances of percolated causal sets.
For this we examine three of the sets of parameters that are used in \cite{Ahmed:2009qm}, which we compile in Table \ref{tab:params}.

\begin{table}
\caption{\label{tab:params}Three sets of parameter values from  \cite{Ahmed:2009qm} which we have examined.}
\centering
\begin{tabularx}{\textwidth}{X X X X X}
\toprule
p & N & d & $\ell$ & m  \\
\midrule
0.03 & $1\,000$ & 3d & $2.331 \pm 0.011$ & $1.046 \pm 0.006$	\\
0.01 & $2\,000$ & 3d & $4.086 \pm 0.028$ & $1.136 \pm 0.006$	\\
0.005 & $15\,000$ & 4d & $ 6.20 \pm 0.12$& $1.710 \pm 0.013$	\\
\bottomrule
\end{tabularx}

\end{table}
\begin{figure}
{\centering
\subfigure[ p=$0.005$ N=$15000$, average size of the interval $\langle N \rangle=1259.1 \pm  47.8$]{\includegraphics{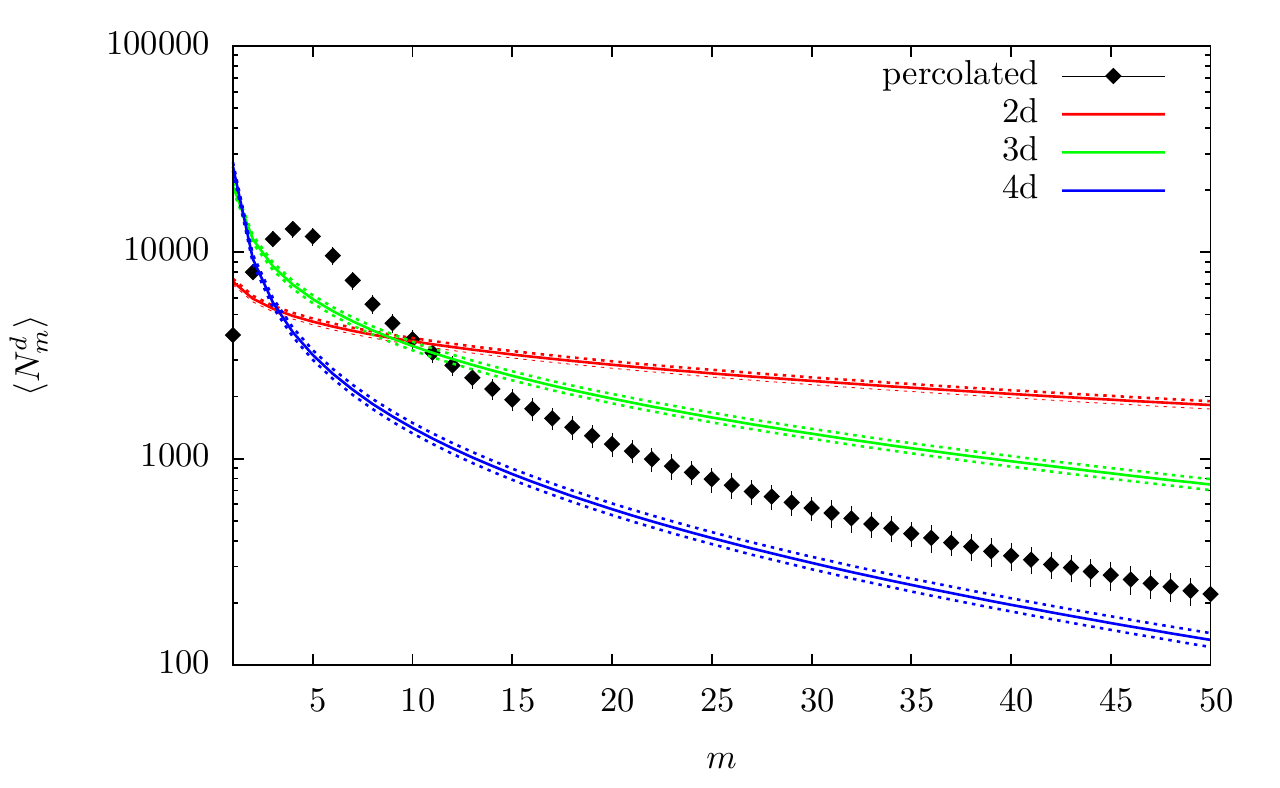}}

}
\subfigure[ p=$0.03$ N=$1000$, average size of the interval \mbox{$\langle N \rangle=335.2 \pm 25.9$} ]{\includegraphics[width=0.49\textwidth]{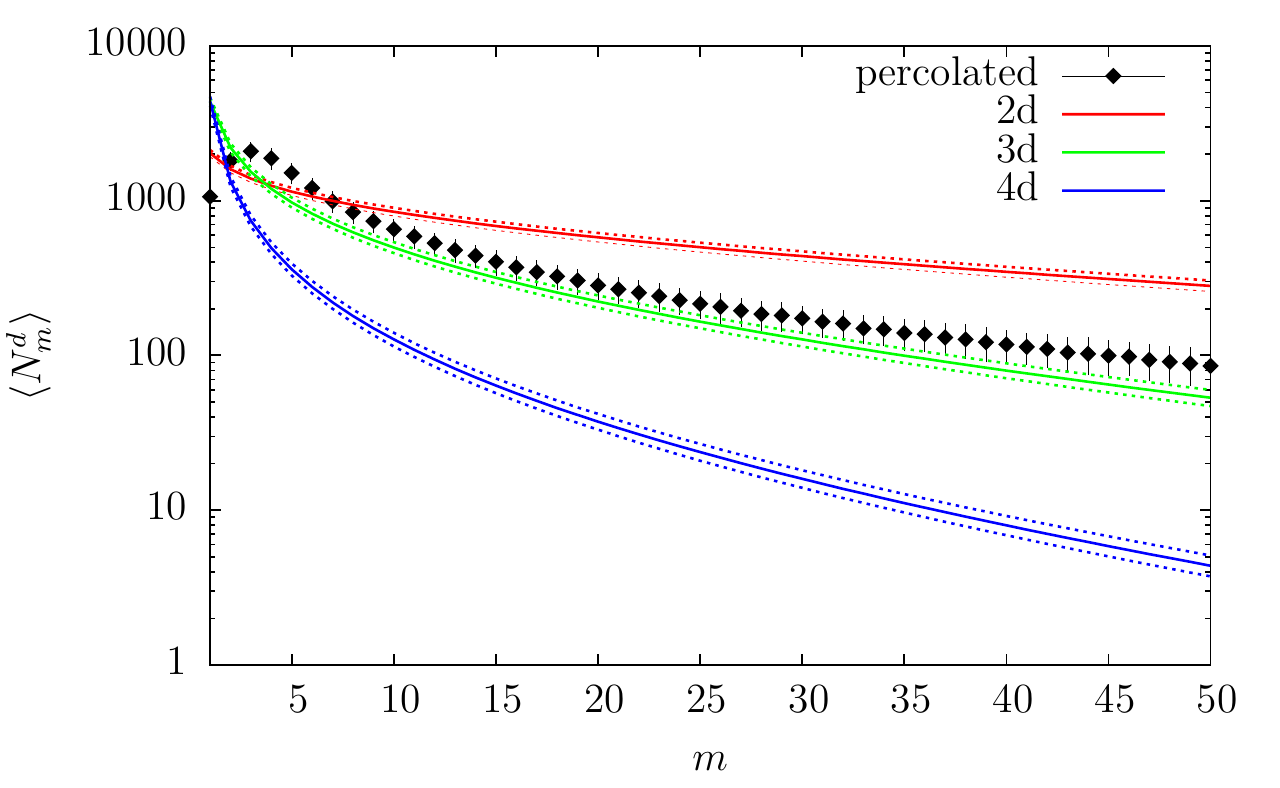}}
\subfigure[ p=$0.01$ N=$2000$, average size of the interval $\langle N \rangle=695.7 \pm 53.5$]{\includegraphics[width=0.49\textwidth]{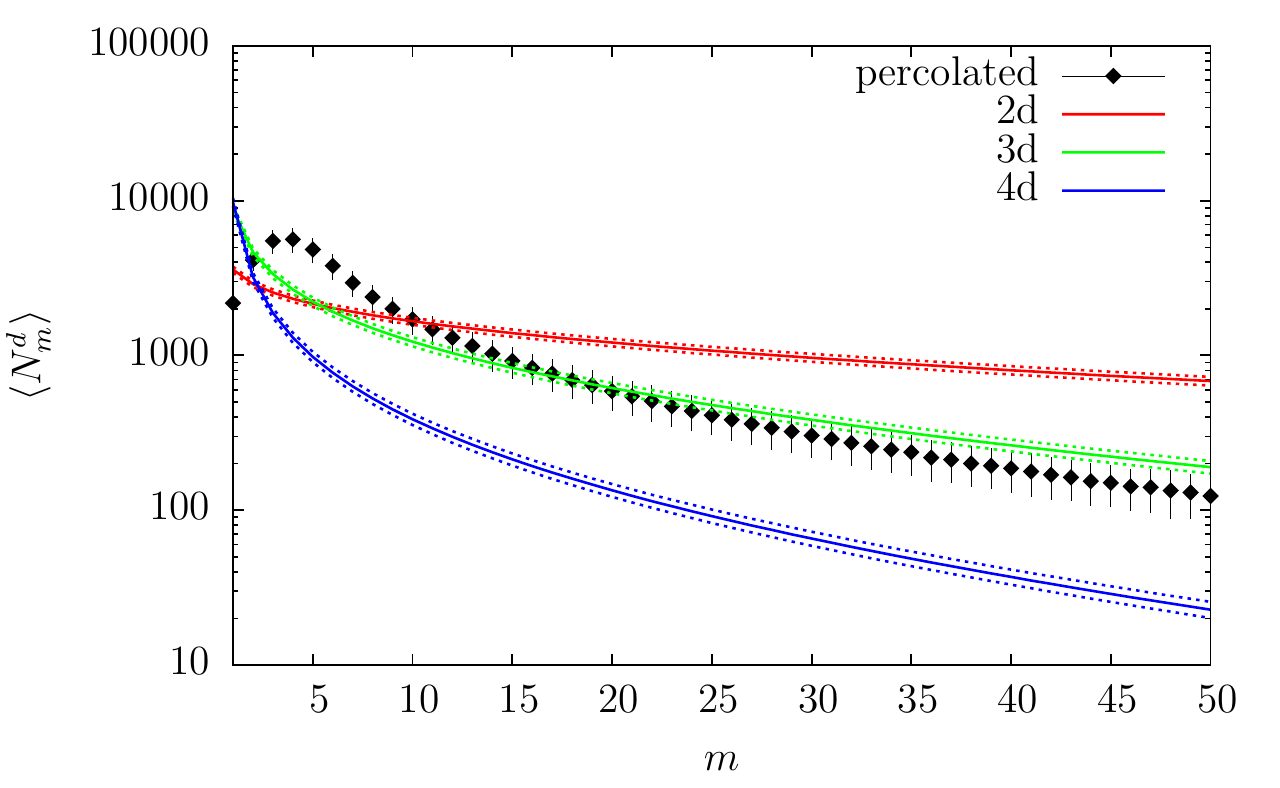}}

\caption{\label{fig:percolation}The $\langle \NmCC \rangle $  for percolated causal sets whose longest path has $20$ elements,
are drawn  in black and compared with the $\Nmd$ for $d=2,3,4$. This figure is taken from  \cite{Glaser:2013pca}. }
\end{figure}

For each of the parameter combinations in Table \ref{tab:params}, we generate $100$ causal sets and measure the average interval abundance for the intervals of largest volume for a given proper time.
In Figure \ref{fig:percolation}, we plot the interval abundances for the largest intervals of proper time length $20$ averaged over these $100$ causal sets.
For comparison we also show the $\Nmd$ for $d=2,3,4$. 
It is then clear that the $\NmCC$ do not follow the Interval abundance for flat space.
The characteristic shape even persists for small intervals, which should reasonably be expected to appear flat.

This is a strong indication that these percolated causal sets are not manifoldlike.

We also examine if a coarse graining of the percolated causal sets might lead to sets with the correct interval abundance, which it does not.
More details on this examination are included in  \cite{Glaser:2013pca}.

\subsubsection{Non-manifoldlike causal sets}
\begin{figure}
\subfigure[\label{fig:chain} $10$ element chain]{\includegraphics[width=0.49\textwidth]{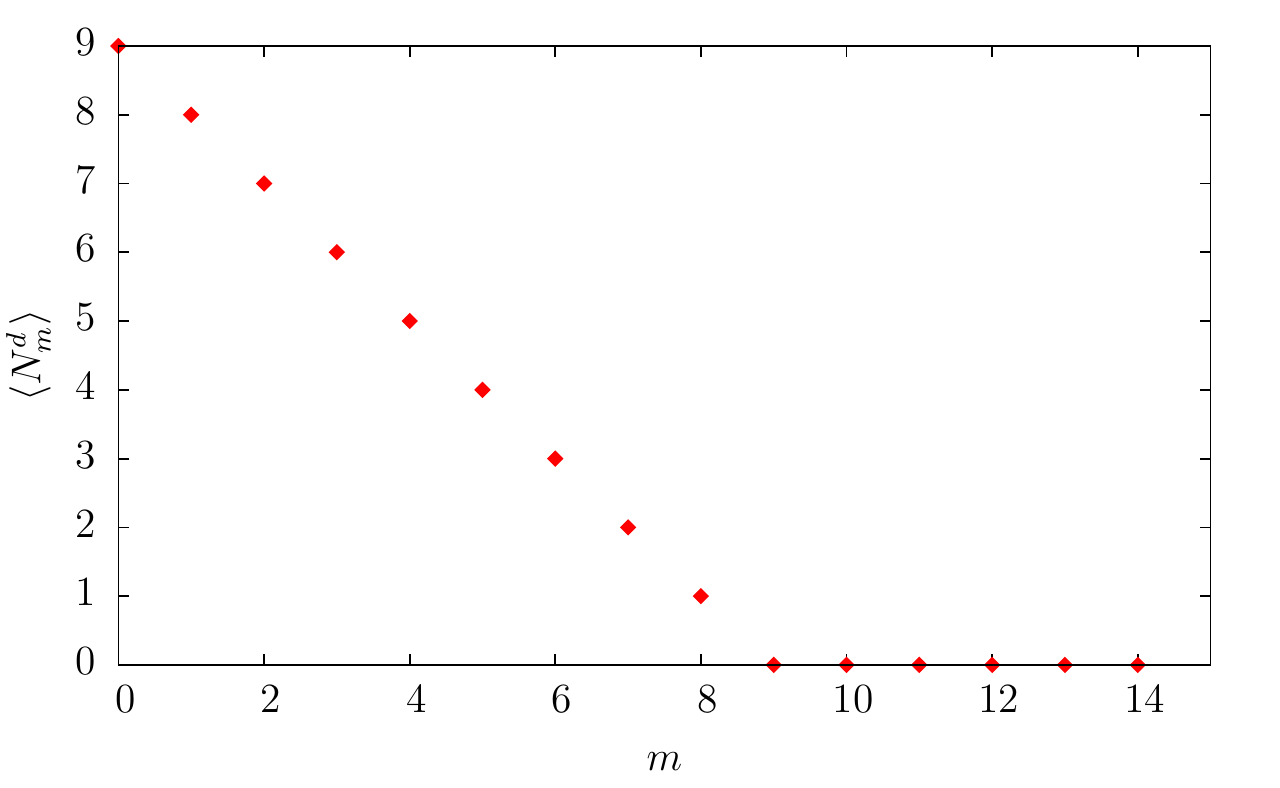}}
\subfigure[\label{fig:KRintervals} $100$ element KR order]{\includegraphics[width=0.49\textwidth]{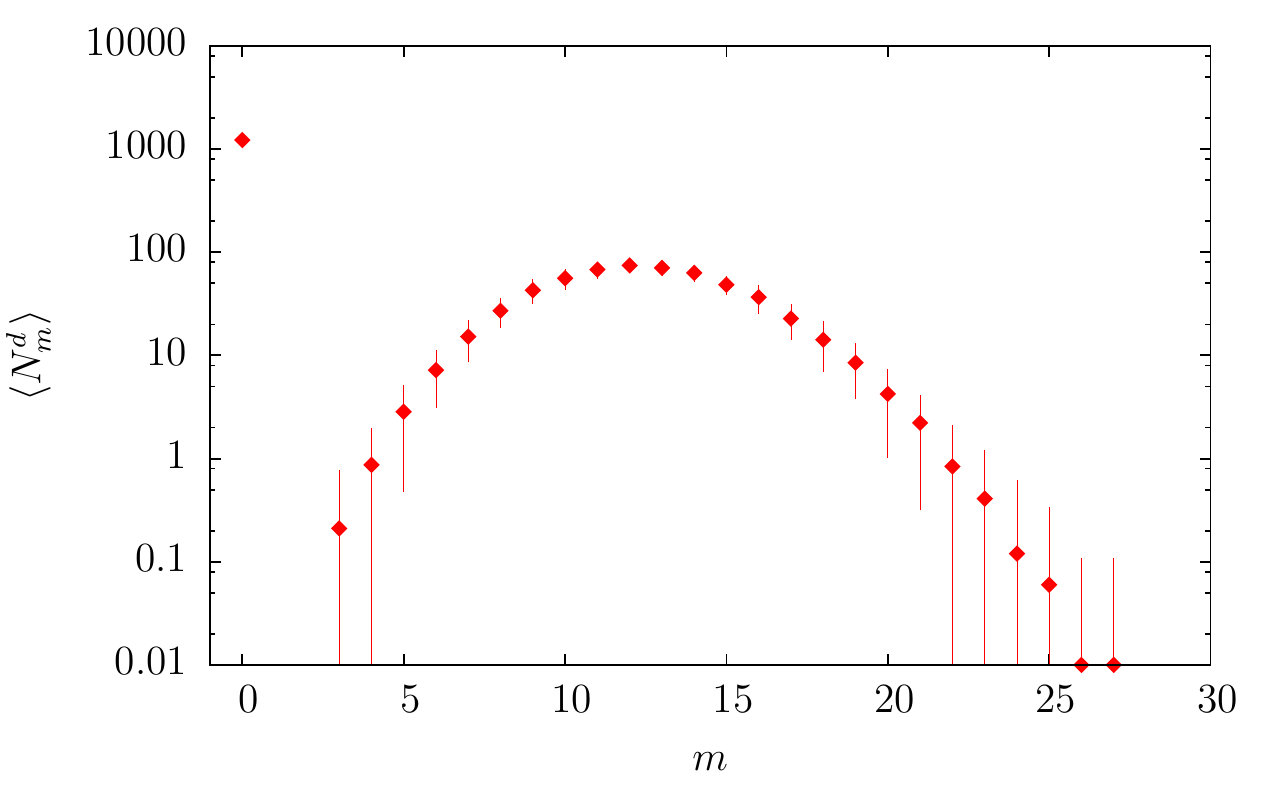}}
\caption{Interval abundance for a $10$ element path on the left hand and average of the interval abundance for $100$ realisations of a $100$ element KR order. This figure is taken from  \cite{Glaser:2013pca}.}
\end{figure}

We have seen that the interval abundance is a useful observable, in manifoldlike space-times.
One important crosscheck is to examine whether causal sets that are not manifoldlike can follow the curve for the abundance by accident.
This is hard to do in general, but we can check it for simple non-manifoldlike causal sets. 
The easiest case is a total anti-chain, which contains absolutely no intervals, and thus has no interval abundance to plot.
In a total chain, which corresponds to a $0+1$d causal set, the number of intervals of size $m$ falls off linearly (c.f. Figure \ref{fig:chain}).

Of special interest is the case of the KR poset. 
If these omnipresent, for our purposes pathological, posets are even weakly d-rigid, the interval abundance would not be of much use.
To test this, we use Cactus to generate $100$ realisations of a $100$ element KR poset and count the average interval abundance.
The result is shown in Figure \ref{fig:KRintervals}.

We can even predict the curve of the interval abundance for KR posets of any size.
The KR posets have $N/4$ elements in the top and bottom layer and $N/2$ elements in the middle.
An element in the top (bottom) layer is then linked to a given element in the middle with probability $1/2$.
On average, a KR poset will contain $N^2/8$ links, that is $0$-intervals, and if it is sufficiently large, it will contain almost no $1$-intervals.
The distribution of $m$-intervals will show a peak around $m= N/8$.

\subsection{Conclusions about the interval abundance}
The interval abundance is a quantity that contains much information about a causal set.
We can use it to determine the dimension of a region in a causal set and to examine if this region is manifoldlike and small.
An obvious extension of these results is to calculate the abundance of intervals for slightly curved space-times using Riemann normal coordinates.
This might then allow us to use the interval abundance to measure the curvature of a region and to examine the manifoldlikeness for curved space-times.

In the past, chains were the objects of choice in trying to examine a causal set. 
They are instrumental in the Myrheim-Meyer dimension, the geodesic distance and can also be used to define the curvature of a causal set  \cite{Roy:2012uz}.
Our exploration of the abundance of intervals shows that intervals might be more useful.
Chains can, by definition, only contain timelike information, while intervals do in some sense also encode spacelike quantities.
The only way to recover spacelike quantities from a causal set is by making use of intervals.

This means, that the abundance of intervals encodes a lot of information about the space-time.
It is very hard to construct causal sets that do on average follow the interval abundance, but are obviously not manifoldlike.
We thus believe that the interval abundance could be a foundation upon which to construct an embedding for a causal set.

\section{The road forward in causal sets}
Causal set theory is an intriguing candidate theory of quantum gravity.
Based on a minimum of structure, the first step in this theory is to show that one can reconstruct a continuum description from it.

Large steps in this direction have been made in \cite{Major:2006hv}, which constructs an algorithm to recover the continuum topology of space-time from a causal set,
and \cite{Henson:2006dk}, where an approximate embedding into Minkowski space was constructed.

In this chapter, new work on the d'Alembertian operator and the interval abundance in causal sets was summarised.

The d'Alembertian operator is an important object to characterise the dynamics of scalar fields on causal sets.
Causal sets are fundamentally non-local, and the nearest neighbours to a point are all points it is linked to.
To control the large contributions this non-locality entails, we introduce a sum over layers and the d'Alembertian
\begin{align}
B^{(d)}\phi(x)=\frac{1}{l^2}\left( \alpha_d \phi(x) +\beta_d \sum\limits_{i=1}^{n}  C^{(d)}_i \sum\limits_{y \in L_{i}}\phi(y) \right) \, .
\end{align}
In this thesis, simple forms for the constants $\alpha_d, \beta_d$ and $C_i^{(d)}$ have been presented.
The d'Alembertian also induces a causal set action.
This so-called Benincasa-Dowker action is at the moment the action of choice when simulating causal sets.

The work presented here proves that the prefactor relating this action and the operator $B^{(d)}$ is a dimension independent constant $- 1/2$.
This action is tested on flat regions of Minkowski space-time, a cylinder space-time, and the topology  changing trousers in $2$d  \cite{Benincasa:2010as}.
Future work will be to test it on other space-times, especially in higher dimensions.
An alternative definition for the causal set action, in terms of chains instead of intervals, is presented in  \cite{Roy:2012uz}.
Comparing these actions on different space-times could lead to further insights into their domain of applicability and possible differences between them.

The interval abundance, as a new observable, can help us determine the dimension of a causal set.
It also allows us to identify small local regions in a causal set that embeds into an arbitrary curved space-time with sufficient density.
This opens the possibility to implement  \cite{Major:2006hv} in a concrete way.
We also conjecture that using the information in the interval abundances will allow us to construct an exact, faithful embedding into flat space.
If this construction can be done, it would prove conjecture \ref{conjecture1}.
It would also prove that no non-manifoldlike causal set can mimic the interval abundance.

To calculate the path integral over causal sets, we have to specify the class of causal sets we sum over.
The most popular choices are, to sum over all causal sets and include an action, or to sum over a class of grown causal sets with their weight determined by the growth process.
Summing over all partial orders presents a challenge in form of the KR orders, which dominate the space of all partial orders.
Any action on the causal set has to suppress these strongly, to avoid them dominating the path integral.
One proposal for an action is the Benincasa-Dowker action, constructed from the d'Alembertian operator.
In $4$d, this action is
\begin{align}
\mc{S}^{(4)}(\mc{C})=\frac{1}{l^2}(N-N_1+9N_2-16 N_3+ 8N_4) \;,
\end{align}
where $N_i$ is the number of exclusive Alexandrov intervals of size $i$.
For a KR order of size $N$ the expected number of links $N_1$ is $N^2/8$, while the expected number of intervals $N_2,N_3,N_4$ is close to zero.
Depending on the discreteness scale $l$ the Benincasa-Dowker action might thus be strong enough to suppress the KR orders.
However, if it is also strong enough to suppress all subleading dominant orders remains to be examined.

\chapter{Conclusion}
The search for a theory of quantum gravity is exploring many paths.
In this thesis, we have peeked into the developments in causal dynamical triangulations and causal set theory.

These approaches are connected by their focus on the Lorentzian structure of space-time and their use of the path integral formalism to quantise gravity.
They do differ fundamentally in their understanding of the underlying structure of space-time.

Causal dynamical triangulations are, in a certain sense, agnostic about the fundamental building blocks.
The discretisation introduced in this approach is a non-physical cut-off to be taken to zero at the end of the calculation.
Causal set theory, on the other hand, assumes that space-time is fundamentally discrete.

Even though both theories focus on the Lorentzian structure of space-time, the way they do this is disparate.
In causal dynamical triangulations, the calculations are conducted in Euclidian space and, at least for analytic results, a Wick rotation to compare to Lorentzian space-times is known.
For this purpose one introduces an anisotropy between time and space, and an explicit time foliation.

The phase diagram of causal dynamical triangulations depends on an analogue of the bare Newton's constant and the anisotropy parameter.
Only a non-trivial choice of the anisotropy leads to the continuum phase C.
The typical geometries in this phase show signs of space-time behaviour.
Their average three-volume profile follows the de Sitter profile and their Hausdorff dimension is $4$ \cite{Ambjorn:2013tki}.

Anisotropy between space and time is also a main ingredient in Ho\v{r}ava-Lifshitz gravity.
In this thesis, we presented arguments to connect causal dynamical triangulations and Ho\v{r}ava-Lifshitz gravity and summarised our own work on this problem.
We showed that 2d causal dynamical triangulations and $2$d projectable Ho\v{r}ava-Lifshitz gravity have the same Hamiltonian and thus describe the same theory \cite{Ambjorn:2013joa}.

This connection opens up new directions of study.
Black holes in Ho\v{r}ava-Lifshitz gravity are one particularly interesting area of research.
The breaking of Lorentz invariance that comes with the anisotropy leads to differences from general relativistic black holes, as described in, for example,  \cite{Kiritsis:2009rx}.
These new black holes might then lend themselves more easily to being implemented in a discretised space-time, like causal dynamical triangulations.
Simulations of these black holes might then allow for the first non-perturbative study of Hawking radiation.
They could also be used to study black hole entropy, the entanglement of fields across the horizon, and the evaporation of black holes.

There are recent developments in causal dynamical triangulations that point in a completely novel direction. 
A new generalisation of causal dynamical triangulations that abolishes the need for a strict foliation has been established.
Instead of the global foliation a local causality condition ensures that the Wick rotation remains well defined. 

In $2$ and $3$ dimensions, simulations of this theory, like standard causal dynamical triangulations, show a de Sitter phase  \cite{Jordan:2013awa}. 
However, there are few analytic results as of yet, and the simulations in this generalisation are harder compared to standard causal dynamical triangulations \cite{Jordan:2013iaa}.

In causal set theory, it is impossible to introduce a Wick rotation.
Since the partial order relations are derived from the causal relations between elements, an Euclidianised version of causal sets makes no sense.

In causal sets, the volume of a region is proportional to the number of elements contained in this region.
The only known way to implement such a volume number correspondence in a Lorentzian space-time of $d>2$, without breaking Lorentz symmetry, is to discretise space-time through a random process.
The distribution of points in a causal set is Poissonian, which implies that it only  depends on the space-time volume.
This ensures that there are no preferred directions associated to any point \cite{Bombelli:2006nm}.

Unfortunately this implies that the causal set has to be a graph of infinite valency.
The high valency arises because the invariant subspace under a Lorentz boost, the light-cone, is non-compact.
Preserving this non-compact space leads to non-local effects.

The non-local effects in causal set theory could be a promising direction to derive phenomenological constraints on quantum gravity.
Despite this, in contexts where we want to reconstruct continuum space-times these effects are problematic.
One example is the construction of a  discrete d'Alembertian operator, to construe scalar field dynamics on a causal set.
 
To define a d'Alembertian on the causal set, we have to introduce a complicated sum over layers that tends to the d'Alembertian on average \cite{Sorkin:2007qi}.
This sum over layers is necessitated by the non-locality.
The fluctuations around the d'Alembertian are large, and to tame them one needs to introduce an intermediate non-locality scale, smearing the layers out. 
It thus  gives rise to a new type of non-locality.
Exploring this intermediate non-locality can lead to new phenomenology that might be experimentally verifiable.

Many methods have been developed to explore the geometry a causal set encodes.
Most of them assume that one applies them to an approximately flat interval.
In a curved space, a local interval is approximately flat.
A non-local interval is one that `stretches' over a region in which curvature varies significantly.
Local or non-local intervals cannot be distinguished by their size, and until recently no way to identify them in a causal set was known.

In a causal set, intervals of size $N>0$ contains smaller sub-intervals.
The number of sub intervals of size $m$ is called the abundance of $m$-intervals.
We presented the derivation of the average interval abundance for general $m$ in $d$-dimensional Minkowski space.
We argued that this quantity can be used as a ruler, to estimate whether a region is approximately flat, thus making it possible to identify local intervals.
It also supplies us with a dimension estimator and an indicator for manifoldlikeness \cite{Glaser:2013pca}.

The curve of interval abundances is certainly a necessary condition for a causal set to embed into flat space-time.
We conjecture that it is even a sufficient condition and could be used to construct an algorithm to embed a causal set into flat space-time.

In this thesis, we have examined the problem of quantum gravity from various sides.
This last paragraph is thus the proper place to dare a prognosis for the future of this area of research.

In the introduction, we mentioned the plethora of approaches to quantum gravity.
Recently a trend of unification and cross-fertilisation between the approaches can be observed.
Several approaches predict the same spectral dimension for space-time  \cite{Horava:2009if,Ambjorn:2005db,Calcagni:2013vsa,Modesto:2008jz}.
A connection between loop quantum gravity and spin foams is proposed, and the details are being currently worked out  \cite{Alesci:2011ia}.
We have here shown signs that causal dynamical triangulations and Ho\v{r}ava-Lifshitz gravity are the same theory in $2$ dimensions  \cite{Ambjorn:2013joa}.
And in turn, it has been shown that Ho\v{r}ava-Lifshitz gravity is related to Einstein-Aether theories  \cite{PhysRevD.81.101502}.
Matrix models have a limit that corresponds to causal dynamical triangulations, and the limiting procedure that connects causal dynamical triangulations and Euclidian dynamical triangulations in this case is well understood  \cite{Ambjorn:2008gk,Loll:2005dr}.

There are many examples of methods developed in one approach being used successfully for another, like the spectral dimension mentioned above or the fact that almost all discretised approaches use some variant of the Regge action  \cite{Regge:1961px}.

An optimistic observer might thus hope that a sum greater than its parts might solve the puzzle of quantum gravity.
The way towards a solution might lie in the unification of approaches.

\bibliographystyle{utphys}
\bibliography{bibliography}

\providecommand{\href}[2]{#2}\begingroup\raggedright\begin{thebibliography}{100}

\bibitem{PDG}
J.~Beringer and {(Particle Data Group)}, ``2013 review of particle physics,''
  {\em Phys. Rev. D} {\bfseries 86} (2012) 010001.

\bibitem{Hack:2012qf}
T.-P. Hack and V.~Moretti, ``{On the Stress-Energy Tensor of Quantum Fields in
  Curved Spacetimes - Comparison of Different Regularization Schemes and
  Symmetry of the Hadamard/Seeley-DeWitt Coefficients},''
  \href{http://dx.doi.org/10.1088/1751-8113/45/37/374019}{{\em J.Phys.}
  {\bfseries A45} (2012) 374019},
\href{http://arxiv.org/abs/1202.5107}{{\ttfamily arXiv:1202.5107 [gr-qc]}}.

\bibitem{Polchinski:1998rq}
J.~Polchinski,
``{String theory. Vol. 1: An introduction to the bosonic string},''.

\bibitem{Polchinski:1998rr}
J.~Polchinski,
``{String theory. Vol. 2: Superstring theory and beyond},''.

\bibitem{Lust:2013koa}
D.~Lust and T.~R. Taylor, ``{Limits on Stringy Signals at the LHC},''
\href{http://arxiv.org/abs/1308.1619}{{\ttfamily arXiv:1308.1619 [hep-ph]}}.

\bibitem{Ashtekar:2004eh}
A.~Ashtekar and J.~Lewandowski, ``{Background independent quantum gravity: A
  Status report},'' \href{http://dx.doi.org/10.1088/0264-9381/21/15/R01}{{\em
  Class.Quant.Grav.} {\bfseries 21} (2004) R53},
\href{http://arxiv.org/abs/gr-qc/0404018}{{\ttfamily arXiv:gr-qc/0404018
  [gr-qc]}}.

\bibitem{lrr-2008-5}
C.~Rovelli, ``Loop quantum gravity,'' 2008.
\newblock \url{http://www.livingreviews.org/lrr-2008-5}.

\bibitem{Thiemann:2004wk}
T.~Thiemann, ``{Reduced phase space quantization and Dirac observables},''
  \href{http://dx.doi.org/10.1088/0264-9381/23/4/006}{{\em Class.Quant.Grav.}
  {\bfseries 23} (2006) 1163--1180},
\href{http://arxiv.org/abs/gr-qc/0411031}{{\ttfamily arXiv:gr-qc/0411031
  [gr-qc]}}.

\bibitem{Brown:1994py}
J.~D. Brown and K.~V. Kuchar, ``{Dust as a standard of space and time in
  canonical quantum gravity},''
  \href{http://dx.doi.org/10.1103/PhysRevD.51.5600}{{\em Phys.Rev.} {\bfseries
  D51} (1995) 5600--5629},
\href{http://arxiv.org/abs/gr-qc/9409001}{{\ttfamily arXiv:gr-qc/9409001
  [gr-qc]}}.

\bibitem{Ashtekar:2006uz}
A.~Ashtekar, T.~Pawlowski, and P.~Singh, ``{Quantum Nature of the Big Bang: An
  Analytical and Numerical Investigation. I.},''
  \href{http://dx.doi.org/10.1103/PhysRevD.73.124038}{{\em Phys.Rev.}
  {\bfseries D73} (2006) 124038},
\href{http://arxiv.org/abs/gr-qc/0604013}{{\ttfamily arXiv:gr-qc/0604013
  [gr-qc]}}.

\bibitem{Barrett:1997gw}
J.~W. Barrett and L.~Crane, ``{Relativistic spin networks and quantum
  gravity},'' \href{http://dx.doi.org/10.1063/1.532254}{{\em J.Math.Phys.}
  {\bfseries 39} (1998) 3296--3302},
\href{http://arxiv.org/abs/gr-qc/9709028}{{\ttfamily arXiv:gr-qc/9709028
  [gr-qc]}}.

\bibitem{lrr-2013-3}
A.~Perez, ``The spin-foam approach to quantum gravity,'' 2013.
\newblock \url{http://www.livingreviews.org/lrr-2013-3}.

\bibitem{Ambjorn:1999nc}
J.~Ambjorn, J.~Jurkiewicz, and R.~Loll, ``{Lorentzian and Euclidean quantum
  gravity: Analytical and numerical results},''
\href{http://arxiv.org/abs/hep-th/0001124}{{\ttfamily arXiv:hep-th/0001124
  [hep-th]}}.

\bibitem{PROP:PROP201300032}
V.~Rivasseau, ``The tensor track, iii,''
  \href{http://dx.doi.org/10.1002/prop.201300032}{{\em Fortschritte der Physik}
  {\bfseries 62} no.~2, (2014) 81--107}.
  \url{http://dx.doi.org/10.1002/prop.201300032}.

\bibitem{Bonzom:2011zz}
V.~Bonzom, R.~Gurau, A.~Riello, and V.~Rivasseau, ``{Critical behavior of
  colored tensor models in the large N limit},''
  \href{http://dx.doi.org/10.1016/j.nuclphysb.2011.07.022}{{\em Nucl.Phys.}
  {\bfseries B853} (2011) 174--195},
\href{http://arxiv.org/abs/1105.3122}{{\ttfamily arXiv:1105.3122 [hep-th]}}.

\bibitem{Ambjorn:2012jv}
J.~Ambjorn, A.~Goerlich, J.~Jurkiewicz, and R.~Loll, ``{Nonperturbative Quantum
  Gravity},'' \href{http://dx.doi.org/10.1016/j.physrep.2012.03.007}{{\em
  Phys.Rept.} {\bfseries 519} (2012) 127--210},
\href{http://arxiv.org/abs/1203.3591}{{\ttfamily arXiv:1203.3591 [hep-th]}}.

\bibitem{Reuter:2012id}
M.~Reuter and F.~Saueressig, ``{Quantum Einstein Gravity},''
  \href{http://dx.doi.org/10.1088/1367-2630/14/5/055022}{{\em New J.Phys.}
  {\bfseries 14} (2012) 055022},
\href{http://arxiv.org/abs/1202.2274}{{\ttfamily arXiv:1202.2274 [hep-th]}}.

\bibitem{Hawking:1976fe}
S.~Hawking, A.~King, and P.~Mccarthy, ``{A New Topology for Curved Space-Time
  Which Incorporates the Causal, Differential, and Conformal Structures},''
\href{http://dx.doi.org/10.1063/1.522874}{{\em J.Math.Phys.} {\bfseries 17}
  (1976) 174--181}.

\bibitem{malament:1399}
D.~B. Malament, ``The class of continuous timelike curves determines the
  topology of spacetime,'' \href{http://dx.doi.org/10.1063/1.523436}{{\em
  Journal of Mathematical Physics} {\bfseries 18} no.~7, (1977) 1399--1404}.
  \url{http://link.aip.org/link/?JMP/18/1399/1}.

\bibitem{Bombelli:1987aa}
L.~Bombelli, J.~Lee, D.~Meyer, and R.~Sorkin, ``{Space-Time as a Causal Set},''
\href{http://dx.doi.org/10.1103/PhysRevLett.59.521}{{\em Phys.Rev.Lett.}
  {\bfseries 59} (1987) 521--524}.

\bibitem{Ambjorn:2012zx}
J.~Ambj{\o}rn, L.~Glaser, A.~G{\"o}rlich, and Y.~Sato, ``{New multicritical
  matrix models and multicritical 2d CDT},''
  \href{http://dx.doi.org/10.1016/j.physletb.2012.04.047}{{\em Phys.Lett.}
  {\bfseries B712} (2012) 109--114},
  \href{http://arxiv.org/abs/1202.4435}{{\ttfamily arXiv:1202.4435 [hep-th]}}.
This article is reproduced in Appendix \ref{app:Dimer}.

\bibitem{Ambjorn:2013joa}
J.~Ambj{\o}rn, L.~Glaser, Y.~Sato, and Y.~Watabiki, ``{2d CDT is 2d
  Horava-Lifshitz quantum gravity},''
  \href{http://dx.doi.org/10.1016/j.physletb.2013.04.006}{{\em Phys.Lett.}
  {\bfseries B722} (2013) 172--175},
  \href{http://arxiv.org/abs/1302.6359}{{\ttfamily arXiv:1302.6359 [hep-th]}}.
This article is reproduced in Appendix \ref{app:HL}.

\bibitem{Ambjorn:2013eha}
J.~Ambj{\o}rn, L.~Glaser, A.~G{\"o}rlich, and J.~Jurkiewicz, ``Euclidian 4d
  quantum gravity with a non-trivial measure term,''
  \href{http://dx.doi.org/10.1007/JHEP10(2013)100}{{\em Journal of High Energy
  Physics} {\bfseries 2013} no.~10, (2013) 1--24}.
  \url{http://dx.doi.org/10.1007/JHEP10%282013%29100}. This article is
  reproduced in Appendix \ref{app:EDT}.

\bibitem{Glaser:2013xha}
L.~Glaser, ``A closed form expression for the causal set d’alembertian,''
  {\em Classical and Quantum Gravity} {\bfseries 31} no.~9, (2014) 095007.
  \url{http://stacks.iop.org/0264-9381/31/i=9/a=095007}. This article is
  reproduced in Appendix \ref{app:Factor}.

\bibitem{Glaser:2013pca}
L.~Glaser and S.~Surya, ``{Towards a Definition of Locality in a Manifoldlike
  Causal Set},'' \href{http://dx.doi.org/10.1103/PhysRevD.88.124026}{{\em
  Phys.Rev.} {\bfseries D88} (2013) 124026},
  \href{http://arxiv.org/abs/1309.3403}{{\ttfamily arXiv:1309.3403 [gr-qc]}}.
This article is reproduced in Appendix \ref{app:Local}.

\bibitem{Feynman:1996kb}
R.~Feynman, F.~Morinigo, W.~Wagner, and B.~Hatfield,
``{Feynman lectures on gravitation},''.

\bibitem{Wald:1984rg}
R.~M. Wald,
``{General Relativity},''.

\bibitem{CardyRG}
J.~Cardy, {\em Scaling and Renormalization in Statistical Physics}.
\newblock Cambridge Lecture Notes in Physics. Cambridge University Press, May,
  1996.

\bibitem{jansbook}
J.~Ambj{\o}rn, B.~Durhuus, and T.~Jonsson, {\em Quantum Geometry: A Statistical
  Field Theory Approach}.
\newblock Cambridge Monographs on Mathematical Physics. Cambridge University
  Press, 2005.
\newblock \url{http://books.google.dk/books?id=6faaPwAACAAJ}.

\bibitem{Gross1992367}
M.~Gross and S.~Varsted, ``Elementary moves and ergodicity in d-dimensional
  simplicial quantum gravity,''
  \href{http://dx.doi.org/http://dx.doi.org/10.1016/0550-3213(92)90012-Z}{{\em
  Nuclear Physics B} {\bfseries 378} no.~1–2, (1992) 367 -- 380}.
  \url{http://www.sciencedirect.com/science/article/pii/055032139290012Z}.

\bibitem{Gorlich:2011ga}
A.~Gorlich, ``{Causal Dynamical Triangulations in Four Dimensions},''
\href{http://arxiv.org/abs/1111.6938}{{\ttfamily arXiv:1111.6938 [hep-th]}}.

\bibitem{Ambjorn:1998xu}
J.~Ambjorn and R.~Loll, ``{Nonperturbative Lorentzian quantum gravity,
  causality and topology change},''
  \href{http://dx.doi.org/10.1016/S0550-3213(98)00692-0}{{\em Nucl.Phys.}
  {\bfseries B536} (1998) 407--434},
\href{http://arxiv.org/abs/hep-th/9805108}{{\ttfamily arXiv:hep-th/9805108
  [hep-th]}}.

\bibitem{Sorkin:2009ka}
R.~D. Sorkin, ``{Is the spacetime metric Euclidean rather than Lorentzian?},''
\href{http://arxiv.org/abs/0911.1479}{{\ttfamily arXiv:0911.1479 [gr-qc]}}.

\bibitem{Ambjorn:2014mra}
J.~Ambjorn, J.~Gizbert-Studnicki, A.~Görlich, and J.~Jurkiewicz, ``{The
  effective action in 4-dim CDT. The transfer matrix approach},''
\href{http://arxiv.org/abs/1403.5940}{{\ttfamily arXiv:1403.5940 [hep-th]}}.

\bibitem{Ambjorn:2012pp}
J.~Ambjorn, J.~Gizbert-Studnicki, A.~Gorlich, and J.~Jurkiewicz, ``{The
  Transfer matrix in four-dimensional CDT},''
  \href{http://dx.doi.org/10.1007/JHEP09(2012)017}{{\em JHEP} {\bfseries 1209}
  (2012) 017},
\href{http://arxiv.org/abs/1205.3791}{{\ttfamily arXiv:1205.3791 [hep-th]}}.

\bibitem{Ambjorn:2010fv}
J.~Ambjorn, A.~Gorlich, J.~Jurkiewicz, and R.~Loll, ``{Geometry of the quantum
  universe},'' \href{http://dx.doi.org/10.1016/j.physletb.2010.05.062}{{\em
  Phys.Lett.} {\bfseries B690} (2010) 420--426},
\href{http://arxiv.org/abs/1001.4581}{{\ttfamily arXiv:1001.4581 [hep-th]}}.

\bibitem{Ambjorn:2004qm}
J.~Ambjorn, J.~Jurkiewicz, and R.~Loll, ``{Emergence of a 4-D world from causal
  quantum gravity},''
  \href{http://dx.doi.org/10.1103/PhysRevLett.93.131301}{{\em Phys.Rev.Lett.}
  {\bfseries 93} (2004) 131301},
\href{http://arxiv.org/abs/hep-th/0404156}{{\ttfamily arXiv:hep-th/0404156
  [hep-th]}}.

\bibitem{Ambjorn:2012ij}
J.~Ambjorn, S.~Jordan, J.~Jurkiewicz, and R.~Loll, ``{Second- and First-Order
  Phase Transitions in CDT},''
  \href{http://dx.doi.org/10.1103/PhysRevD.85.124044}{{\em Phys.Rev.}
  {\bfseries D85} (2012) 124044},
\href{http://arxiv.org/abs/1205.1229}{{\ttfamily arXiv:1205.1229 [hep-th]}}.

\bibitem{Ambjorn:2011cg}
J.~Ambjorn, S.~Jordan, J.~Jurkiewicz, and R.~Loll, ``{A Second-order phase
  transition in CDT},''
  \href{http://dx.doi.org/10.1103/PhysRevLett.107.211303}{{\em Phys.Rev.Lett.}
  {\bfseries 107} (2011) 211303},
\href{http://arxiv.org/abs/1108.3932}{{\ttfamily arXiv:1108.3932 [hep-th]}}.

\bibitem{Ambjorn:2013tki}
J.~Ambjorn, A.~Goerlich, J.~Jurkiewicz, and R.~Loll, ``{Quantum Gravity via
  Causal Dynamical Triangulations},''
\href{http://arxiv.org/abs/1302.2173}{{\ttfamily arXiv:1302.2173 [hep-th]}}.

\bibitem{ben2000diffusion}
D.~Ben-Avraham and S.~Havlin, {\em Diffusion and reactions in fractals and
  disordered systems}.
\newblock Cambridge University Press, 2000.

\bibitem{Ambjorn:2005db}
J.~Ambjorn, J.~Jurkiewicz, and R.~Loll, ``{Spectral dimension of the
  universe},'' \href{http://dx.doi.org/10.1103/PhysRevLett.95.171301}{{\em
  Phys.Rev.Lett.} {\bfseries 95} (2005) 171301},
\href{http://arxiv.org/abs/hep-th/0505113}{{\ttfamily arXiv:hep-th/0505113
  [hep-th]}}.

\bibitem{Horava:2009if}
P.~Horava, ``{Spectral Dimension of the Universe in Quantum Gravity at a
  Lifshitz Point},''
  \href{http://dx.doi.org/10.1103/PhysRevLett.102.161301}{{\em Phys.Rev.Lett.}
  {\bfseries 102} (2009) 161301},
\href{http://arxiv.org/abs/0902.3657}{{\ttfamily arXiv:0902.3657 [hep-th]}}.

\bibitem{Lauscher:2005qz}
O.~Lauscher and M.~Reuter, ``{Fractal spacetime structure in asymptotically
  safe gravity},'' \href{http://dx.doi.org/10.1088/1126-6708/2005/10/050}{{\em
  JHEP} {\bfseries 0510} (2005) 050},
\href{http://arxiv.org/abs/hep-th/0508202}{{\ttfamily arXiv:hep-th/0508202
  [hep-th]}}.

\bibitem{Ambjorn:priv}
J.~Ambj{\o}rn, ``private communication.''.

\bibitem{Kazakov:1986hu}
V.~Kazakov, ``{Ising model on a dynamical planar random lattice: Exact
  solution},''
\href{http://dx.doi.org/10.1016/0375-9601(86)90433-0}{{\em Phys.Lett.}
  {\bfseries A119} (1986) 140--144}.

\bibitem{Staudacher:1989fy}
M.~Staudacher, ``{The Yang-lee Edge Singularity on a Dynamical Planar Random
  Surface},''
\href{http://dx.doi.org/10.1016/0550-3213(90)90432-D}{{\em Nucl.Phys.}
  {\bfseries B336} (1990) 349}.

\bibitem{PhysRevLett.24.1412}
O.~J. Heilmann and E.~H. Lieb, ``Monomers and dimers,''
  \href{http://dx.doi.org/10.1103/PhysRevLett.24.1412}{{\em Phys. Rev. Lett.}
  {\bfseries 24} (Jun, 1970) 1412--1414}.
  \url{http://link.aps.org/doi/10.1103/PhysRevLett.24.1412}.

\bibitem{Kurtze:1979zz}
D.~A. Kurtze and M.~E. Fisher, ``{Yang-Lee edge singularities at high
  temperatures},''
\href{http://dx.doi.org/10.1103/PhysRevB.20.2785}{{\em Phys.Rev.} {\bfseries
  B20} (1979) 2785--2796}.

\bibitem{Durhuus:2009sm}
B.~Durhuus, T.~Jonsson, and J.~F. Wheater, ``{On the spectral dimension of
  causal triangulations},''
\href{http://arxiv.org/abs/0908.3643}{{\ttfamily arXiv:0908.3643 [math-ph]}}.

\bibitem{DiFrancesco:1993nw}
P.~Di~Francesco, P.~H. Ginsparg, and J.~Zinn-Justin, ``{2-D Gravity and random
  matrices},'' \href{http://dx.doi.org/10.1016/0370-1573(94)00084-G}{{\em
  Phys.Rept.} {\bfseries 254} (1995) 1--133},
\href{http://arxiv.org/abs/hep-th/9306153}{{\ttfamily arXiv:hep-th/9306153
  [hep-th]}}.

\bibitem{Ambjorn:2008gk}
J.~Ambjorn, R.~Loll, Y.~Watabiki, W.~Westra, and S.~Zohren, ``{A New continuum
  limit of matrix models},''
  \href{http://dx.doi.org/10.1016/j.physletb.2008.11.003}{{\em Phys.Lett.}
  {\bfseries B670} (2008) 224--230},
\href{http://arxiv.org/abs/0810.2408}{{\ttfamily arXiv:0810.2408 [hep-th]}}.

\bibitem{Glaser:2012ej}
L.~Glaser, ``{Coupling Dimers to CDT},''
\href{http://arxiv.org/abs/1210.4063}{{\ttfamily arXiv:1210.4063 [hep-th]}}.

\bibitem{Atkin:2012yt}
M.~R. Atkin and S.~Zohren, ``{An Analytical Analysis of CDT Coupled to
  Dimer-like Matter},''
  \href{http://dx.doi.org/10.1016/j.physletb.2012.05.017}{{\em Phys.Lett.}
  {\bfseries B712} (2012) 445--450},
\href{http://arxiv.org/abs/1202.4322}{{\ttfamily arXiv:1202.4322 [hep-th]}}.

\bibitem{Ambjorn:2014voa}
J.~Ambjorn, B.~Durhuus, and J.~Wheater, ``{A restricted dimer model on a
  2-dimensional random causal triangulation},''
\href{http://arxiv.org/abs/1405.6782}{{\ttfamily arXiv:1405.6782 [hep-th]}}.

\bibitem{Ambjorn:2010hu}
J.~Ambjorn, A.~Gorlich, S.~Jordan, J.~Jurkiewicz, and R.~Loll, ``{CDT meets
  Horava-Lifshitz gravity},''
  \href{http://dx.doi.org/10.1016/j.physletb.2010.05.054}{{\em Phys.Lett.}
  {\bfseries B690} (2010) 413--419},
\href{http://arxiv.org/abs/1002.3298}{{\ttfamily arXiv:1002.3298 [hep-th]}}.

\bibitem{Horava:2009uw}
P.~Horava, ``{Quantum Gravity at a Lifshitz Point},''
  \href{http://dx.doi.org/10.1103/PhysRevD.79.084008}{{\em Phys.Rev.}
  {\bfseries D79} (2009) 084008},
\href{http://arxiv.org/abs/0901.3775}{{\ttfamily arXiv:0901.3775 [hep-th]}}.

\bibitem{Sotiriou:2010wn}
T.~P. Sotiriou, ``{Horava-Lifshitz gravity: a status report},''
  \href{http://dx.doi.org/10.1088/1742-6596/283/1/012034}{{\em
  J.Phys.Conf.Ser.} {\bfseries 283} (2011) 012034},
\href{http://arxiv.org/abs/1010.3218}{{\ttfamily arXiv:1010.3218 [hep-th]}}.

\bibitem{Arnowitt:1962hi}
R.~L. Arnowitt, S.~Deser, and C.~W. Misner, ``{The Dynamics of general
  relativity},''
\href{http://arxiv.org/abs/gr-qc/0405109}{{\ttfamily arXiv:gr-qc/0405109
  [gr-qc]}}.

\bibitem{Donnelly:2011df}
W.~Donnelly and T.~Jacobson, ``{Hamiltonian structure of Horava gravity},''
  \href{http://dx.doi.org/10.1103/PhysRevD.84.104019}{{\em Phys.Rev.}
  {\bfseries D84} (2011) 104019},
\href{http://arxiv.org/abs/1106.2131}{{\ttfamily arXiv:1106.2131 [hep-th]}}.

\bibitem{Loll:2005dr}
R.~Loll, W.~Westra, and S.~Zohren, ``{Taming the cosmological constant in 2-D
  causal quantum gravity with topology change},''
  \href{http://dx.doi.org/10.1016/j.nuclphysb.2006.06.033}{{\em Nucl.Phys.}
  {\bfseries B751} (2006) 419--435},
\href{http://arxiv.org/abs/hep-th/0507012}{{\ttfamily arXiv:hep-th/0507012
  [hep-th]}}.

\bibitem{Sotiriou:2009}
T.~P. Sotiriou, M.~Visser, and S.~Weinfurtner, ``Quantum gravity without
  lorentz invariance,'' {\em Journal of High Energy Physics} {\bfseries 2009}
  no.~10, (2009) 033. \url{http://stacks.iop.org/1126-6708/2009/i=10/a=033}.

\bibitem{Charmousis:2009tc}
C.~Charmousis, G.~Niz, A.~Padilla, and P.~M. Saffin, ``{Strong coupling in
  Horava gravity},''
  \href{http://dx.doi.org/10.1088/1126-6708/2009/08/070}{{\em JHEP} {\bfseries
  0908} (2009) 070},
\href{http://arxiv.org/abs/0905.2579}{{\ttfamily arXiv:0905.2579 [hep-th]}}.

\bibitem{Jordan:2013iaa}
S.~Jordan and R.~Loll, ``{De Sitter Universe from Causal Dynamical
  Triangulations without Preferred Foliation},''
  \href{http://dx.doi.org/10.1103/PhysRevD.88.044055}{{\em Phys.Rev.}
  {\bfseries D88} (2013) 044055},
\href{http://arxiv.org/abs/1307.5469}{{\ttfamily arXiv:1307.5469 [hep-th]}}.

\bibitem{Jordan:2013awa}
S.~Jordan and R.~Loll, ``{Causal Dynamical Triangulations without Preferred
  Foliation},''
\href{http://arxiv.org/abs/1305.4582}{{\ttfamily arXiv:1305.4582 [hep-th]}}.

\bibitem{PhysRevD.81.101502}
T.~Jacobson, ``Extended ho\ifmmode \check{r}\else \v{r}\fi{}ava gravity and
  einstein-aether theory,''
  \href{http://dx.doi.org/10.1103/PhysRevD.81.101502}{{\em Phys. Rev. D}
  {\bfseries 81} (May, 2010) 101502}.
  \url{http://link.aps.org/doi/10.1103/PhysRevD.81.101502}.

\bibitem{Laiho:2011zz}
J.~Laiho and D.~Coumbe, ``{Asymptotic safety and lattice quantum gravity},''
{\em PoS} {\bfseries LATTICE2011} (2011) 005.

\bibitem{Coumbe:2012qr}
D.~Coumbe and J.~Laiho, ``{Exploring the Phase Diagram of Lattice Quantum
  Gravity},'' {\em PoS} {\bfseries LATTICE2011} (2011) 334,
\href{http://arxiv.org/abs/1201.2864}{{\ttfamily arXiv:1201.2864 [hep-lat]}}.

\bibitem{Ambjorn:1993vz}
J.~Ambjorn, S.~Jain, and G.~Thorleifsson, ``{Baby universes in 2-d quantum
  gravity},'' \href{http://dx.doi.org/10.1016/0370-2693(93)90188-N}{{\em
  Phys.Lett.} {\bfseries B307} (1993) 34--39},
\href{http://arxiv.org/abs/hep-th/9303149}{{\ttfamily arXiv:hep-th/9303149
  [hep-th]}}.

\bibitem{Ambjorn:1993sy}
J.~Ambjorn, S.~Jain, J.~Jurkiewicz, and C.~Kristjansen, ``{Observing 4-d baby
  universes in quantum gravity},''
  \href{http://dx.doi.org/10.1016/0370-2693(93)90109-U}{{\em Phys.Lett.}
  {\bfseries B305} (1993) 208--213},
\href{http://arxiv.org/abs/hep-th/9303041}{{\ttfamily arXiv:hep-th/9303041
  [hep-th]}}.

\bibitem{Rindlisbacher:2013gka}
T.~Rindlisbacher and P.~de~Forcrand, ``{Euclidean Dynamical Triangulation
  revisited: is the phase transition really first order?},''
\href{http://arxiv.org/abs/1311.4712}{{\ttfamily arXiv:1311.4712 [hep-lat]}}.

\bibitem{Anderson:2011bj}
C.~Anderson, S.~J. Carlip, J.~H. Cooperman, P.~Horava, R.~K. Kommu, {\em
  et~al.}, ``{Quantizing Horava-Lifshitz Gravity via Causal Dynamical
  Triangulations},'' \href{http://dx.doi.org/10.1103/PhysRevD.85.044027,
  10.1103/PhysRevD.85.049904}{{\em Phys.Rev.} {\bfseries D85} (2012) 044027},
\href{http://arxiv.org/abs/1111.6634}{{\ttfamily arXiv:1111.6634 [hep-th]}}.

\bibitem{Ambjorn:2014gsa}
J.~Ambjorn, A.~Goerlich, J.~Jurkiewicz, A.~Kreienbuehl, and R.~Loll,
  ``{Renormalization Group Flow in CDT},''
\href{http://arxiv.org/abs/1405.4585}{{\ttfamily arXiv:1405.4585 [hep-th]}}.

\bibitem{Myrheim:1978ce}
J.~Myrheim,
``{STATISTICAL GEOMETRY},''.

\bibitem{meyer_dimension_1988}
D.~A. D.~A. Meyer, {\em The dimension of causal sets}.
\newblock Thesis, Massachusetts Institute of Technology, 1988.
\newblock \url{http://dspace.mit.edu/handle/1721.1/14328}.
\newblock Thesis (Ph. D.)--Massachusetts Institute of Technology, Dept. of
  Mathematics, 1989.

\bibitem{Sorkin:2003bx}
R.~D. Sorkin, ``{Causal sets: Discrete gravity},''
\href{http://arxiv.org/abs/gr-qc/0309009}{{\ttfamily arXiv:gr-qc/0309009
  [gr-qc]}}.

\bibitem{crownsets}
G.~Brightwell and P.~Winkler, ``Sphere orders,''
  \href{http://dx.doi.org/10.1007/BF00563524}{{\em Order} {\bfseries 6} no.~3,
  (1989) 235--240}. \url{http://dx.doi.org/10.1007/BF00563524}.

\bibitem{Rideout:2009zh}
D.~Rideout and P.~Wallden, ``{Emergence of spatial structure from causal
  sets},'' \href{http://dx.doi.org/10.1088/1742-6596/174/1/012017}{{\em
  J.Phys.Conf.Ser.} {\bfseries 174} (2009) 012017},
\href{http://arxiv.org/abs/0905.0017}{{\ttfamily arXiv:0905.0017 [gr-qc]}}.

\bibitem{Saravani:2014gza}
M.~Saravani and S.~Aslanbeigi, ``{On the Causal Set-Continuum
  Correspondence},''
\href{http://arxiv.org/abs/1403.6429}{{\ttfamily arXiv:1403.6429 [hep-th]}}.

\bibitem{Bombelli1989226}
L.~Bombelli and D.~A. Meyer, ``The origin of lorentzian geometry,''
  \href{http://dx.doi.org/http://dx.doi.org/10.1016/0375-9601(89)90474-X}{{\em
  Physics Letters A} {\bfseries 141} no.~5–6, (1989) 226 -- 228}.
  \url{http://www.sciencedirect.com/science/article/pii/037596018990474X}.

\bibitem{Major:2006hv}
S.~Major, D.~Rideout, and S.~Surya, ``{On Recovering continuum topology from a
  causal set},'' \href{http://dx.doi.org/10.1063/1.2435599}{{\em J.Math.Phys.}
  {\bfseries 48} (2007) 032501},
\href{http://arxiv.org/abs/gr-qc/0604124}{{\ttfamily arXiv:gr-qc/0604124
  [gr-qc]}}.

\bibitem{Henson:2006dk}
J.~Henson, ``{Constructing an interval of Minkowski space from a causal set},''
  \href{http://dx.doi.org/10.1088/0264-9381/23/4/L02}{{\em Class.Quant.Grav.}
  {\bfseries 23} (2006) L29--L35},
\href{http://arxiv.org/abs/gr-qc/0601069}{{\ttfamily arXiv:gr-qc/0601069
  [gr-qc]}}.

\bibitem{LV}
D.~Mattingly, ``Modern tests of lorentz invariance,'' 2005.
\newblock \url{http://www.livingreviews.org/lrr-2005-5}.

\bibitem{Bombelli:2006nm}
L.~Bombelli, J.~Henson, and R.~D. Sorkin, ``{Discreteness without symmetry
  breaking: A Theorem},''
  \href{http://dx.doi.org/10.1142/S0217732309031958}{{\em Mod.Phys.Lett.}
  {\bfseries A24} (2009) 2579--2587},
\href{http://arxiv.org/abs/gr-qc/0605006}{{\ttfamily arXiv:gr-qc/0605006
  [gr-qc]}}.

\bibitem{Sorkin:1990bj}
R.~D. Sorkin,
``{Space-time and causal sets},''.

\bibitem{Ahmed:2002mj}
M.~Ahmed, S.~Dodelson, P.~B. Greene, and R.~Sorkin, ``{Everpresent lambda},''
  \href{http://dx.doi.org/10.1103/PhysRevD.69.103523}{{\em Phys.Rev.}
  {\bfseries D69} (2004) 103523},
\href{http://arxiv.org/abs/astro-ph/0209274}{{\ttfamily arXiv:astro-ph/0209274
  [astro-ph]}}.

\bibitem{MR0369090}
D.~J. Kleitman and B.~L. Rothschild, ``Asymptotic enumeration of partial orders
  on a finite set,''
  \href{http://dx.doi.org/10.1090/S0002-9947-1975-0369090-9}{{\em Trans. Amer.
  Math. Soc.} {\bfseries 205} (1975) 205--220}.

\bibitem{0264-9381-15-11-009}
A.~Daughton, ``An investigation of the symmetric case of when causal sets can
  embed into manifolds,'' {\em Classical and Quantum Gravity} {\bfseries 15}
  no.~11, (1998) 3427. \url{http://stacks.iop.org/0264-9381/15/i=11/a=009}.

\bibitem{Dhar:1980}
D.~Dhar, ``Asymptotic enumeration of partially ordered sets,''
  \href{http://dx.doi.org/10.2140/pjm.1980.90.299}{{\em Pacific J. Math.}
  {\bfseries 90 (2)} (1980) 299--305}.

\bibitem{Sorkin:1998hi}
R.~D. Sorkin, ``{Indications of causal set cosmology},''
  \href{http://dx.doi.org/10.1023/A:1003629312096}{{\em Int.J.Theor.Phys.}
  {\bfseries 39} (2000) 1731--1736},
\href{http://arxiv.org/abs/gr-qc/0003043}{{\ttfamily arXiv:gr-qc/0003043
  [gr-qc]}}.

\bibitem{Martin:2000js}
X.~Martin, D.~O'Connor, D.~P. Rideout, and R.~D. Sorkin, ``{On the
  'renormalization' transformations induced by cycles of expansion and
  contraction in causal set cosmology},''
  \href{http://dx.doi.org/10.1103/PhysRevD.63.084026}{{\em Phys.Rev.}
  {\bfseries D63} (2001) 084026},
\href{http://arxiv.org/abs/gr-qc/0009063}{{\ttfamily arXiv:gr-qc/0009063
  [gr-qc]}}.

\bibitem{Ahmed:2009qm}
M.~Ahmed and D.~Rideout, ``{Indications of de Sitter Spacetime from Classical
  Sequential Growth Dynamics of Causal Sets},''
  \href{http://dx.doi.org/10.1103/PhysRevD.81.083528}{{\em Phys.Rev.}
  {\bfseries D81} (2010) 083528},
\href{http://arxiv.org/abs/0909.4771}{{\ttfamily arXiv:0909.4771 [gr-qc]}}.

\bibitem{Surya:2011du}
S.~Surya, ``{Evidence for a Phase Transition in 2D Causal Set Quantum
  Gravity},'' \href{http://dx.doi.org/10.1088/0264-9381/29/13/132001}{{\em
  Class.Quant.Grav.} {\bfseries 29} (2012) 132001},
\href{http://arxiv.org/abs/1110.6244}{{\ttfamily arXiv:1110.6244 [gr-qc]}}.

\bibitem{Philpott:2008vd}
L.~Philpott, F.~Dowker, and R.~D. Sorkin, ``{Energy-momentum diffusion from
  spacetime discreteness},''
  \href{http://dx.doi.org/10.1103/PhysRevD.79.124047}{{\em Phys.Rev.}
  {\bfseries D79} (2009) 124047},
\href{http://arxiv.org/abs/0810.5591}{{\ttfamily arXiv:0810.5591 [gr-qc]}}.

\bibitem{Johnston:2008za}
S.~Johnston, ``{Particle propagators on discrete spacetime},''
  \href{http://dx.doi.org/10.1088/0264-9381/25/20/202001}{{\em
  Class.Quant.Grav.} {\bfseries 25} (2008) 202001},
\href{http://arxiv.org/abs/0806.3083}{{\ttfamily arXiv:0806.3083 [hep-th]}}.

\bibitem{Afshordi:2012jf}
N.~Afshordi, S.~Aslanbeigi, and R.~D. Sorkin, ``{A Distinguished Vacuum State
  for a Quantum Field in a Curved Spacetime: Formalism, Features, and
  Cosmology},'' \href{http://dx.doi.org/10.1007/JHEP08(2012)137}{{\em JHEP}
  {\bfseries 1208} (2012) 137},
\href{http://arxiv.org/abs/1205.1296}{{\ttfamily arXiv:1205.1296 [hep-th]}}.

\bibitem{Sorkin:2007qi}
R.~D. Sorkin, ``{Does locality fail at intermediate length-scales},''
\href{http://arxiv.org/abs/gr-qc/0703099}{{\ttfamily arXiv:gr-qc/0703099
  [GR-QC]}}.

\bibitem{Dowker:2013vba}
F.~Dowker and L.~Glaser, ``Causal set d'alembertians for various dimensions,''
  {\em Classical and Quantum Gravity} {\bfseries 30} no.~19, (2013) 195016.
  \url{http://stacks.iop.org/0264-9381/30/i=19/a=195016}.

\bibitem{Benincasa:2010ac}
D.~M. Benincasa and F.~Dowker, ``{The Scalar Curvature of a Causal Set},''
  \href{http://dx.doi.org/10.1103/PhysRevLett.104.181301}{{\em Phys.Rev.Lett.}
  {\bfseries 104} (2010) 181301},
\href{http://arxiv.org/abs/1001.2725}{{\ttfamily arXiv:1001.2725 [gr-qc]}}.

\bibitem{Benincasa:2010as}
D.~M. Benincasa, F.~Dowker, and B.~Schmitzer, ``{The Random Discrete Action for
  2-Dimensional Spacetime},''
  \href{http://dx.doi.org/10.1088/0264-9381/28/10/105018}{{\em
  Class.Quant.Grav.} {\bfseries 28} (2011) 105018},
\href{http://arxiv.org/abs/1011.5191}{{\ttfamily arXiv:1011.5191 [gr-qc]}}.

\bibitem{Aslanbeigi:2014zva}
S.~Aslanbeigi, M.~Saravani, and R.~D. Sorkin, ``{Generalized Causal Set
  d'Alembertians},''
\href{http://arxiv.org/abs/1403.1622}{{\ttfamily arXiv:1403.1622 [hep-th]}}.

\bibitem{Roy:2012uz}
M.~Roy, D.~Sinha, and S.~Surya, ``{The Discrete Geometry of a Small Causal
  Diamond},'' \href{http://dx.doi.org/10.1103/PhysRevD.87.044046}{{\em
  Phys.Rev.} {\bfseries D87} (2013) 044046},
\href{http://arxiv.org/abs/1212.0631}{{\ttfamily arXiv:1212.0631 [gr-qc]}}.

\bibitem{dr-cactus}
G.~Allen, T.~Goodale, F.~L{\"o}ffler, D.~Rideout, E.~Schnetter, and E.~L.
  Seidel, ``Component specification in the cactus framework: The cactus
  configuration language,'' 2010.
\newblock \url{http://arxiv.org/abs/1009.1341}.

\bibitem{cactus2}
T.~Goodale, G.~Allen, G.~Lanfermann, J.~Mass, T.~Radke, E.~Seidel, and
  O.~Shalf, {\em The Cactus framework and toolkit: Design and applications},
  2003.

\bibitem{Rideout:1999ub}
D.~Rideout and R.~Sorkin, ``{A Classical sequential growth dynamics for causal
  sets},'' \href{http://dx.doi.org/10.1103/PhysRevD.61.024002}{{\em Phys.Rev.}
  {\bfseries D61} (2000) 024002},
\href{http://arxiv.org/abs/gr-qc/9904062}{{\ttfamily arXiv:gr-qc/9904062
  [gr-qc]}}.

\bibitem{Kiritsis:2009rx}
E.~B. Kiritsis and G.~Kofinas, ``{On Horava-Lifshitz 'Black Holes'},''
  \href{http://dx.doi.org/10.1007/JHEP01(2010)122}{{\em JHEP} {\bfseries 1001}
  (2010) 122},
\href{http://arxiv.org/abs/0910.5487}{{\ttfamily arXiv:0910.5487 [hep-th]}}.

\bibitem{Calcagni:2013vsa}
G.~Calcagni, A.~Eichhorn, and F.~Saueressig, ``{Probing the quantum nature of
  spacetime by diffusion},''
  \href{http://dx.doi.org/10.1103/PhysRevD.87.124028}{{\em Phys.Rev.}
  {\bfseries D87} no.~12, (2013) 124028},
\href{http://arxiv.org/abs/1304.7247}{{\ttfamily arXiv:1304.7247 [hep-th]}}.

\bibitem{Modesto:2008jz}
L.~Modesto, ``{Fractal Structure of Loop Quantum Gravity},''
  \href{http://dx.doi.org/10.1088/0264-9381/26/24/242002}{{\em
  Class.Quant.Grav.} {\bfseries 26} (2009) 242002},
\href{http://arxiv.org/abs/0812.2214}{{\ttfamily arXiv:0812.2214 [gr-qc]}}.

\bibitem{Alesci:2011ia}
E.~Alesci, T.~Thiemann, and A.~Zipfel, ``{Linking covariant and canonical LQG:
  New solutions to the Euclidean Scalar Constraint},''
  \href{http://dx.doi.org/10.1103/PhysRevD.86.024017}{{\em Phys.Rev.}
  {\bfseries D86} (2012) 024017},
\href{http://arxiv.org/abs/1109.1290}{{\ttfamily arXiv:1109.1290 [gr-qc]}}.

\bibitem{Regge:1961px}
T.~Regge, ``{GENERAL RELATIVITY WITHOUT COORDINATES},''
\href{http://dx.doi.org/10.1007/BF02733251}{{\em Nuovo Cim.} {\bfseries 19}
  (1961) 558--571}.

\end{thebibliography}\endgroup

\begin{appendix}

\chapter{\label{app:papers}Papers}
In this appendix, the papers on which this thesis is based are collected.

\section[New multicritical matrix models and multicritical 2d CDT]{\label{app:Dimer}New multicritical matrix models\\ and multicritical 2d CDT}
Jan Ambj{\o}rn, Lisa Glaser, Andrzej G{\"o}rlich, and Yuki Sato. \\
\newblock {\em Phys.Lett.}, B712:109--114, 2012.
\href{http://arxiv.org/abs/arXiv:1202.4435}{arXiv:1202.4435}

\section[A CDT Hamiltonian from H\texorpdfstring{\v{o}}{o}rava-Lifshitz gravity]{\label{app:HL}A CDT Hamiltonian \\from H\v{o}rava-Lifshitz gravity}
Jan Ambj{\o}rn, Lisa Glaser, Yuki Sato, and Yoshiyuki Watabiki.\\
\newblock {\em Phys.Lett.}, B722:172--175, 2013.
\href{http://arxiv.org/abs/arXiv:1302.6359}{arXiv:1302.6359}

\section[Euclidian 4d quantum gravity with a non-trivial measure term]{\label{app:EDT}Euclidian 4d quantum gravity\\ with a non-trivial measure term}
J.~Ambj{\o}rn, L.~Glaser, A.~G{\"o}rlich, and J.~Jurkiewicz.\\
\newblock {\em Journal of High Energy Physics}, 2013(10):1--24, 2013.
\href{http://arxiv.org/abs/arXiv:1307.2270}{arXiv:1307.2270}

\section[A closed form expression for the causal set d’Alembertian]{\label{app:Factor}A closed form expression \\for the causal set d’Alembertian}
Lisa Glaser.\\
\newblock {\em Classical and Quantum Gravity}, 31(9):095007, 2014.
\href{http://arxiv.org/abs/arXiv:1311.1701}{arXiv:1311.1701}

\section[Towards a definition of locality in a manifoldlike causal set]{\label{app:Local}Towards a Definition of Locality\\ in a Manifoldlike Causal Set}

Lisa Glaser and Sumati Surya.\\
\newblock {\em Phys.Rev.}, D88:124026, 2013.
\href{http://arxiv.org/abs/arXiv:1309.3403}{arXiv:1309.3403}

\end{appendix}

\thispagestyle{empty}
\blankpage

\vspace{350pt}
\begin{flushright}
\begin{minipage}[l]{0.7\textwidth}
\begin{shadequote}[c]{Faust II, Vers 11507 ff. / Faust}
Auf strenges Ordnen, raschen Flei\ss \\ Erfolgt der allersch\"{o}nste Preis; \\Dass sich das Werk vollende, \\ Gen\"{u}gt ein Geist f\"{u}r tausend H\"{a}nde.
\end{shadequote}
\blankpage
\end{minipage}
\end{flushright}

\end{document}